\documentstyle[12pt]{article}
\parskip=\medskipamount

\def\appendix#1{
\addtocounter{section}{1}
\setcounter{equation}{0}
\renewcommand{\thesection}{\Alph{section}}
\section*{Appendix \thesection\protect\indent #1}
\addcontentsline{toc}{section}{Appendix \thesection\ \ \ #1}
}
\newcommand{\tr}[1]{\,{\rm tr}\,#1\,}

\def\e{\varepsilon}

\def\be{\begin{equation}}
\def\la{\label}
\def\ee{\end{equation}}
\def\bea{\begin{eqnarray}}
\def\eea{\end{eqnarray}}
\def\eps{\varepsilon}
\def\a{\alpha}
\def\b{\beta}

\def\n{\nabla}
\def\S{\Sigma}

\def\g{\gamma}
\def\d{\delta}

\def\l{\left(}
\def\r{\right)}
\def\p{\partial}

\newcommand{\k}{{\kappa}}

\newcommand{\cL}{{\cal L }}
\newcommand{\N}{{\cal N }}
\begin{document}
\begin{titlepage}
\thispagestyle{empty}
\title{
\large{
\begin{flushright}
LMU-TPW 99-23\\
UAHEP996\\
hep-th/9912210
\end{flushright} }
\vskip 0.5cm
{\bf \huge{Scalar Quartic Couplings in Type IIB Supergravity on 
$AdS_5\times S^5$} }}  
\author{G.Arutyunov$^{a\, c}$
\thanks{arut@theorie.physik.uni-muenchen.de} \mbox{}
 and \mbox{} S.Frolov$^{b\,c}$
\thanks{frolov@bama.ua.edu 
\newline
$~~~~~$$^c$On leave of absence from 
Steklov Mathematical Institute, Gubkin str.8, GSP-1, 117966, Moscow, Russia
}
\vspace{0.4cm} \mbox{} \\
\small {$^a$ Sektion Physik,}
\vspace{-0.1cm} \mbox{} \\
\small {Universit\"at M\"unchen,}
\small {Theresienstr. 37,}
\vspace{-0.1cm} \mbox{} \\
\small {D-80333 M\"unchen, Germany}
\vspace{0.4cm} \mbox{} \\
\small {$^b$ Department of Physics and Astronomy,}
\vspace{-0.1cm} \mbox{} \\
\small {University of Alabama, Box 870324,}
\vspace{-0.1cm} \mbox{} \\
\small {Tuscaloosa, Alabama 35487-0324, USA}
\mbox{}
}
\date {}
\maketitle
\begin{abstract}
All quartic couplings of scalar fields $s^I$ that  
are dual to extended chiral primary operators 
in ${\cal N}=4$ SYM$_4$ are derived by using the covariant 
equations of motion for type IIB supergravity on $AdS_5\times S^5$.
It is shown that despite some expectations 
if one keeps the structure of the cubic terms untouched,
the quartic action
obtained contains terms with $two$ and $four$ derivatives. 
It is shown that the quartic action vanishes on shell in
the extremal case, e.g. $k_1=k_2+k_3+k_4$.
Consistency of the truncation of the quartic couplings to the 
massless multiplet of the ${\cal N}=8$, $d=5$ supergravity 
is proven and the explicit values of the couplings are found. 
It is argued that the consistency of the KK reduction implies 
non-renormalization of $n$-point functions of $n-1$ operators dual
to the fields from the massless multiplet and one operator dual to
a field from a massive multiplet. 
\end{abstract}
\end{titlepage}

\section{Introduction and Summary}
The AdS/CFT correspondence \cite{M,GKP,W} provides a powerful method
of studing correlation functions in conformal field theories, in particular,
in $D=4$, ${\cal N}=4$ supersymmetric Yang-Mills theory (SYM$_4$).
According to the proposal by \cite{GKP,W}, the generating
functional of Green functions in SYM$_4$
at large $N$ and at strong 't Hooft coupling $\lambda$ coincides with 
the on-shell value of the 
type IIB supergravity action on  $AdS_5\times S^5$. Thus,
the computation of an $n$-point Green function requires the knowledge of
the supergravity action up to the $n$-th order. 
In particular,  the quadratic \cite{AF3} and cubic actions 
\cite{LMRS,AF5,Lee} for physical fields of type IIB supergravity 
determine the normalization constants 
of two- and three-point Green functions \cite{AV}-\cite{DFMMR}. 
In principle, by using the quadratic action \cite{AF3} and 
the covariant equations of motion for type IIB supergravity \cite{S,SW,HW} 
one can easily compute $any$ three-point function of gauge invariant operators
in SYM$_4$ in the supergravity approximation, the problem
that can be hardly solved in perturbative SYM$_4$ even at the one-loop 
approximation.

The problem of computing four-point functions 
\cite{LT1}-\cite{HMMR} is obviously 
much more involved, and consists in general of two independent parts -- 
one first has to derive the relevant part of the supergravity action up 
to the fourth order, and then to find the minimum of the action 
that amounts to computing the corresponding exchange and contact
Feynman diagrams. However, as was pointed out in \cite{LT1}, in 
the simplest cases of massless modes of dilaton and 
axion fields, the relevant part of the supergravity action was known. 
Computing the corresponding 4-point functions was initiated in
\cite{LT1}, and completed in \cite{HFMMR}.  
Unfortunately, these modes correspond to rather complicated 
operators $\tr (F^2+\cdots )$ and $\tr (F\tilde{F} +\cdots )$,
and not much seems to be known about their 
four-point functions in perturbative SYM$_4$. Nevertherless, 
the analysis of the four-point functions of these operators performed in
\cite{HMMR} allows one to conclude that at strong 't Hooft
coupling all operators with large anomalous dimensions, which are
dual to massive string states, decouple.

The important and simplest operators in SYM$_4$ are single-trace
\footnote{Throughout the paper we refer to a single-trace CPO as
a CPO. Multi-trace operators may also belong to the same short 
representation of the supersymmetry algebra as was shown in \cite{AFer}.}
chiral primary operators (CPOs) \cite{LMRS} 
that are of the form $O_k^{I}=\tr (\phi^{(i_1}\cdots\phi^{i_k)})$.     
It is well known that all other operators in SYM$_4$ 
corresponding to type IIB supergravity fields are 
descendents of CPOs.  
 
Type IIB supergravity on $AdS_5\times S^5$
contains in its particle spectrum \cite{KRN,GM} scalar 
fields $s^I$ that are mixtures of the five form field strength on 
$S^5$ and the 
trace of the graviton on $S^5$. 
At the linear approximation to the supergravity, namely these scalars 
correspond to CPOs, as one can see from 
their transformation properties
with respect to the superconformal group of SYM$_4$.

Although the correlation functions of CPOs are the simplest 
ones to compute in SYM$_4$, the corresponding calculation in 
the supergravity approximation is nontrivial due to
the absence of a relevant action for the scalars $s^I$.
The quadratic and cubic actions for the scalars $s^I$ 
have been found and used to calculate all three-point functions of 
normalized CPOs in \cite{LMRS}.\footnote{To get rid of 
higher-derivative terms in the equations of motion for scalars 
$s^I$ a derivative-dependent  field redefinition 
was made in \cite{LMRS}. By this reason the scalars $s^I$ used in
\cite{LMRS} correspond not to CPOs but to extended CPOs  
involving  products of CPOs and their descendants \cite{AF5}. However,
three-point functions of the extended CPOs  coincide with 
the ones of CPOs for generic values of conformal dimensions of CPOs.} 

To compute four-point functions of CPOs one first has to know all 
cubic terms that involve two scalar fields $s^I$,
and the $s^I$-dependent quartic terms, and then to find the on-shell
value of the supergravity action. 

In the present paper as the first step in this direction we 
determine all necessary cubic and quartic terms 
by using the quadratic action \cite{AF3} and 
the covariant equations of motion for type IIB 
supergravity \cite{S,SW,HW}. 
It is clear that one can consider only the sector of 
type IIB supergravity 
that depends on the graviton and the four-form potential.
There are four different types of cubic vertices 
describing interaction 
of two scalars $s^I$
with $(i)$ symmetric tensor fields coming 
from the $AdS_5$ components
of the graviton, $(ii)$ with vector fields, $(iii)$ with scalar 
fields coming from the $S^5$ 
components of the graviton, and $(iiii)$ with 
scalar fields $t^I$ that are mixtures of the 
trace of the graviton on the sphere and the five 
form field strength on the sphere.
Although all the cubic terms were recently found in 
\cite{AF5,Lee}, we will see that the derivation done in the papers 
should be reconsidered. The reason is that this time, since we are
interested in $cubic$ corrections to
equations of motion for scalars $s^I$,
dealing with quadratic terms,
we have to take into account quadratic corrections to equations of 
motion for the gravity fields. 

Actually, it is straightforward, 
although cumbersome, to find
cubic corrections to equations of motion by decomposing the covariant 
equations of motion up to the third order, and keeping 
only relevant terms.
The main problem in deriving the quartic couplings 
comes from the fact that the equations of motion 
such obtained are $non-Lagrangian$, and one should perform a very
complicated and fine analysis to reduce the equations of motion to  
a Lagrangian form. In particular, although the original equations 
contain terms with $six$ derivatives, we will show that one can remove
these terms completely by means of a chain of field redefinitions,
and by using nontrivial identities between spherical harmonics of
different types. Even after removing the terms with six derivatives,
the resulting equations containing terms with four and two
derivatives (and without derivatives, of course) 
are still non-Lagrangian. To make the equations Lagrangian one should 
again redefine the scalar fields, and use the identities. 
Since any mistake in the computation would destroy the possibility of
obtaining Lagrangian equations of motion, we are pretty sure that
the quartic couplings we found are correct. 

The fact that the covariant equations of motion are non-Lagrangian, 
and one has to perform nontrivial field redefinitions to reduce them to
a Lagrangian form, explicitly shows that the gravity fields entering the covariant equations of motion $cannot$ correspond to $any$ operator
in SYM$_4$. The quartic action presented in the next Section is in fact
written for the scalar fields $s^I$ corresponding not to CPOs, but to
$extended$ CPOs, as was discussed in \cite{AF5}. In principle it seems 
possible to find an action for the scalars dual to 
CPOs by performing the field
redefinitions reversed to the ones used to reduce the equations of
motion to a Lagrangian form
\footnote{Note that the reversed transformations should be
made at the level of the quartic action, but not at the level of equations
of motion.}.
However, the resulting action for 
the new scalars will be much more complicated and will contain 
higher-derivative terms with six derivatives. 
It's worth noting that the equations of motion derived from the new action
certainly differ from the original ones despite the fact
that one made reversed transformations. 

We show that despite some expectations, if we keep the structure
of the cubic terms untouched,
the action obtained contains quartic terms with $two$ and $four$ 
derivatives, and there is no field redefinition allowing one to 
remove these terms. Thus, the problem of computing the four-point
functions of CPOs will require computing two new types of 
Feynman diagrams: $(i)$ exchange diagrams involving massive
tensor fields of second rank, and $(ii)$ contact diagrams
with four-derivative quartic vertices. All other necessary diagrams were 
computed in \cite{HF,HFMMR,HFR}. 

In our previous paper \cite{AF5} we argued that quartic couplings
of the scalars $s^I$ had to vanish in the extremal case when, say, $k_1=k_2+k_3+k_4$. This conjecture was based on the fact that 
all exchange Feynman diagrams vanished and 
contact Feynman diagrams had singularity in the extremal case, 
thus non-vanishing quartic couplings would contradict to
the AdS/CFT correspondence. Although the vanishing of the quartic 
couplings obtained in the present paper is not manifest, we show that 
this important property does take place after an additional field
redefinition. This means that 4-point extremal correlators
of extended CPOs vanish, and 
also implies the non-renormalization theorem \cite{DFMMR} 
for the corresponding extremal correlators of single-trace CPOs. 
It is clear that since the quartic couplings vanish then 
there should exist such a representation of the quartic couplings, 
that makes the vanishing explicit. We, however, have not looked for such
a representation yet.

The quartic couplings we found allow us to study the problem of
the consistency of the Kaluza-Klein (KK) reduction down to five dimensions.\footnote{
For a recent discussion of the consistency problem see 
\cite{NVN1,NVN2}, and references therein.}
It is customarily believed that the $S^5$ compactification of type IIB 
supergravity admits a consistent truncation to the massless multiplet, 
which can be identified with the field 
content of the gauged ${\cal N}=8$, $d=5$ supergravity \cite{PPN,GRW}.
Consistency means that there is no term linear in 
massive KK modes in the untruncated supergravity action, 
so that all massive KK fields can be put to zero without any
contradiction with equations of motion.
From the AdS/CFT correspondence point of view the consistent 
truncation implies that $any$ $n$-point correlation function 
of $n-1$ operators dual to the fields from the massless multiplet
and one operator dual to a massive KK field vanishes because, as one can 
easily see there is no exchange Feynman diagram in this case.   

It is obvious that
the cubic couplings found in \cite{AF5, Lee}
obey the consistency condition allowing therefore  
truncation to the fields from the massless multiplet at the level 
of the cubic action.   
In this paper we show that after an additional simple 
field redefinition
the quartic vertices we found indeed vanish 
when one of the four fields is not from the massless multiplet,
proving thereby the consistency of the reduction at the level of the  
quartic scalar couplings. This in particular provides an additional
argument that the scalars $s^I$ (and, in general, any supergravity field) 
correspond not to CPOs but rather to extended CPOs.
Indeed, if we assume that the consistent truncation takes place at
all orders in gravity fields, we get that correlators of the form
$\langle O_2^{I_1}O_2^{I_2}\cdots O_2^{I_{n-1}}O_k^{I_n}\rangle$ 
vanish for $k\ge 3$.
This is certainly not the case for single-trace CPOs, and we are forced
to conclude once more that supergravity fields are in general dual
to extended operators which are admixtures of single-trace operators 
and multi-trace ones.\footnote{Note that the lowest modes $s_2$ may be dual
only to single-trace CPOs. It is possible that 
any field from the massless supergravity multiplet is dual to a single-trace
operator.} 
Since an extended operator is uniquely determined
by a single-trace one, it is natural to assume that if
a correlation function of extended operators vanishes then
there exists a kind of a non-renormalization theorem for an analogous 
correlation function of single-trace operators. 
If we further assume that type IIB string theory on $AdS_5\times S^5$ 
respects the consistent truncation, then the vanishing of $n$-point 
correlation functions of $n-1$ extended operators dual to
the supergravity modes from the massless multiplet, and one 
extended operator dual to a massive KK mode seems to imply that 

{\it  at large $N$ the $n$-point functions of the 
corresponding single-trace 
operators are independent of 't Hooft coupling $\lambda =g_{YM}^2 N$.}

If the consistent truncation is valid at quantum level, that 
seems to be plausible because of a large amount of supersymmetry,
then these $n$-point functions are independent of $g_{YM}$ for any $N$.

In particular this conjecture is applied to 
$n$-point functions of $n-1$ CPOs $O_2$ and a CPO $O_4$. It would 
be interesting to check this in perturbation theory.
    
We also use the quartic couplings to find quartic action for
the scalars $s_2^I$ from the massless multiplet. 
The 4-derivative terms vanish in this case, and we arrive at an action 
with 2-derivative and non-derivative quartic couplings. We do not compare
the action obtained with the one of the gauged ${\cal N}=8$, $d=5$
supergravity on the $AdS_5$ background. This problem will be considered
together with the problem of computing 4-point functions of CPOs $O_2$
dual to the scalars from the massless multiplet in a latter paper.  

The plan of the paper is as follows. 
In Section 2 we recall
equations of motion for the graviton and the four-form potential, 
introduce notations, and represent the action obtained.
In Section 3 we
prove that there is the consistent reduction of type IIB 
supergravity on $AdS_5\times S^5$ down to five 
dimensions at the level of the quartic action. In Section 4 we show that 
the quartic couplings vanish in the extremal case.
In Section 5 we discuss the structure of the
cubic corrections to the equations of motion due to the
contributions of the gravity fields. In Section 6 we explain
what steps one should undertake to reduce the equations of motion to
a Lagrangian form. In Conclusion we disscuss unsolved problems. 
In Appendix A we present the values of the quartic 
couplings obtained, and in the Appendix B we 
summarize several important
identities involving spherical harmonics of different kinds. 
\section{Quartic Couplings}
\setcounter{equation}{0}
Quartic couplings of scalars $s^I$ may be derived from cubic
corrections to the covariant equations of motion \cite{S,SW,HW} 
for type IIB supergravity. Only the graviton and the four-form potential
give relevant contributions to cubic terms.  
The equations of motion of  the metric and the 4-form
potential are
\bea
F_{M_1...M_5}&=&\frac{1}{5!}\eps_{M_1...M_{10}}F^{M_6...M_{10}},
\la{ffe}\\
R_{MN}&=&\frac{1}{3!}F_{MM_1...M_4}F_N^{~~M_1...M_4}.
\la{gre}
\eea
Here $M,N,\ldots ,=0,1,\ldots 9$ and we use the 
following notation
$$F_{M_1\ldots M_5}=5\partial_{[M_1}A_{M_2\ldots M_5]}=
\partial_{M_1}A_{M_2\ldots M_5}+4~{\mbox{ terms}} ,$$
i.e all antisymmetrizations are with "weight"1. 
The dual forms are defined as
\bea
&&\e_{01\ldots 9}=\sqrt{-G},\quad \
e^{01\ldots 9}=-\frac{1}{\sqrt{-G}},
\nonumber\\
&&\e^{M_1\ldots M_{10}}=
G^{M_1N_1}\cdots G^{M_{10}N_{10}}\e_{N_1\ldots N_{10}},
\nonumber\\
&& (F^*)_{M_1\ldots M_k}=
\frac{1}{k!}\e_{M_1\ldots M_{10}}F^{M_{k+1}\ldots M_{10}}
=\frac{1}{k!}\e^{N_1\ldots N_{10}}G_{M_1N_1}\cdots
G_{M_{k}N_{k}}F_{N_{k+1}\ldots N_{10}}.
\nonumber
\eea
In the units in which the radius of $S^5$ is set to be unity, 
the $AdS_5\times S^5$ background solution looks as
\bea
&&ds^2=\frac{1}{x_0^2}(dx_0^2+\eta_{ij}dx^idx^j)+d\Omega_5^2=
g_{MN}dx^Mdx^N
\nonumber\\
&&R_{abcd}=-g_{ac}g_{bd}+g_{ad}g_{bc};
\quad R_{ab}=-4g_{ab}\nonumber\\
&&R_{\a\b\g\d}=g_{\a\g}g_{\b\d}-g_{\a\d}g_{\b\g};
\quad R_{\a\b}=4g_{\a\b}\nonumber\\
&&\bar{F}_{abcde}=\e_{abcde};\quad \bar{F}_{\a\b\g\d\e}=
\e_{\a\b\g\d\e},
\la{back}
\eea
where $a,b,c,\ldots $ and $\a ,\b ,\g ,\ldots$ are the AdS and 
the sphere indices respectively and $\eta_{ij}$ is the 
$4$-dimensional Minkowski metric.
We represent the gravitational field and the 4-form potential as
$$G_{MN}=g_{MN}+h_{MN};\quad A_{MNPQ}=\bar{A}_{MNPQ}+a_{MNPQ};
\quad F=\bar{F} +f.$$
The gauge symmetry of the equations of motion allows one to 
impose the de Donder gauge:
\bea
\n^\a h_{a\a }=\n^\a h_{(\a\b )}=\n^\a a_{M_1M_2M_3\a}=0;\quad  
h_{(\a\b )}\equiv h_{\a\b }-\frac{1}{5}g_{\a\b}h_\g^\g .
\la{ga}
\eea
This gauge choice does not remove all the gauge symmetry of the 
theory, for a detailed discussion of the residual symmetry 
see \cite{KRN}. As was shown in \cite{KRN}, the gauge 
condition (\ref{ga}) implies that the components of 
the 4-form potential of the form $a_{\a\b\g\d}$ and $a_{a\a\b\g}$ can be represented as follows:
\be
a_{\a\b\g\d}=\e_{\a\b\g\d\e}\n^\e b ;\quad
a_{a\a\b\g}=\e_{\a\b\g\d\e}\n^\d\phi_{a}^{\e}.
\la{phib}
\ee
It is also convenient to introduce the dual 1- and 2-forms for 
$a_{abcd}$ and $a_{abc\a}$:
\be
a_{abcd}=-\e_{abcde}Q^e;\quad
a_{abc\a}=-\e_{abcde}\phi_\a^{de}.
\la{aa}
\ee
Then the solution of the first-order self-duality equation
can be written as
\be
Q^a=\n^ab,\quad \phi_\a^{ab}=\n^{[a}\phi_\a^{b]}.
\la{qphi}
\ee
To write down the action for scalars $s^I$ that can be used to
compute 4-point correlation functions of CPOs in SYM$_4$ we need to
expand fields in spherical harmonics\footnote{Here and in what 
follows we suppose that the spherical harmonics of all types
are orthogonal with "weight" 1, i.e. $\int Y^IY^J=\d^{IJ}$, 
$\int Y_\a^IY_\a^J=\d^{IJ}$, $\int Y_{(\a\b )}^IY_{(\a\b )}^J=\d^{IJ}$, and summation over $\a ,\b$ is assumed. Namely this normalization was 
used in \cite{AF3}.}
\bea
&&h_\a^\a(x,y)=\sum\, \pi^{I_1}(x)Y^{I_1}(y);\quad 
b(x,y)=\sum\, b^{I_1}(x)Y^{I_1}(y),
\nonumber\\
&&h_{ab}(x,y)=\sum\, h_{ab}^{I_1}(x)Y^{I_1}(y); 
\quad \n_\b^2Y^k=-k(k+4)Y^k=-f(k)Y^k
\nonumber\\
&&h_{a\a}(x,y)=\sum\, h_a^{I_5}(x)Y_\a^{I_5}(y);\quad 
\phi_{a\a}(x,y)=\sum\, \phi_a^{I_5}(x)Y_\a^{I_5}(y); \nonumber\\
&&(\n_\b ^2-4)Y_\a^k=-(k+1)(k+3)Y_\a^k,
\nonumber\\
&&h_{(\a\b)}(x,y)=\sum\, \phi^{I_{14}}(x)Y_{(\a\b)}^{I_{14}}(y);\quad 
(\n_\g^2-10)Y_{(\a\b)}^k=-(k^2+4k+8)Y_{(\a\b)}^k,
\nonumber
\eea
We also need to make a number of fields redefinitions, the simplest ones
required to diagonalize the linear equations of motion are
\footnote{We often denote
$\pi^{I_1}$ as $\pi_k$ or as $\pi_1$, and a 
similar notation for other fields.}
\bea
&&\pi_k=10ks_k+10(k+4)t_k;\quad b_k=-s_k+t_k
\la{redbpi}\\
&&h_{ab}^k=\varphi_{ab}^k + g_{ab}\eta_k +\n_a\n_b\zeta_k ,
\la{redh}\\
&&\zeta_k =\frac{4}{k+1}s_k +\frac{4}{k+3}t_k .
\la{zeta}\\
&&\eta_k =-\frac{2k(k-1)}{k+1}s_k -\frac{2(k+4)(k+5)}{k+3}t_k
\la{eta}\\
&&A_a^k=h_a^k -4(k+3)\phi_a^k;\quad C_a^k=h_a^k +4(k+1)\phi_a^k
\la{redhphi}
\eea
Note that we use the off-shell shift of $h_{ab}$ (\ref{redh}) 
that was used to
find the quadratic action for type IIB supergravity on 
$AdS_5\times S^5$ in \cite{AF3}. It differs from the on-shell 
shift \cite{KRN} by higher-order terms.  

The field redefinitions that are needed to make the equations of 
motion Lagrangian and to remove 
higher-derivative terms from quadratic terms in the equations of 
motion will be discussed in the next Sections.

Then the action for the scalars $s^I$ may be written in the form
\bea
S(s)=\frac{4N^2}{(2\pi )^5}\int d^{5}x\sqrt{-g_a} ~ &\biggl (&
\cL_2(s)+\cL_2(t)+\cL_2(\phi )+
\cL_2(\varphi_{ab} )+\cL_2(A_a)+\cL_2(C_a)
\nonumber\\
&+&\cL_3(s)+\cL_3(t)+\cL_3(\phi )+\cL_3(\varphi_{ab} )
+\cL_3(A_a)+\cL_3(C_a)\nonumber\\
&+&\cL_4^{(0)}+\cL_4^{(2)}+\cL_4^{(4)}\biggr ) .
\la{action}
\eea
Here the quadratic terms are given by \cite{AF3}
\bea
\cL_2(s)&=&\sum ~ 
\frac{32k(k-1)(k+2)}{k+1}\left( -\frac12 \n_as_k\n^as_k
-\frac12 m^2s_k^2\right),
\la{as2}\\
\cL_2(t)&=&\sum ~ 
\frac{32(k+2)(k+4)(k+5)}{k+3}
\left( -\frac12 \n_at_k\n^at_k
-\frac12 m_t^2t_k^2\right), 
\la{at2}\\
\cL_2(\phi )&=&\sum ~ 
\left( -\frac14 \n_a\phi_k\n^a\phi_k -\frac14 f(k)\phi_k^2\right) ,
\la{aphi2}\\
\cL_2(\varphi_{ab})&=&\sum ~\left( 
-\frac{1}{4}\n_c\varphi_{ab}^k\n^c\varphi^{ab}_k+
\frac{1}{2}\n_a\varphi^{ab}_k\n^c\varphi_{cb}^k-
\frac{1}{2}\n_a\varphi_c^{ck}\n_b\varphi^{ba}_k
\right.\nonumber\\
&+&\frac{1}{4}\n_c\varphi_a^{ak}\n^c\varphi^{bk}_b
+\left.\frac{1}{4}(2-f(k))\varphi_{ab}^k\varphi^{ab}_k+
\frac{1}{4}(2+f(k))(\varphi_{a}^{ak})^2\right) ,
\la{agr2}\\
\cL_2(A_a)&=&\sum ~  
\frac{k+1}{2(k+2)}\left(-\frac14 (F_{ab}(A^k))^2
-\frac12 m_A^2(A_a^k)^2\right) ,
\la{aA2}\\
\cL_2(C_a)&=&\sum  ~
\frac{k+3}{2(k+2)}\left(-\frac14 (F_{ab}(C^k))^2
-\frac12 m_C^2(C_a^k)^2\right) ,
\la{aC2}
\eea
where the masses of the particles are 
\bea
&&m^2= k(k-4),\quad m_t^2=(k+4)(k+8), 
\quad m_\phi^2=m_\varphi^2=f(k)=k(k+4),\nonumber\\
&&m_A^2=k^2-1, \quad m_C^2= (k+3)(k+5)\nonumber
\eea
and $F_{ab}(A)=\p_aA_b-\p_bA_a$. 

The cubic terms were found in \cite{LMRS,AF5,Lee}, and may be written 
as follows
\bea
&&\cL_3(s)= S_{I_1I_2I_3}~s^{I_1}s^{I_2}s^{I_3},
\nonumber\\
\la{sss}
&&S_{I_1I_2I_3}=
a_{123}\frac{2^7\S ((\frac12 \S )^2-1)((\frac12 \S )^2-4) \a_1\a_2\a_3}
{3(k_1+1)(k_2+1)(k_3+1)},\\
\nonumber
&&\cL_3(t)=T_{I_1I_2I_3}~s^{I_1}s^{I_2}t^{I_3},\\
\la{sst}
&&T_{I_1I_2I_3}=
a_{123}\frac{2^7(\S+4)(\a_1+2)(\a_2+2)\a_3(\a_3-1)(\a_3-2)(\a_3-3)(\a_3-4)}
{(k_1+1)(k_2+1)(k_3+3)},\\
&&\cL_3(\phi )=\Phi_{I_1I_2I_3}~s^{I_1}s^{I_2}\phi^{I_3},\nonumber\\
\la{ssphi}
&&\Phi_{I_1I_2I_3}=\frac{4 p_{123}\S (\S+2)}
{(k_1+1)(k_2+1)}(\a_3-1)(\a_3-2),\\
&&\cL_3(\varphi_{ab})=G_{I_1I_2I_3}~\biggl ( 
\n^a s^{I_1}\n^b s^{I_2}\varphi_{ab}^{I_3} 
-\frac12 \left( \n^a s^{I_1}\n_a s^{I_2}+
\frac12 (m^2_1+m^2_2-f_3) s^{I_1}s^{I_2}\right)\varphi_c^{cI_3}\biggr ),\nonumber\\
\la{ssgr}
&&G_{I_1I_2I_3}=\frac{4(\S+2)(\S+4)\a_3(\a_3-1)}{(k_1+1)(k_2+1)} a_{123},\\
&&\cL_3(A_a)= A_{I_1I_2I_3}~
s^{I_1}\n^a s^{I_2}A_a^{I_3},\nonumber\\
\la{ssA}
&&A_{I_1I_2I_3}=
\frac{2(k_3+1)(\a_3-1/2)(\S-1)(\S+1)(\S+3)}
{(k_1+1)(k_2+1)(k_3+2)}t_{123},\\
&&\cL_3(C_a)= C_{I_1I_2I_3}~
s^{I_1}\n^a s^{I_2}C_a^{I_3},\nonumber\\
\la{ssC}
&&C_{I_1I_2I_3}=
\frac{8(k_3+3)(\a_3-1/2)(\a_3-3/2)(\a_3-5/2)(\S+3)}
{(k_1+1)(k_2+1)(k_3+2)}t_{123}.
\eea
Here the summation over $I_1,~I_2,~I_3$ is assumed, 
and we use the following notations
\bea
&&\a_1=\frac12 (k_2+k_3-k_1),\quad 
\a_2=\frac12 (k_1+k_3-k_2),\quad 
\a_3=\frac12 (k_1+k_2-k_3),\quad 
\S =k_1+k_2+k_3 ,\nonumber\\
&&a_{123}=\int~Y^{I_1}Y^{I_2}Y^{I_3},\quad
 p_{123}=\int~\n^\a Y^{I_1}\n^\b Y^{I_2}Y_{(\a\b )}^{I_3},\quad
t_{123}=\int~\n^\a Y^{I_1}Y^{I_2}Y_\a^{I_3},
\nonumber
\eea
and for any function $f_i\equiv f(k_i)$.

The quartic terms represent our main result and are given by
\bea
&&\cL_4^{(0)}=S_{I_1I_2I_3I_4}^{(0)}~s^{I_1}s^{I_2}s^{I_3}s^{I_4};
\la{ssss0}\\
&&\cL_4^{(2)}=
\l S_{I_1I_2I_3I_4}^{(2)}
+A_{I_1I_2I_3I_4}^{(2)}\r s^{I_1}\n_as^{I_2}s^{I_3}\n^as^{I_4},
\nonumber\\
&&S_{I_1I_2I_3I_4}^{(2)}=S_{I_2I_1I_3I_4}^{(2)}=S_{I_3I_4I_1I_2}^{(2)},
\la{ssss2}\\
&&A_{I_1I_2I_3I_4}^{(2)}=-A_{I_2I_1I_3I_4}^{(2)}=A_{I_3I_4I_1I_2}^{(2)};
\la{aaaa2}\\
&&\cL_4^{(4)}=
\l S_{I_1I_2I_3I_4}^{(4)}
+A_{I_1I_2I_3I_4}^{(4)}\r s^{I_1}\n_as^{I_2}\n_b^2(s^{I_3}\n^as^{I_4}),
\nonumber\\
&&S_{I_1I_2I_3I_4}^{(4)}=S_{I_2I_1I_3I_4}^{(4)}=S_{I_3I_4I_1I_2}^{(4)},
\la{ssss4}\\
&&A_{I_1I_2I_3I_4}^{(4)}=-A_{I_2I_1I_3I_4}^{(4)}=A_{I_3I_4I_1I_2}^{(4)}.
\la{aaaa4}
\eea
The explicit values of the quartic couplings are collected in Appendix A.
\section{Reduction to the gauged ${\cal N}=8$ 5-dimensional supergravity}
\setcounter{equation}{0}
There is much evidence that the $S^5$ compactification of the IIB 
supergravity admits a consistent truncation to the massless graviton multiplet, 
which can be identified with the field 
content of the gauged ${\cal N}=8$, $d=5$ supergravity \cite{PPN,GRW}.
Consistency means that there is no term linear in massive 
KK modes in the untruncated action so that all 
massive KK fields can be put to zero. As was noted in \cite{NVN1,NVN2}
the cubic couplings (\ref{sss})-(\ref{ssC}) 
obviously obey this condition allowing therefore the consistent 
truncation to the massless gravity multiplet.  

In this Section we show that after an additional field redefinition
the found quartic vertices (\ref{ssss0})-(\ref{aaaa4}) indeed vanish 
when one of the four fields is not from the massless multiplet,
proving thereby the consistency of the reduction at the level of the  
quartic scalar couplings. 

Recall that the gauged ${\cal N}=8$ five-dimensional supergravity 
has in particular
42 scalars with 20 of them forming the singlet of the global 
invariance group $SL(2,{\bf R})$. 
These 20 scalars comprise the ${\bf 20}$ irrep. of $SO(6)$ and 
correspond 
to the IIB supergravity fields\footnote{In this Section 
we use the explicit notation $s_k^I$ for a scalar transforming 
in $I=(0,k,0)$-irrep.} $s_k^I$ with $k=2$. 
The five-dimensional scalar Lagrangian consists of the kinetic 
energy and the potential. 
The maximal number of derivatives appearing in the Lagrangian is 
two and that is due to 
the non-linear sigma model type kinetic energy. We have however found 
the quartic 4-derivative vertices that can not be shifted away by any 
field redefinition. Thus, a highly non-trivial check of the relation
between the compactification of the ten-dimensional theory and the gauged
supergravity in five dimensions as well of the results obtained    
involves showing that the 4-derivative vertices vanish for the modes from 
the massless multiplet.

We start to analyse the consistency of the truncation with 
the quartic couplings of 4-derivative vertices and assume the fields 
$s_2^{I_2},s_2^{I_3},s_2^{I_4}$ 
belong to the massless multiplet. Upon substituting 
$k_2=k_3=k_4=2$ the couplings  $(A_0)^{(4)}_{I_1I_2,I_3I_4}$ and $(A_{-1})^{(4)}_{I_1I_2,I_3I_4}$
turn to zero. The other couplings are non-zero
and, therefore, the only possibility is that their sum should vanish. 
Moreover, according to the above discussion this vanishing should 
hold regardless of the fact if the remaing field $s_1$ is in a massive or 
in the massless graviton multiplet.

Among the couplings we consider there is a distinguished one, namely,
$(A_{t2})^{(4)}_{I_1I_2,I_3I_4}$ since it involves 
another type of $SO(6)$ tensors. To deal with this coupling we note that for $k_3=k_4=2$
there exists only two values of $k_5$ for which $t_{345}$ does not vanish,
namely, $k_5=1$ and  $k_5=3$. Then one represents
\bea
(f_5-1)^2t_{125}t_{345}=\biggl( (f_5-5)(f_5-21)
+24(f_5-1)-80\biggr) t_{125}t_{345},
\eea
so that $(f_5-5)(f_5-21)$ vanishes for both  $k_5=1$ and  $k_5=3$. 
By using relations (\ref{TT}) and (\ref{TTF}) the remaing terms in the last formula  
may be now reduced to involve the same type of the $SO(6)$ tensors as the rest
of the 4-derivative quartic couplings.

Finally, we sum up all couplings assuming that three
fields are from massless multiplet and the fourth one is $s_k^J$. 
The resulting expression $\L_4$ contains the tensor 
$(a_{JI_4I_5}a_{I_2I_3I_5}-a_{J I_3I_5}a_{I_2I_4I_5})$ 
as a multiplier\footnote{We do not assume here a summation 
over the index $I_5$}, which for given three fields from the massless multiplet
restricts a number of possible fields $s_k^I$ 
to a finite number. Namely, $k$ can be equal only to $2,4,6$.
The case $k=6$ is the most simple one, 
since in this case the only  value 
of $k_5$ for which the tensor does not vanish is $4$. Thus,
we can extend the summation index over the whole set and use the fact that 
\bea
\la{sym}
\sum_{I_5}a_{JI_2I_5}a_{I_3I_4I_5}
=\sum_{I_5}a_{JI_4I_5}a_{I_2I_3I_5}=\sum_{I_5}a_{J I_3I_5}a_{I_2I_4I_5}.
\eea
Hence, for $k=6$ the sum of the couplings vanishes.

If $k=4$ then there are two possible values of $k_5$: $k_5=2,4$. 
Evaluating $\L_4$ for these values of $k_5$ we find 
\bea
\nonumber
\L_4=\frac{128}{3}\sum_{k_5}(a_{JI_4I_5}a_{I_2I_3I_5}-a_{J I_3I_5}a_{I_2I_4I_5}),
\eea
where sum is over $k_5=2$ and $k_5=4$. Thus, for $k=4$ the sum $\L_4$
vanishes by virtue of (\ref{sym}).

Finally we have $k=2$ that allows for $k_5$ three values: $k_5=0,2,4$ and 
corresponds to the case when all fields are from the massless multiplet.
Substituting $k=2$ we find
\bea
\L_4=\frac{1}{324}\sum_{I_5}
(a_{JI_4I_5}a_{I_2I_3I_5}-a_{J I_3I_5}a_{I_2I_4I_5})(k_5-2)(k_5-4)f_5(k_5+6)(k_5+8).
\eea
For these values of $k_5$ the r.h.s. here   
vanishes identically. Thus, we have shown that there is no 
4-derivative linear couplings of the massive fields with the massless ones
and that the 4-derivative vertices are absent for  
fields from the massless multiplet.  

The analysis of the couplings of the 2-derivative and non-derivative vertices
proceeds in the same manner. 
For the $SO(6)$ tensors involving vector spherical harmonics one can use 
the formula 
\bea
(f_5-1)^3t_{125}t_{345}=\biggl( (f_5+23) (f_5-5)(f_5-21)+
496(f_5-1)-1920\biggr) t_{125}t_{345}
\eea
to make the nonreducible part vanishing when three of four fields are 
from the massless multiplet. 
For the $SO(6)$ tensors involving tensor spherical harmonics
the corresponding representations look as
\bea
f_5^2p_{125}p_{345}=\biggl( (f_5-12)^2
+24(f_5-12)+ 144\biggr)p_{125}p_{345}
\eea
\bea
f_5^3p_{125}p_{345}&=&\biggl( f_5(f_5-12)^2
+24f_5(f_5-12)+144(f_5-12)+1728\biggr) p_{125}p_{345}
\eea
and they are based on the fact that for $k_3=k_4=2$ tensor $p_{345}$ is nonzero 
only for $k_5=2$.

The relevant part of the quartic 
Lagrangian involving two derivatives can be written as follows 
\bea
\L_2 = \sum_{I_2,I_3,I_4}\biggl( S_{JI_2I_3I_4}
\n_a (s_k^Js_2^{I_2})\n^a (s_2^{I_3}s_2^{I_4}) +
A_{JI_2I_3I_4}
(\n_a s_k^Js_2^{I_2}- s_k^J\n_a s_2^{I_2})
(\n^as_2^{I_3}s_2^{I_4}-s_2^{I_3}\n^as_2^{I_4})\biggr)\nonumber
\eea

In this Section and throughout the paper we often use the following
notations
$$l=a_{125}a_{345},\quad m=a_{145}a_{235},\quad
n=a_{135}a_{245},$$ where we do not assume summation over the index 5.

Calculating the coefficients in $\L_2$, we see that they have the structure
$$S=S^l\cdot l+S^m\cdot m+S^n\cdot n,\quad
A=A^n\cdot (n-m)$$
Omitting total-derivative terms, taking into account the symmetry of 
the coefficients in $I_3,I_4$, and using linear equations of motion,
one can rewrite this Lagrangian in the form
\bea
\L_2 &=& \sum_{I_2,I_3,I_4}a_{JI_2I_5}a_{I_3I_4I_5}s_k^J\biggl( 
(-2S^l_{JI_2I_3I_4}+4A^n_{JI_2I_3I_4})
s_2^{I_2}\n^as_2^{I_3}\n^as_2^{I_4}\nonumber\\
&+& 
(-2S^m_{JI_2I_3I_4}-2S^n_{JI_2I_3I_4}-4A^n_{JI_2I_3I_4})
\n^as_2^{I_2}s_2^{I_3}\n^as_2^{I_4}\biggr)\nonumber\\
&+& \sum_{I_2,I_3,I_4}8S_{JI_2I_3I_4}s_k^Js_2^{I_2}s_2^{I_3}s_2^{I_4} 
\nonumber
\eea
To remove the 2-derivative terms, we make the shift
$$s_k^J\to s_k^J + \frac{1}{\k_k} J_{JI_2I_3I_4}s_2^{I_2}s_2^{I_3}s_2^{I_4}
 +  \frac{1}{\k_k} L_{JI_2I_3I_4}s_2^{I_2}s_2^{I_3}s_2^{I_4},$$
where 
$$ J=\frac12 (S^m+S^n+2A^n)\cdot l,\quad
 L=(-A^n+\frac16 (2S^l-S^m-S^n))\cdot l .
$$
Computing $L$ we see that $L$ is completely symmetric in $I_2,I_3,I_4$ for
$k=4,6$. After the shift all 2-derivative terms are removed and we get 
the following Lagrangian
\bea
\L_2=
\sum_{I_2,I_3,I_4}\l 8S_{JI_2I_3I_4}
+(-12-m_k^2)(J_{JI_2I_3I_4}+L_{JI_2I_3I_4})\r 
s_k^Js_2^{I_2}s_2^{I_3}s_2^{I_4}. 
\nonumber
\eea
This Lagrangian should be summed up with the non-derivative terms 
of the quartic Lagrangian whose couplings will be denoted by 
$S^{(0)}_{JI_2I_3I_4}$. Thus the complete Lagrangian for 
non-derivative terms is given by
\bea
\L_0=
\sum_{I_2,I_3,I_4}\l 4S^{(0)}_{JI_2I_3I_4}
+8S_{JI_2I_3I_4} +(-12-m_k^2)(J_{JI_2I_3I_4}+L_{JI_2I_3I_4})\r
s_k^Js_2^{I_2}s_2^{I_3}s_2^{I_4} 
\nonumber
\eea
One can easily check that for $k=6$ this Lagrangian vanishes, because 
in this case $k_5$ can be equal only to 4.

In the case of $k=4$ there are two possible values of $k_5$: $k_5=2,4$, 
and we get at first sight a nonzero result which has 
the form
$$\L_0=\biggl(\alpha \sum_{k(I_5)=2}a_{JI_2I_5}a_{I_3I_4I_5} +
\b \sum_{k(I_5)=4}a_{JI_2I_5}a_{I_3I_4I_5}\biggr)
s_k^Js_2^{I_2}s_2^{I_3}s_2^{I_4}.$$

However, now we can use identity (\ref{imre}) to show that
$$\sum_{k(I_5)=4}a_{JI_2I_5}a_{I_3I_4I_5}s_4^{J}s_2^{I_2}s_2^{I_3}s_2^{I_4} =
\frac87 \sum_{k(I_5)=2}a_{JI_2I_5}a_{I_3I_4I_5}s_4^{J}s_2^{I_2}s_2^{I_3}s_2^{I_4}$$
Taking into account this relation we find that non-derivative 
Lagrangian $\L_0$ vanishes too.

Thus we have shown that at least at the level of the quartic action for scalars
$s^I$ there is a consistent dimensional reduction of the type IIB supergravity to 
the gauged supergravity on the $AdS_5$ background. We would like to stress again that this reduction  
requires non-trivial redefinitions of the fields. 

Now we can compute the quartic couplings for the case when all four fields are 
from the massless multiplet. Summing up the nonvanishing quartic couplings of the two-derivative vertices 
we find out that the answer contains the terms involving the $SO(6)$ tensors 
of the form $a_{I_1I_3I_5}a_{I_2I_4I_5}$ and  $a_{I_1I_4I_5}a_{I_2I_3I_5}$. Integrating 
by parts it 
is possible to convert the tensor indices to the normal order, namely to $a_{I_1I_2I_5}a_{I_3I_4I_5}$
This also leads to the additional contributions to the quartic couplings of the non-derivative vertices.
Recall that $k(I_5)$ runs now the set $0,2,4$.

Again it is useful to note that identity (\ref{imre}) implies a number of relations 
between the Lagrangian terms involving tensors $a_{125}a_{345}$. In particular, when all the fields
are from the massless multiplet one finds the relation
$$
\sum_{k(I_5)=4}a_{I_1I_2I_5}a_{I_3I_4I_5}s_2^{I_1}s_2^{I_2}s_2^{I_3}s_2^{I_4} =
\frac14\sum_{k(I_5)=2}a_{I_1I_2I_5}a_{I_3I_4I_5}s_2^{I_1}s_2^{I_2}s_2^{I_3}s_2^{I_4}
+\sum_{k(I_5)=0}a_{I_1I_2I_5}a_{I_3I_4I_5}s_2^{I_1}s_2^{I_2}s_2^{I_3}s_2^{I_4}.
$$
Actually there is no sum over $k(I_5)=0$ since in this case $I_5$ is just the trivial representation.
Analogously, multiplying both sides of (\ref{imre}) by $s_2^{I_1}\n_a s_2^{I_2} s_2^{I_3}\n^as_2^{I_4}$ 
and then integrating by parts and using the previous relation 
one obtains the following formula:
\bea
\sum_{k(I_5)=0,2,4}a_{I_1I_2I_5}a_{I_3I_4I_5}s_2^{I_1}\n_as_2^{I_2}s_2^{I_3}\n^as_2^{I_4}&=&
\nonumber
\frac{8}{3}\sum_{k(I_5)=0}a_{I_1I_2I_5}a_{I_3I_4I_5}s_2^{I_1}s_2^{I_2}s_2^{I_3}s_2^{I_4}\\
\nonumber
&+&
\frac{5}{3}\sum_{k(I_5)=2}a_{I_1I_2I_5}a_{I_3I_4I_5}s_2^{I_1}s_2^{I_2}s_2^{I_3}s_2^{I_4}.
\eea
These relations allow one to exclude from the Lagrangian for the 
scalar fields from 
the massless multiplet  the contributions 
of the representations $I_5$ with $k(I_5)=4$.

In this way we find the following values of the  quartic couplings 
of the 2-derivative vertex 
\bea
\la{AdS52}
\cL^{(2)}_{AdS5}&=&\frac{5^2\cdot 2^9}{27}
\sum_{k(I_5)=2}a_{I_1I_2I_5}a_{I_3I_4I_5}\n_a(s_2^{I_1}s_2^{I_2})\n^a(s_2^{I_3}s_2^{I_4})\\
\nonumber
&+&\frac{2^{13}}{27}
\sum_{k(I_5)=0}a_{I_1I_2I_5}a_{I_3I_4I_5}\n_a(s_2^{I_1}s_2^{I_2})\n^a(s_2^{I_3}s_2^{I_4})
\eea
and of the non-derivative vertex
\bea
\la{AdS50}
\cL^{(0)}_{AdS5}=-\frac{5^2\cdot 2^{11}}{9}
\sum_{k(I_5)=2}a_{I_1I_2I_5}a_{I_3I_4I_5}s_2^{I_1}s_2^{I_2}s_2^{I_3}s_2^{I_4}.
\eea  
Note that the contribution of the trivial representation completely dissappears 
from the non-derivative quartic coupling.

The quartic action can be further simplified by substituting the integrals of spherical harmonics
for their explicit value. By using (\ref{aexp}) one gets 
\bea
\nonumber
\sum_{k(I_5)=2}a_{I_1I_2I_5}a_{I_3I_4I_5}
=\frac{2^5\cdot 3}{5^2\pi^3}\sum_{I_5}C^{I_1I_2I_5}C^{I_3I_4I_5},
\eea
where $C^{I_1I_2I_3}=\langle C^{I_1}C^{I_2}C^{I_3}\rangle $.
One can easily establish the following summation formula 
$$
\sum_{I}C^I_{ij}C^I_{kl}=\frac{1}{2}\d_{ik}\d_{jl}+\frac{1}{2}\d_{il}\d_{jk}-\frac{1}{6}\d_{ij}\d_{kl}
$$
that steams from the fact that the l.h.s. of the expression above is a fourth rank tensor 
of $SO(6)$,
symmetric and traceless both in $(ij)$ and $(kl)$ indices with the normalization
condition $C^{I}_{ij}C_{ij}^I=20$. Applying this formula one gets
\bea
\la{sumC}
\sum_{k(I_5)=2}a_{I_1I_2I_5}a_{I_3I_4I_5}
=\frac{2^4\cdot 3}{5^2\pi^3}\l C^{I_1I_2I_3I_4}+C^{I_1I_2I_4I_3}-\frac{1}{3}\d^{I_1I_2}\d^{I_3I_4}\r ,
\eea 
where the shorthand notation 
$C^{I_1I_2I_3I_4}=C^{I_1}_{i_1i_2}C^{I_2}_{i_2i_3}C^{I_3}_{i_3i_4}C^{I_4}_{i_4i_1}$
for the trace product of four matrices $C^{I}$ was introduced.
It remains to note that for $k_5=0$
the normalization condition for scalar spherical harmonics gives 
$\sum_{k(I_5)=0}a_{I_1I_2I_5}a_{I_3I_4I_5}=\frac{1}{\pi^3}\d^{I_1I_2}\d^{I_3I_4}$. 
By exploting this formula together with (\ref{sumC}) the two-derivative Lagrangian may be reduced
to the following form:
\bea
\la{AdS52m}
\cL^{(2)}_{AdS5}&=&\frac{2^{14}}{9\pi^3}
C_{I_1I_2I_3I_4}\n_a(s_2^{I_1}s_2^{I_2})\n^a(s_2^{I_3}s_2^{I_4})
\eea 
One can also introduce the fields $s_{ij}\equiv C^I_{ij}s^I$ that
provide the natural parametrization of the coset space 
$SL(6,{\bf R})/SO(6)$.
Although it is clear that namely this form of the action 
should be compared 
to the one of the gauged ${\cal N}=8$, $d=5$ supergravity, we will 
not do this here. 
We only note that the comparison requires 
to perform the field redefinitions of the type 
$s^{I_1}\to s^{I_1}+j_{I_1I_2I_3I_4}s^{I_2}s^{I_3}s^{I_4}$ to convert 
the Lagrangian terms with two derivatives to the form (\ref{AdS52m}).
\section{Quartic couplings in the extremal case}
\setcounter{equation}{0}
In our previous paper \cite{AF5} we conjectured that  
quartic couplings of scalars $s^I$ vanish 
in the extremal case when, say, $k_1=k_2+k_3+k_4$. In this Section we
show that the quartic couplings we found do satisfy the property
after an additional shift of the fields.
In principle by using the shift one can find 
such a representation of the quartic couplings, that makes 
the vanishing explicit.

The vanishing of the quartic couplings means that correlation
functions of extended CPOs vanish in the extremal case \cite{AF5} and   
also implies the non-renormalization theorem \cite{DFMMR} 
for the corresponding extremal correlators of single-trace CPOs. 

To prove the vanishing we find convenient to use different 
4-derivative vertices. Namely, one can easily show that the following 
relations are valid on-shell
\bea
A^{(4)}_{1234}\int s_1\n_a s_2 \n_b^2(s_3\n^a s_4)&=&
-2A^{(4)}_{1234}\int \n_a s_1\n_b s_2 \n^a s_3\n^b s_4 \\
\nonumber
&-&4A^{(4)}_{1234} \int s_1\n_a s_2 s_3\n^a s_4\\\nonumber
&-&\frac{1}{4}A^{(4)}_{1234}(m_1^2-m_2^2)(m_3^2-m_4^2) 
\int s_1 s_2 s_3 s_4.\\\nonumber
S^{(4)}_{1234}\int s_1\n_a s_2 \n_b^2(s_3\n^a s_4)&=&
-S^{(4)}_{1234}\int \n_a s_1\n^a s_2 \n_b s_3\n^b s_4 \\
\nonumber
&+&S^{(4)}_{1234}\l m_1^2+m_2^2+m_3^2+m_4^2-4 \r
\int s_1\n_a s_2 s_3\n^a s_4 \\
\la{1der}
&+&\frac{1}{4}S^{(4)}_{1234}(m_1^2+m_2^2)(m_3^2+m_4^2) 
\int s_1 s_2 s_3 s_4.
\eea
Thus we replace all 4-derivative vertices by the ones with only 
one derivative on each field. This representation has also the advantage
that in this case the Hamiltonian reformulation of the quartic action
is straightforward, and, therefore, as was shown in \cite{AF1}, there is 
no need to add boundary terms. 

We assume for definiteness that $k_1=k_2+k_3+k_4$. It is easy to
show, by using the description of spherical harmonics as restrictions
of functions, vectors and tensors on the ${\bf R}^6$ in which the sphere $S^5$ 
is embedded \cite{LMRS,AF5}, that the tensors (we do not assume
summation over $I_5$ here)
$$t_{125}t_{345},\quad\mbox{and}\quad p_{125}p_{345}$$ 
vanish in the extremal case, and that the tensors $a_{125}a_{345}$,
$a_{135}a_{245}$ and $a_{145}a_{235}$
differ from zero only if $k_5=k_3+k_4$, $k_5=k_2+k_4$, $k_5=k_2+k_3$
respectively. Thus in all vertices we 
can replace $k_5$ by a corresponding function of $k_2,k_3,k_4$, and,
then the only dependence on $k_5$ is in tensors 
$a_{125}a_{345}$, $a_{135}a_{245}$ and $a_{145}a_{235}$ which
are obviously symmetric in 2, 3, 4.

We single out the field $s^{I_1}$ and write the relevant part
of the quartic 4-derivative vertices in the form
\bea
\L_{ext}^{(4)} = 4\sum_{I_2,I_3,I_4}\biggl( -S_{I_1I_2I_3I_4}^{(4)}
-A_{I_1I_3I_2I_4}^{(4)}+A_{I_1I_4I_3I_2}^{(4)}\biggr)
\n_a s^{I_1}\n^as^{I_2}\n_bs^{I_3}\n^bs^{I_4}, \nonumber
\eea
where we sum over the representations satisfying the extremality
condition. Now, we substitute the values of $k_5$ discussed above, and
$k_1=k_2+k_3+k_4$ in the quartic couplings, and obtain zero.

To analyze 2-derivative terms we represent the 2-derivative Lagrangian
as follows
\bea
\L_{ext}^{(2)} &=& 4\sum_{I_2,I_3,I_4}\biggl( 
\l -\frac12\tilde{S}_{I_1I_3I_2I_4}^{(2)}
+\tilde{A}_{I_1I_2I_3I_4}^{(2)}\r 
s^{I_1}\n^as^{I_2}s^{I_3}\n_as^{I_4} \nonumber\\
&+&\frac14\l \tilde{A}_{I_1I_2I_3I_4}^{(2)}(m_4^2-m_3^2)-
\tilde{S}_{I_1I_2I_3I_4}^{(2)}(m_4^2+m_3^2)\r 
s^{I_1}s^{I_2}s^{I_3}s^{I_4}\biggr) ,
\nonumber
\eea
where using (\ref{1der}) we define
\bea
\tilde{S}_{I_1I_2I_3I_4}^{(2)}&=&S_{I_1I_2I_3I_4}^{(2)}
+S^{(4)}_{I_1I_2I_3I_4}\l m_1^2+m_2^2+m_3^2+m_4^2-4 \r ,\nonumber\\
\tilde{A}_{I_1I_2I_3I_4}^{(2)}&=&A_{I_1I_2I_3I_4}^{(2)}
-4A^{(4)}_{I_1I_2I_3I_4}.
\nonumber
\eea
This time substituting $k_5$ and $k_1$ and symmetrizing the expression 
obtained in $I_2$ and $I_4$, we get a non-zero function which is, however,
completely symmetric in $I_2$, $I_3$ and $I_4$. Thus we can remove the
2-derivative term by using the shift
$$
s^{I_1}\to s^{I_1} -\frac{2}{3\kappa_1} \l 
-\frac12\tilde{S}_{I_1I_3I_2I_4}^{(2)}
+\tilde{A}_{I_1I_2I_3I_4}^{(2)}\r s^{I_2}s^{I_3}s^{I_4} .$$
This shift also produces an additional contribution to the non-derivative 
terms which is equal to
$$ -\frac23 \l 
-\frac12\tilde{S}_{I_1I_3I_2I_4}^{(2)}
+\tilde{A}_{I_1I_2I_3I_4}^{(2)}\r (m_2^2+m_3^2+m_4^2-m_1^2)
s^{I_1}s^{I_2}s^{I_3}s^{I_4}.
$$
After accounting this contribution the non-derivative terms aquire the form
\bea
\L_{ext}^{(0)} &=& 4\sum_{I_2,I_3,I_4}\biggl( S_{I_1I_2I_3I_4}^{(0)}-
\frac16
\l -\frac12\tilde{S}_{I_1I_3I_2I_4}^{(2)}
+\tilde{A}_{I_1I_2I_3I_4}^{(2)}\r (m_2^2+m_3^2+m_4^2-m_1^2) \nonumber\\
&+&\frac14\l \tilde{A}_{I_1I_2I_3I_4}^{(2)}(m_4^2-m_3^2)-
\tilde{S}_{I_1I_2I_3I_4}^{(2)}(m_4^2+m_3^2)\r \cr
&+&\frac{1}{4}S^{(4)}_{1234}(m_1^2+m_2^2)(m_3^2+m_4^2)
-\frac{1}{4}A^{(4)}_{1234}(m_1^2-m_2^2)(m_3^2-m_4^2) 
\biggr)
s^{I_1}s^{I_2}s^{I_3}s^{I_4}.
\nonumber
\eea
Substituting $k_5$ and $k_1$ and symmetrizing the coefficient 
obtained in $I_2$, $I_3$ and $I_4$ we end up with zero.
\section{Equations of motion}
\setcounter{equation}{0}
The equations of motion that follow from the action (\ref{action}) 
have the form
\bea
&&\k_1(\n_a^2-m_1^2)s_1=-3S_{125}s_2s_5-2T_{125}s_2t_5  
-2\Phi_{125}s_2\phi_5\nonumber\\
&&\qquad+G_{125}\left( 2\n^a(\n^bs_2\varphi_{ab}^5)-\n^a(\n_as_2\varphi^c_{c5})
+\frac12 (m_1^2+m_2^2-f_5)s_2\varphi^c_{c5})\right)\nonumber\\
&&\qquad-A_{125}(2\n_as_2 A^a_5+s_2\n_aA^a_5)-
C_{125}(2\n_as_2 C^a_5+s_2\n_aC^a_5) - \frac{\d S_4}{\d s_1},
\la{eqs}\\
&&\frac{32(k_5+2)(k_5+4)(k_5+5)}{k_5+3}
\left( \n_a^2t_5-m_t^2t_5 \right)=-T_{125}s_1s_2,
\la{eqt}\\
&&\n_a^2\phi_5 - m_\phi^2\phi_5 =-2\Phi_{125}s_1s_2,
\la{eqphi}\\
&&Eq_{ab}(\varphi)\equiv -\frac{1}{2}\n_c^2\varphi_{ab}^5+
\frac{1}{2}\n_a\n^c\varphi_{cb}^5
+\frac{1}{2}\n_b\n^c\varphi_{ca}^5+(\frac{1}{2}f_5-1)\varphi_{ab}^5\\
\nonumber
&&\qquad-\frac{1}{2}g_{ab}\n_c\n_d\varphi^{cd}_5
-\frac{1}{2}\n_a\n_b \varphi_{c5}^c
+\frac{1}{2}g_{ab}\n_c^2\varphi_{d5}^d
-g_{ab}(\frac{1}{2}f_5+1)\varphi_{c5}^c\nonumber\\
&&\qquad=G_{125}\l \n_a s_1\n_b s_2-\frac12 g_{ab}\n_c s_1\n^c s_2 
-\frac14 g_{ab}(m^2_1+m^2_2-f_5)s_1 s_2 \r ,
\la{eqgr}\\ 
&&\frac{k_5+1}{2(k_5+2)}\left(\n_b^2A^5_a-\n^b\n_aA_b^5-m_A^2A_a^5\right)
=-A_{125}s_1\n_as_2 ,
\la{eqA}\\
&&\frac{k_5+3}{2(k_5+2)}\left(\n_b^2C^5_a-\n^b\n_aC_b^5-m_C^2C_a^5\right)
=-C_{125}s_1\n_as_2 .
\la{eqC}
\eea
Here $\k\equiv \frac{32k(k-1)(k+2)}{k+1}$, the summation over 
2 and 5 is assumed, and the masses of all fields except $s$ depend on
$k_5$. 

To obtain these equations of motion from the covariant equations
(\ref{ffe}) and (\ref{gre})  we first need to decompose them up to the 
third order, and then to perform a number of fields redefinitions 
to make the equations Lagrangian. It is convenient to begin by
considering quadratic terms in the covariant equations because 
as we will see they also give contributions to cubic terms.
It is also useful to single out contributions coming from different 
fields. 
\subsection{Contribution of scalars $s$}
We begin by considering the contribution of the scalars $s$ coming from
quadratic terms in the equations of motion for $s$ that are obtained from
the covariant equations (\ref{ffe}) and (\ref{gre}).
Decomposing these equations up to the second order in fields, and keeping
only terms quadratic in $s$, one can represent the equation for $s$ 
in the following form\footnote{We do not present the explicit values
of the coefficients here and below because they are pretty complicated
and not very instructive.} 
\bea
\nonumber
(\n_a^2-m_1^2)s_1&=&D^s_{125}s_2s_5+
E^s_{125}\n_as_2 \n^a s_5+
F^s_{125}\n_a\n_bs_2 \n^a\n^b s _5 \\
&+&R^s_{125}s_2(\n_b^2-m_5^2)s_5
+T^s_{125}\n_as_2\n^a (\n_b^2-m_5^2)s_5
\la{eqss}
\eea
We see that the r.h.s of (\ref{eqss}) contains terms proportional
to linear part of the equations of motion: $(\n_b^2-m^2_5)s_5$. 
Although such terms do not give contributions to quadratic terms,
and by this reason were neglected in \cite{LMRS}, they $do$ contribute
to cubic terms. To remove the higher-derivative terms on the 
first line of (\ref{eqss}) one should make
the field redefinition \cite{LMRS}
\bea
s_1={s'}_1+J^s_{125}s'_2s'_5+L^s_{125}\n_as'_2\n^as'_5,
\la{reds1}
\eea
where
\bea
&&2L^s_{125}=F^s_{125},\quad 
2J^s_{125}+L^s_{125}(m_2^2+m_5^2-m_1^2-8)=E^s_{125}.
\nonumber
\eea
Then (\ref{eqss}) takes the form
\bea
(\n_a^2-m_1^2)s'_1=V^s_{125}s'_2s'_5 + cubic~ terms,
\nonumber
\eea
where
\bea
V^s_{125}=D^s_{125}-J^s_{125}(m_2^2+m_5^2-m_1^2)=-\frac{3}{\k_1}S_{125}.
\la{Vs}
\eea
To simplify the form of the cubic terms we make an additional 
shift\footnote{Here and in what follows we omit the primes on
redefined fields.}
\bea
s_1\to {s}_1&+&F^s_{125}\n^as_2 \n_a (J^s_{534}s_3s_4+
L^s_{534}\n_a s_3\n^a s_4)\nonumber\\
&+&2J^s_{125}J^s_{534}s_2s_3s_4+
2J^s_{125}L^s_{534}s_2\n^a s_3\n_a s_4
\nonumber
\eea
and represent the equation in the form
\bea
(\n_a^2-m_1^2)s_1&=&V^s_{123}s_2s_3+(s0)_{1234}s_2s_3s_4+
(s2a)_{1234}\n^as_2s_3\n_a s_4+
(s2b)_{1234}s_2\n^a s_3\n_a s_4\nonumber\\
\nonumber
&+&(s4a)_{1234}\n^as_2\n^b s_3\n_a\n_b s_4
+(s4b)_{1234}s_2\n^b\n^b s_3\n_a\n_b s_4 \\
\la{eqssf}
&+&(s6a)_{1234}\n^as_2\n^b\n^c s_3\n_a\n_b\n_c s_4,
\eea
where the coefficients are given by
\bea
\nonumber
(s0)_{1234}&=&2V^s_{125}J^s_{534}+(R^s_{125}-2J^s_{125})D^s_{534}\\
\nonumber
(s2a)_{1234}&=&2(T^s_{125}-2L^s_{125})D^s_{534} \\
\nonumber
(s2b)_{1234}&=&2V^s_{125}L^s_{534}+(R^s_{125}-2J^s_{125})E^s_{534}\\
\nonumber
(s4a)_{1234}&=&2(T^s_{125}-2L^s_{125})E^s_{534} \\
\nonumber
(s4b)_{1234}&=&(R^s_{125}-2J^s_{125})F^s_{534} \\
\la{coefs}
(s6a)_{1234}&=&2(T^s_{125}-2L^s_{125})F^s_{534}. 
\eea
The coefficients $s_{1234}$ in this equation depend on $SO(6)$ tensors 
of the form 
$$\frac{k_5^n}{k_5(k_5-1)(k_5+1)(k_5+2)}a_{125}a_{345},\quad n\ge 0 .
$$

\subsection{Contribution of scalars $t$}
To obtain the contribution of the scalars $t$ to cubic terms 
in the equations of motion for $s$ we need to decompose 
the covariant equations (\ref{ffe}) and (\ref{gre}) up to 
the second order in fields, and to keep the terms of the form $st$ 
in the equation for $s$ and terms quadratic in $s$ in the equation for $t$.
We represent the equations for $s$ and $t$ in the following form 
\bea
(\n_a^2-m_1^2)s_1&=&K^t_{125}s_2t_5+
N^t_{125}\n_as_2 \n^a t_5+
P^t_{125}\n_a\n_bs_2 \n^a\n^b t_5 \nonumber\\
\la{eqst}
&+&R^t_{125}s_2(\n_b^2-m_t^2)t_5
+T^t_{125}\n_as_2\n^a (\n_b^2-m_t^2)t_5,\\
(\n_a^2-m_t^2)t_5&=&D^t_{345}s_3s_4+
E^t_{345}\n_as_3 \n^a s_4+
F^t_{345}\n_a\n_bs_3 \n^a\n^b s _4. 
\la{eqtt}
\eea
To get rid of the higher-derivative terms in (\ref{eqst}) 
we perform the following redefinition of the fields $s$:
\bea
s_1\to {s}_1+J^s_{t125}s_2t_5+L^s_{t125}\n_as_2\n^at_5,
\nonumber
\eea
where
\bea
&&2L^s_{t125}=P^t_{125},\quad 2J^s_{t125}+L^s_{t125}(m_2^2+m_t^2-m_1^2-8)
=N^t_{125}.
\nonumber
\eea
Then eq.(\ref{eqst}) takes the form
\bea
(\n_a^2-m_1^2)s_1&=&V^s_{t125}s_2t_5
+(R^t_{125}-J^s_{t125})s_2(\n_b^2-m_t^2)t_5\nonumber\\
&+&(T^t_{125}-L^s_{t125})\n_as_2\n^a (\n_b^2-m_t^2)t_5,
\la{eqst1}
\eea
where
\bea
V^s_{t125}=K^t_{125}-J^s_{t125}(m_2^2+m_t^2-m_1^2)=-\frac{2}{\k_1}T_{125}.
\la{Vst}
\eea
To take into account the terms proportional to the linear 
part of the equation for $t$, one should use eq.(\ref{eqtt}).
We also need to reduce eq.(\ref{eqtt}) to the canonical form (\ref{eqt}).
To this end we perform the shift of the field $t$ \cite{AF5}
\bea
t_5\to {t}_5+J^t_{345}s_3s_4+L^t_{345}\n_as_3\n^as_4.
\nonumber
\eea
This shift removes all terms with derivatives, and we end up with 
eq.(\ref{eqt}). Finally we make the same shift of $t$ in 
eq.(\ref{eqst1}), and represent the equation in the form
\bea
(\n_a^2-m_1^2)s_1&=&V^s_{t125}s_2t_5+(t0)_{1234}s_2s_3s_4+
(t2a)_{1234}\n^as_2s_3\n_a s_4+
(t2b)_{1234}s_2\n^a s_3\n_a s_4\nonumber\\
\nonumber
&+&(t4a)_{1234}\n^as_2\n^b s_3\n_a\n_b s_4
+(t4b)_{1234}s_2\n^b\n^b s_3\n_a\n_b s_4 \\
\la{eqstf}
&+&(t6a)_{1234}\n^as_2\n^b\n^c s_3\n_a\n_b\n_c s_4,
\eea
where 
\bea
\nonumber
(t0)_{1234}&=&V^s_{t125}J^t_{345}+(R^t_{125}-J^s_{t125})D^t_{345}\\
\nonumber
(t2a)_{1234}&=&2(T^t_{125}-L^s_{t125})D^t_{345} \\
\nonumber
(t2b)_{1234}&=&V^s_{t125}L^t_{345}+(R^t_{125}-J^s_{t125})E^t_{345}\\
\nonumber
(t4a)_{1234}&=&2(T^t_{125}-L^s_{t125})E^t_{345} \\
\nonumber
(t4b)_{1234}&=&(R^t_{125}-J^s_{t125})F^t_{345} \\
\la{coeft}
(t6a)_{1234}&=&2(T^t_{125}-L^s_{t125})F^t_{345} 
\eea
The coefficients $t_{1234}$ in this equation depend on $SO(6)$ tensors 
of the form 
$$\frac{k_5^n}{(k_5+2)(k_5+3)(k_5+4)(k_5+5)}a_{125}a_{345},\quad n\ge 0 .
$$
Summing up the contributions of scalars $s^I$ and $t^I$ we find that
the resulting coefficients only depend on the following $SO(6)$ tensors
$$f_5^na_{125}a_{345},\quad n\ge -1,\quad \frac{1}{f_5+3}a_{125}a_{345},\quad
\frac{1}{f_5-5}a_{125}a_{345}.
$$
\subsection{Contribution of scalars $\phi$}
Decomposing 
the covariant equations (\ref{ffe}) and (\ref{gre}) up to 
the second order in fields, and keeping the terms of the form $s\phi$ 
in the equation for $s$ and terms quadratic in $s$ in the equation 
for $\phi$, we represent the equations for $s$ and 
$\phi$ as follows
\bea
(\n_a^2-m_1^2)s_1&=&K^\phi_{125}s_2\phi_5
+N^\phi_{125}\n_as_2\n^a\phi_5
+P^\phi_{125}\n_a\n_b s_2\n^a\n^b \phi_5 \nonumber\\
\la{eqsphi}
&+&R^\phi_{125}s_2(\n_a^2-m_\phi^2)\phi_5,\\
(\n_a^2-m_\phi^2)\phi_5&=&D^\phi_{345}s_3s_4+E^\phi_{345}\n_as_3\n^a s_4
+F^\phi_{345}\n_a\n_b s_3\n^a\n^b s_4.
\la{eqphiphi}
\eea
Following the same steps as above, we make the following 
field redefinitions to
remove higher-derivative terms, and to reduce (\ref{eqphiphi}) to the 
canonical form (\ref{eqphi})
\bea
&&s_1\to s_1+J^s_{\phi 125}s_2\phi_5+L^s_{\phi 125}\n_as_2\n^a\phi_5,
\nonumber\\
&&\phi_5\to {\phi}_5+J^\phi_{345}s_3s_4+L^\phi_{345}\n_as_3\n^a s_4,
\la{redphi}
\eea
where
\bea
J^\phi_{345}=\frac{2(-4+k_3^2+4k_3k_4+k_4^2-f_5)}{(k_3+1)(k_4+1)}p_{345},
\quad
L^\phi_{345}=-\frac{4p_{345}}{(k_3+1)(k_4+1)}.
\la{jlphi}
\eea
Then eq.(\ref{eqsphi}) takes the form
\bea
(\n_a^2-m_1^2)s_1&=&V^s_{\phi 125}s_2\phi_3+(\phi 0)_{1234}s_2s_3s_4+
(\phi 2a)_{1234}\n^as_2s_3\n_a s_4+
(\phi 2b)_{1234}s_2\n^a s_3\n_a s_4\nonumber\\
\nonumber
&+&(\phi 4a)_{1234}\n^as_2\n^b s_3\n_a\n_b s_4
+(\phi 4b)_{1234}s_2\n^b\n^b s_3\n_a\n_b s_4 \\
\la{eqsphif}
&+&(\phi 6a)_{1234}\n^as_2\n^b\n^c s_3\n_a\n_b\n_c s_4,
\eea
where 
\bea
\nonumber
(\phi 0)_{1234}&=&V^s_{\phi 125}J^\phi _{345}+
(R^\phi _{125}-J^s_{\phi 125})D^\phi _{345}\\
\nonumber
(\phi 2a)_{1234}&=&-2L^s_{\phi 125}D^\phi_{345} \\
\nonumber
(\phi 2b)_{1234}&=&V^s_{\phi 125}L^\phi _{345}+
(R^\phi_{125}-J^s_{\phi 125})E^\phi _{345}\\
\nonumber
(\phi 4a)_{1234}&=&-2L^s_{\phi 125}E^\phi_{345} \\
\nonumber
(\phi 4b)_{1234}&=&(R^\phi _{125}-J^s_{\phi 125})F^\phi_{345} \\
\la{coefphi}
(\phi 6a)_{1234}&=&-2L^s_{\phi 125}F^\phi_{345}. 
\eea
The coefficients $\phi_{1234}$ in this equation depend on $SO(6)$ tensors 
of the form 
$$f_5^np_{125}p_{345},\quad n=0,1,2,3.
$$
The surprizing fact is that this 
equation is Lagrangian. Namely, it can be derived from the 
following Lagrangian:
\bea
{\cal L}_\phi&=&{\cal L}_2(s)+
\frac14 \n_a (J^\phi_{125}s_1s_2+L^\phi_{125}\n_b s_1\n^b s_2)
\n^a (J^\phi_{345}s_3s_4+L^\phi_{345}\n_c s_3\n^c s_4)\nonumber\\
&+&\frac14 f_5 (J^\phi_{125}s_1s_2+L^\phi_{125}\n_b s_1\n^b s_2)
(J^\phi_{345}s_3s_4+L^\phi_{345}\n_c s_3\n^c s_4)\nonumber\\
&+&\Phi_{125} s_1s_2(J_{345}s_3s_4+L^\phi_{345}\n_b s_3\n^b s_4),
\label{lphi}
\eea
where $J^\phi_{125}$ and $L^\phi_{125}$ are given by (\ref{jlphi}).
One can easily see that if one makes a redefinition of $\phi$ inverse to
(\ref{redphi}): 
$\phi_5\to\phi_5 -J^\phi_{125}s_1s_2-L^\phi_{125}\n_a s_1\n^a s_2$, then 
the quartic terms in (\ref{lphi}) will be removed from the action, 
but we obtain additional cubic higher-derivative terms. 
\subsection{Contribution of massive gravitons}
We loosely refer to symmetric tensor fields coming from the $AdS_5$ components of the graviton as massive gravitons. 
To account for the massive graviton contribution we first need
to derive equations of motion for the massive gravitons. In principle 
to obtain these equations one should consider the Einstein
equations (\ref{gre}) not only with the indices $(a,b)$, 
but also with the indices $(a,\a )$ and $(\a ,\b )$. The reason is that 
the equations for $\n^a\varphi_{ab}$ and $\varphi_a^a$ that are constraints
and, therefore should be a consequence of a true equation, 
do not follow from (\ref{gre}) if one restricts oneself 
by considering only indices $(a,b)$. To find the true equation for 
the massive gravitons, it is convenient to replace (\ref{gre}) by the 
following equivalent equation
\bea
R_{MN}-\frac{1}{2}g_{MN}R=\frac{1}{3!}\l 
F_{MM_2...M_4} F_N^{M_2...M_4} 
-\frac{1}{5}g_{ab}F_{M_1...M_5}F^{M_1...M_5}\r .
\la{gree}
\eea
Namely this equation one would derive from the usual Lagrangian for 
the metric and the nonchiral five-form in ten dimensions. 
An important property of the equation is that after performing the 
off-shell shift of $h_{ab}$ (\ref{redh}),
its linear part coincides with the linear part of (\ref{eqgr})
\bea
\la{pl}
\l R_{ab}-\frac{1}{2}g_{ab}R-\frac{1}{3!}\l F_aF_b-\frac{1}{5}g_{ab}F_{M_1...M_5}F^{M_1...M_5}\r\r^{(1)}=
Eq_{ab}(\varphi).
\eea
Decomposing eq. (\ref{gree}) up to 
the second order in fields, and keeping the terms quadratic in 
$s$, we can represent the equations for $\varphi_{ab}$ in the following 
form
\bea
Eq_{ab}(\varphi)&+&\a_{123}\n_as_1\n_bs_2
+\b_{123}\n_c\n_as_1\n^c\n_bs_2
+\g_{123}\n_c\n^d\n_as_1\n^c\n_d\n_bs_2 \nonumber \\
\nonumber
&+& \mu_{123}(\n_a(\n_b s_1s_2)+\n_b(\n_a s_1s_2))
+\nu_{123}(\n_a(\n^c s_1\n_b\n_c s_2)+\n_b(\n^c s_1\n_a\n_c s_2) )
\\&+&\rho_{123}\n_a\n_b(s_1s_2) 
+\sigma_{123}\n_a\n_b(\n_c\n_ds_1 \n^c\n^d s_2) \nonumber\\
&+&\delta_{123}(\n_a(\n^c\n^d s_1\n_c\n_d\n_b s_2)
+\n_b(\n^c\n^d s_1\n_c\n_d\n_a s_2) )+g_{ab}C=0.
\la{eqgrgr}
\eea
Here $C$ denotes the following contribution:
\bea
C=T^1_{123}s_1s_2+T^2_{123}\n_as_1\n^as_2+T^3_{123}\n_a\n_b s_1\n^a\n^b s_2
-\frac{1}{2}L_{123}\n_a\n_b\n_cs_1\n^a\n^b\n^c s_2.
\eea 
To remove the higher-derivative and total-derivative terms we perform
the following shift of the massive gravitons:
\bea
\varphi_{ab}^3= {\varphi'}_{ab}^3+\n_a\xi_b^3+
\n_b\xi_a^3+g_{ab}\eta^3+J_{123}\n_as_1\n_bs_2
+L_{123}\n_c\n_a s_1\n^c\n_b s_2.
\la{redgr}
\eea
Here $J_{123}$ and $L_{123}$ depend on 
the coefficients $\a ,\b ,\g$ as follows
$$L_{123}=\g_{123},\quad J_{123}=\b_{123}
-\frac12 L_{123}(m_1^2+m_2^2-f_3-18)$$
and the cubic vertex $G_{123}$ is expressed through them as
$$G_{123}=-\a_{123}+\frac12 J_{123}(m_1^2+m_2^2-f_3-6)-
L_{123}(m_1^2+m_2^2).$$
To get rid of the total derivative terms with the coefficients 
$\mu,\nu,\rho,\sigma,\delta$
one also has to impose the following relation 
\bea
\la{xi}
\frac{1}{2}f_3\xi_a^3-\frac{3}{4}\n_a\eta^3 +U_{123}s_2\n_a s_1+H_{123}
\n_c s_1\n^c\n_a s_2=0,
\eea
where 
\bea
U_{123}&=&\frac{1}{2}(m_2^2J_{123}-3m_2^2L_{123}
+2\rho_{123}+2\mu_{123}+2m_2^2 \sigma_{123}), \nonumber\\
\nonumber
H_{123}&=&\frac{1}{2}(m_1^2-3)L_{123}+\nu_{123}-\sigma_{123}.
\eea
Thus only the coefficient $\eta$ has not been fixed yet. Actually, 
a change of the coefficient (with the simultaneous change of $\xi$ 
according to (\ref{xi})) results only in a change of the interaction
of the trace $\varphi^a_a$ of the massive graviton with the 
scalars $s$. In particular, one can choose $\eta$ in such a way
that only traceless part of a massive graviton interacts with
the scalars $s$. However, this choice leads to the appearance of 
quartic couplings with 6 derivatives. Terms with 6 derivatives are absent
only if we choose the cubic vertex as in eq.(\ref{ssgr}). This vertex 
is a natural generalization of the interaction vertex of a massless 
graviton with scalar fields. To determine $\eta$ we take the trace 
of eq.(\ref{eqgrgr}) and represent the resulting equation as 
\bea
&&-\frac{3}{2}\n_a(\n_b \varphi^{ab}-\n^a\varphi_c^c)
-2(f+3)\varphi_c^c+\tilde{\a}_{123}\n_as_1\n^as_2
 \nonumber\\
&&+\tilde{\b}_{123}\n_a\n_bs_1\n^a\n^b s_2+
\tilde{\g}_{123}\n_a\n_b\n_cs_1\n^a\n^b\n^c s_2+
\tilde{\d}_{123}s_1s_2=0.
\la{eqtrgr}
\eea
Then, by requiring that after the shift
(\ref{redgr}) eq.(\ref{eqtrgr}) coincides with the trace of (\ref{eqgr}) 
and assuming that $\eta$ has the form
$$\eta_3=A_{123}s_1s_2+B_{123}\n_as_1\n^as_2
+C_{123}\n_a\n_bs_1\n^a\n^b s_2,$$
we find the following relations 
\bea
&& 8H_{123}+\l\frac{3}{2}J_{123} - (\frac{33}{2}+2f_3)K_{123}\r 
+\tilde{\b}_{123}-10(f_3+3)C_{123} =0, \\
\nonumber
&&
 8H_{123}(m_2^2-4)+8U_{123}-2(f_3+6)J_{123}
+(6-\frac{3}{2}(m_1^2-4)(m_2^2-4)K_{123})\\
\nonumber
&&+\tilde{\a}_{123}
-10(f_3+3)B_{123} =\frac32 G_{123}, \\
&&
\ 8U_{123}m_1^2 +\frac{3}{2}m_1^2m_2^2(3K_{123}-J_{123})+\tilde{\d}_{123}
-10(f_3+3)A_{123} =\frac34 (m_1^2+m_2^2-f_3)G_{123}.\nonumber
\eea
In particular, one can show that $C_{123}=0$.

To find the massive graviton contribution to the equations for $s$ we also
need to know equations of motion for 
$\n^b\varphi_{ab}$ ($\n^b\varphi'_{ab}$) and $\varphi_a^a$ (${\varphi'}_a^a$).
The simplest way to derive the equations is first to take into account 
that $\varphi'_{ab}$ satisfies eq.(\ref{eqgr}), and then to use the 
graviton redefinition (\ref{redgr}) to find 
$\n^a\varphi_{ab}$ and $\varphi_a^a$. 
Differentiating and taking the trace
of (\ref{eqgr}) one can easily obtain 
\bea
{\varphi'}_{a3}^a&=&-\frac{3G_{123}}{2f_3(f_3+3)}\l 
-\frac14 (m_1^2-m_2^2)^2-\frac16 f_3(m_1^2+m_2^2)+
\frac{5}{12}f_3^2\r s_1s_2,\nonumber\\
\n^b{\varphi'}^3_{ab}&=&\frac{G_{123}}{f_3}(m_2^2-m_1^2+f_3)\n_as_1s_2
+\n_a{\varphi'}_{b3}^b.\nonumber
\eea
These equations also explain why we took the interaction vertex of massive
gravitons in the form (\ref{ssgr}), namely, there are no terms with 
derivatives in the r.h.s. of the equation for ${\varphi'}_a^a$ under 
this choice.

Now we can proceed with the massive graviton contribution to
the equations for $s$. To find the equations we should decompose
the covariant equations of motion up to the second order, make the shift
(\ref{redh}), and keep only the terms of the form $s\varphi$. To simplify
the consideration we also find convenient not to shift the trace of
the original gravitons $h^a_a$ (however we do not decompose $h_{ab}$ 
in the sum of a traceless tensor and a trace)
but to take into account the contribution
of $h_a^a$ later. This can be easily done because the equation of
motion for $h_a^a$ follows from the Einstein equation (\ref{gre})
with indices $(\a ,\b )$. Then the equations for $s$ take the form
\bea
(\n_a^2-m_1^2)s_1=V^g_{123}\n^a\n^bs_2\varphi_{ab}^3
+R^g_{123}\n^as_2\n^b\varphi_{ab}^3+
T^g_{123}s_2\varphi_a^{a3},
\la{eqsgr1}
\eea
where $V^g_{123}=\frac{2}{\k_1}G_{123}$.

Finally, substituting the massive graviton redefinition (\ref{redgr}) in
(\ref{eqsgr1}), and performing the following redefinition of $s$ to simplify
the equation
$$
s_1\to s_1+V^g_{123}\n^bs_2\xi_b^3,
$$ 
we represent the contribution of the massive gravitons in the form
\bea
(\n_a^2-m_1^2)s_1&=&
V^g_{125}\left( \n^a(\n^bs_2{\varphi'}_{ab}^5)
-\frac12\n^a(\n_as_2{\varphi'}^c_{c5})
+\frac14 (m_1^2+m_2^2-f_5)s_2{\varphi'}^c_{c5})\right)\nonumber\\
&+&(g0)_{1234}s_2s_3s_4+(g2a)_{1234}\n^a s_2s_3\n_a s_4+
(g2b)_{1234}s_2\n_a s_3\n^a s_4
\nonumber\\
&+&
(g4a)_{1234}\n^a s_2\n^b s_3\n_a\n_b s_4
+(g4b)_{1234}s_2\n_a\n_b s_3\n^a\n^b s_4 \nonumber\\&+&
(g4c)_{1234}\n^a\n^bs_2\n_a s_3\n_b s_4
\nonumber\\&+&(g6a)_{1234}\n^as_2\n^b\n^c s_3\n_a\n_b\n_c s_4 \nonumber\\
\la{eqsgrf}
&+&(g6c)_{1234}\n^a\n^bs_2\n_c\n_a s_3\n^c\n_b s_4.
\eea
The coefficients $g_{1234}$ in this equation depend on $SO(6)$ tensors 
of the form 
$$f_5^na_{125}a_{345},\quad n\ge -1,\quad \frac{1}{f_5+3}a_{125}a_{345}.
$$
\subsection{Contribution of the trace of massive gravitons}
It is known \cite{KRN} that at linear order the graviton trace $h_a^a$
is equal to $-\frac35 \pi$. By using the Einstein equation 
(\ref{gre}) with indices $(\a ,\b )$ one can easily find that 
the combination
$$\bar{\varphi}^a_a=h_a^a+\frac35 \pi$$
is equal to
\bea
{\bar{\varphi}_a^a}=\Omega^1_{123}s_1s_2+\Omega^2_{123}\n_as_1\n^as_2
+\Omega^2_{123}\n_a\n_b s_1\n^a\n^b s_2.
\nonumber
\eea
Taking into account that the terms of the form $s\bar{\varphi}^a_a$ 
enter the equation for $s$ as follows
\bea
(\n_a^2-m_a^2)s_1=\mu_{123}s_2\bar{\varphi}_{a3}^a+
\nu_{123}\n_as_2\n^a\bar{\varphi}_{c3}^c
\nonumber
\eea
we obtain the contribution of  $\bar{\varphi}^a_a$
\bea
(\n_a^2-m_1^2)s_1&=&(tr0)_{1234}s_2s_3s_4+
(tr2a)_{1234}\n^as_2s_3\n_a s_4+
(tr2b)_{1234}s_2\n^a s_3\n_a s_4\nonumber\\
\nonumber
&+&(tr4a)_{1234}\n^as_2\n^b s_3\n_a\n_b s_4
+(tr4b)_{1234}s_2\n^b\n^b s_3\n_a\n_b s_4 \\
\la{coeftr}
&+&(tr6a)_{1234}\n^as_2\n^b\n^c s_3\n_a\n_b\n_c s_4.
\eea
The coefficients $tr_{1234}$ in this equation depend on $SO(6)$ tensors 
of the form 
$$f_5^na_{125}a_{345},\quad n\ge -1,\quad
\frac{1}{f_5-5}a_{125}a_{345}.
$$

\subsection{Contribution of vector fields}
 In this subsection $V_a$ denotes either the vector field
$A_a$ or $C_a$. The contribution of the vector fields to the equations of 
motion for the scalars $s$, and to the equations of motion
for the vector fields may be written in the form
\bea
\nonumber
&&(\n_a^2-m^2)s^1= K^V_{125}\n^as^2V^5_a
+N^V_{125}\n^a\n^b s^2 \n_aV_b^5\\
&&\qquad +R^V_{125}s^2 \n^aV_a^5
+T^V_{125}\n^a s^2 (\n_b^2 V_a^5-\n^b\n_a V_b^5-m^2_V V_b^5),
\la{eqsv}
\\
\nonumber
&&\n_a^2V_b^5-\n^a\n_bV_a^5-m_V^2V_a^5=\n_a V+
D^V_{125}s_1\n_as_2+E^V_{125}\n^bs_1\n_a\n_bs_2\\
&&\qquad +F^V_{125}\n^b\n^cs_1\n_a\n_b\n_c s_2,
\la{eqv}
\eea
where the constants $D,E,F$ are antisymmetric in 1, 2, 
and $V$ has the following dependence on $s$
$$
V=Q^V_{123}s_1s_2+H^V_{123}\n^as_1\n_a s_2.
$$ 
To get rid of higher-derivative terms in eq.(\ref{eqsv}) 
we perform the following shift
\bea
&&s_1\to {s}_1+\frac{1}{2}N^V_{123}\n^as_2 V_a^3.  
\nonumber
\eea
Then the equation acquires the form
\bea
\nonumber
(\n_a^2-m^2)s_1&=& V_{123}\n^as_2V^3_a
+R_{123}s_2 \n^aV_a^3-\frac12 N_{123}\n^bs_2 \n_b(\n^aV_a^3)  \\
\nonumber
&+&(T_{123}-
\frac12N_{123})\n^a s_2 (\n_b^2 V_a^3-\n^b\n_a V_b^3-m^2_V V_b^3).
\la{eqsv1}
\eea 
To remove higher-derivative and total-derivative term from 
eq.({\ref{eqv}), and to reduce it to the canonical form we make the 
following fields redefinition \cite{AF5}
\bea
V_a^3={V'}_a^3-\frac{1}{m_V^2}\n_a\bar{V}+J^V_{453} s_4\n_a s_5
+L^V_{453} \n^bs_4\n_a\n_bs_5.
\nonumber
\eea
Finally we should substitute the redefinition in eq.(\ref{eqsv1}), and 
represent it in the form
\bea
\nonumber
(\n_a^2-m^2)s_1&=& V_{125}\l\n^a s_2{V}'^5_a 
+\frac{1}{2}s_2\n^a {V}'^5_a \r 
+(V0)_{1234}s_2s_3s_4 \\
\nonumber
&+&
(V2a)_{1234}\n^a s_2s_3\n_a s_4+(V2b)_{1234}s_2\n^as_3\n_as_4 \\
\nonumber
&+&(V4a)_{1234}\n^a s_2\n^b s_3\n_a\n_b s_4
+(V4b)_{1234}s_2\n^a\n^bs_3\n_a\n_b s_4 \\
\nonumber
&+&
(V6a)_{1234}\n^a s_2\n^b\n^c s_3\n_a\n_b\n_c s_4. 
\la{eqsvf} 
\eea 
Summing up the contributions of the vectors $A$ and $C$ we see that 
the coefficients $V_{1234}$ depend on the $SO(6)$ tensors 
of the form 
$$f_5^nt_{125}t_{345},\quad n\ge 0,\quad\frac{1}{f_5-5}t_{125}t_{345}.
$$
\subsection{Contribution of contact terms}
Finally we have to take into account the contribution of contact
terms that appear when we decompose the covariant equations of
motion (\ref{ffe}) and (\ref{gre}) up to the third order in
the fields, and keep only terms cubic in the scalars $s$. 
This contribution has the form 
\bea
(\n_a^2-m_1^2)s_1&=&(c0)_{1234}s_2s_3s_4+
(c2a)_{1234}\n^as_2 s_3 \n_a s_4+
(c2b)_{1234}s_2\n_a s_3 \n^a s_4\nonumber\\
&+&(c4a)_{1234}\n^as_2\n^bs_3\n_a\n_bs_4
+(c4b)_{1234}s_2\n^a\n^bs_3\n_a\n_b s_4
\nonumber\\
&+&(c4c)_{1234}\n_a\n_b s_2 \n^as_3\n^bs_4\nonumber\\
&+&(c6a)_{1234}\n^a s_2\n^b\n^c s_3\n_a\n_b\n_c s_4\nonumber\\
&+&(c6c)_{1234}\n^a\n^bs_2\n_b\n_c s_3\n_a\n^cs_4.
\la{eqscf}
\eea
The coefficients $c_{1234}$ in this equation depend on $SO(6)$ tensors 
of the form 
$$f_5^na_{125}a_{345},\quad n\ge -1.
$$
\section{Analysis of the equations}
\setcounter{equation}{0}
In this Section we explain what steps one should undertake to
obtain Lagrangian equations of motion from the original ones. 
Looking at the contributions derived in the previous section 
we see that the cubic corrections to the equations of motion 
for the scalars $s^I$ have the form
\bea
(\n_a^2-m_1^2)s_1&=&(w0)_{1234}s_2s_3s_4+
(w2a)_{1234}\n^as_2 s_3 \n_a s_4\nonumber\\
&+&(w4a)_{1234}\n^as_2\n^bs_3\n_a\n_bs_4
+(w4b)_{1234}s_2\n^a\n^bs_3\n_a\n_b s_4
\nonumber\\
&+&(w6a)_{1234}\n^a s_2\n^b\n^c s_3\n_a\n_b\n_c s_4\nonumber\\
&+&(w6c)_{1234}\n^a\n^bs_2\n_b\n_c s_3\n_a\n^cs_4
\la{eqsor}
\eea
The coefficients $w_{1234}$ in this equation may in general depend 
on $SO(6)$ tensors of the form 
$$f_5^na_{125}a_{345},\quad f_5^nt_{125}t_{345},\quad
f_5^np_{125}p_{345},\quad n\ge 0$$
$$f_5^{-1}a_{125}a_{345},\quad \frac{1}{f_5+3}a_{125}a_{345},\quad
\frac{1}{f_5-5}a_{125}a_{345},\quad\frac{1}{f_5-5}t_{125}t_{345},
$$
and tensors obtained from them by permutation of the indices 1, 2, 3, 4. 

However, by using the identities (\ref{TT}), (\ref{TTF}), (\ref{PP}) and 
(\ref{PPF}) from the Appendix, we may reduce
the tensors  $t_{125}t_{345}$ and  $f_5t_{125}t_{345}$ to the tensors
$f_5^na_{125}a_{345},~ n\ge -1$, and 
$p_{125}p_{345}$ and  $f_5p_{125}p_{345}$
to the tensors
$f_5^na_{125}a_{345},~ n\ge -1$,
$f_5^2t_{125}t_{345}$, 
 $\frac{1}{f_5-5}a_{125}a_{345}$, and $\frac{1}{f_5-5}t_{125}t_{345}$, and 
tensors obtained from them by permutation of the indices 1, 2, 3, 4. 
Then we find that (after the additional shift)
the tensors of the form $\frac{1}{f_5+3}a_{125}a_{345},~ 
\frac{1}{f_5-5}t_{125}t_{345}$ 
completely disappear from the total contribution, and the tensor
$\frac{1}{f_5-5}a_{125}a_{345}$ occures only in the terms without
derivatives, and with two derivatives.
\subsection{6-derivative terms}
We begin our analysis of (\ref{eqsor}) with the six-derivative terms. 
We see that the equation contains in particular 
the following term 
coming from the vectors contribution after the use of the 
identities (\ref{TT}) and (\ref{TTF}) 
\bea
(\n_a^2-m_1^2)s_1=w_{1234}f_5(a_{135}a_{245}-a_{145}a_{235})
\n^as_2\n_b\n^cs_3\n_a\n_b\n_cs_4.
\la{w6au}
\eea
All other terms in the coefficients $w6a$ and $w6c$ only depend on tensors
$f_5^na_{125}a_{345},~n=0,1,2$. To compare (\ref{w6au}) 
with the other contributions we perform the shift 
\bea
s_1\to s_1+J_{1234}\n^as_2\n^bs_3\n_a\n_bs_4,\la{shift4}
\eea
and choose
$$J_{1234}=j_{1234}f_5(a_{135}a_{245}-a_{145}a_{235}),
$$
where
$$
j_{1234}=\frac 12 \left( \frac13 (w_{1234}+w_{1324})+
w_{1234}-w_{1324}\right) =\frac23 w_{1234}-\frac13 w_{1324}.
$$
This results in the following change of eq.(\ref{w6au}) 
\bea
\nonumber
(\n_a^2-m_1^2)s_1&=&
-2(j_{1324}-j_{1432})f_5a_{125}a_{345}\n^a\n_bs_2\n^b\n^cs_3\n_a\n_cs_4\\
\nonumber
&+&j_{1324}(f_1+f_2+f_3+f_4-3f_5)
a_{125}a_{345}\n^as_2\n^b\n_c s_3\n_a\n_b\n_cs_4\\
\nonumber
&+&(dw4a)_{1234}\n^as_2\n^bs_3\n_a\n_bs_4
+(dw2a)_{1234}\n^as_2 s_3 \n_a s_4, 
\eea
where
\bea
(dw4a)_{1234}&=&- 
f_5(a_{135}a_{245}-a_{145}a_{235})
(j_{1234}(m_2^2+m_3^2+m_4^2-m_1^2-18)-2j_{1243})\cr
&-&2j_{1423}
f_5(a_{125}a_{245}-a_{135}a_{245}),  
\cr
(dw2a)_{1234}&=&2( (j_{1234}+j_{1243})(a_{135}a_{245}-a_{145}a_{235})
+j_{1324}(a_{125}a_{345}-a_{145}a_{235})  )f_5 m_3^2 \nonumber
\eea
are the additional contributions to the coefficients $w4a$ and $w2a$. 
Now we can use the symmetry of the 4-derivative term  and 
the 2-derivative term  under the permutation 
of the indices 2, 3, and 2, 4 respectively, and identity 
(\ref{imre}) from the Appendix to express 
$f_5a_{145}a_{235}$ and the whole right hand side of the equation 
only through $a_{125}a_{345}$. Then the coefficients $dw4a$ and $dw2a$ 
acquire the form
\bea
\nonumber
(dw4a)_{1234}&=&-\left[ f_5 \biggl( 
(2j_{1324}+j_{1234})(m_2^2+m_3^2+m_4^2-m_1^2-18)\right.
\\
&-& 2(2j_{1342}-j_{1423}+j_{1432}+j_{1243}) \biggr)
 \cr
\nonumber 
&-& \left.   (f_1+f_2+f_3+f_4)( j_{1234}(m_2^2+m_3^2+m_4^2-m_1^2-18)-2j_{1243}  )  \right]  a_{125}a_{345}\\
\nonumber
(dw2a)_{1234}&=&2\left[  
(j_{1234}+j_{1243})(f_1+f_2+f_3+f_4-f_5)\right.\\
&+&\left. (j_{1324}-j_{1342}-2j_{1432}-2j_{1423})f_5
\right] m_3^2 a_{125}a_{345} \nonumber
\eea
Thus we reduced all 6-derivative terms to terms which depend only on 
the tensors $f_5^na_{125}a_{345}$.

Now summing up the vector fields contribution with the contributions 
of all the other fields we get that the term $w6c$ has 
the following structure
\bea
(w6c)_{1234}=(w6c0)_{1234}a_{125}a_{345}+
(w6c1)_{1234}f_5a_{125}a_{345}-\frac{f_5^2}{4\rho}a_{125}a_{345}, 
\la{w6c}
\eea
where $(w6c1)_{1234}$ is a function symmetric under permutation of
 2, 3, and 4, and we denote
$\rho = (k_1-1)k_1(k_1+2)(k_2+1)(k_3+1)(k_4+1)$.
Thus we may use the identity
$$
w_{1234}f_5a_{125}a_{345}=\frac13 w_{1234}(f_1+f_2+f_3+f_4)a_{125}a_{345},
$$
valid for any function symmetric in 2, 3, 4.
So, we see that $w6c$ does not depend on the tensor 
$f_5a_{125}a_{345}$, and, moreover $wc60$ should be symmetrized 
in 2, 3, 4 because it is multiplied by a symmetric tensor
$\n_a\n^bs_2\n_b\n^cs_3\n^a \n_cs_4$. 

Looking at the term $w6a$ we see that  
this term has the same form (\ref{w6c}), and moreover, the coefficient
$w6a1$ is also symmetric in 2, 3, 4, and, therefore, can be reduced
to $w6a0$. The term $wa62$ proportional to 
$f_5^2a_{125}a_{345}$ comes from massive gravitons, 
vectors and scalars $\phi$ and is given explicitly
\bea
\la{fqus}
(\n_a^2-m_1^2)s_1&=&
-\frac{f_5^2}{2\rho}a_{145}a_{235} \n^as_2\n^b\n^cs_3\n_a \n_b\n_c s_4.
\eea
We may reduce the term to the structure 
$\n_a\n^bs_2\n_b\n^cs_3\n^a \n_c s_4$  by performing the shift 
$$
s_1\to s_1-\frac{f_5^2}{8\rho}a_{145}a_{235}\n^bs_2\n^cs_3\n_b\n_cs_4
$$
that results in
\bea
\la{bsh1}
(\n_a^2-m_1^2)s_1&=&
\frac{f_5^2}{4\rho}a_{125}a_{345}\n_a\n^bs_2\n_b\n^cs_3\n^a \n_c s_4 \\
\nonumber
&+&(\frac{f_5^2}{2\rho}a_{125}a_{345}
+\frac{f_5^2}{8\rho}a_{145}a_{235}(m_2^2+m_3^2+m_4^2-m_1^2-18) ) 
\n^as_2\n^b s_3\n_a \n_b s_4 \\
\nonumber
&-&m_3^2 \frac{f_5^2}{2\rho}a_{125}a_{345} \n^as_2 s_3 \n_a s_4
+m_2^2 \frac{f_5^2}{4\rho}a_{125}a_{345} s_2 \n^a s_3 \n_a s_4.
\eea
Summing up the coefficient on the first line of (\ref{bsh1}) with 
$w6c$ we obtain that the final contribution does not depend on 
the tensor $f_5^2a_{125}a_{345}$. So the new coefficient $w6c$ depends
only on the tensor $a_{125}a_{345}$, and we can easily reduce 
it to $w6a$ by means of the shift
$$
s_1\to s_1+\frac12 (w6c)_{1234}\n^bs_2\n^cs_3\n_b\n_cs_4
$$
This results in 
\bea
\nonumber
(\n_a^2-m_1^2)s_1&=&-2 (w6c)_{1234}\n^as_2\n^b\n^cs_3\n_a\n_b\n_cs_4 \\
\nonumber
&-&\l 2 (w6c)_{1234}+\frac{1}{2}
(w6c)_{1234}(m_2^2+m_3^2+m_4^2-m_1^2-18)\r 
\n^as_2\n^bs_3 \n_a \n_b s_4 \\
\nonumber
&+&(w6c)_{1234}m_3^2 \n^as_2 s_3 \n_a s_4.
\eea
Adding the coefficient $-2(w6c)$ from the first line of the equation to $w6a$
we obtain a new coefficient that is symmetric in 2, 3, and, therefore, 
the first term on the r.h.s. of the equation 
can be transformed to the structure 
$\n_a\n^bs_2\n_b\n^cs_3\n^a \n_c s_4$ by using (\ref{shift4}). 
Symmetrizing the coefficient in front of the 6-derivative term 
we obtain zero. This shift also produces additional contributions to 
the coefficients $w4a$ and $w2a$.

Thus we have shown that all 6-derivative terms could be shifted away.
\subsection{4-derivative terms}
We proceed with 4-derivative terms for which, 
we take into account all the additional contributions 
appeared in the previous subsection due to our way of 
working with 6-derivative terms. The coefficient $w4a$ contains
the term $\frac{16(f_5-1)^2}{\rho}t_{125}t_{345}$ that gives 
Lagrangian contribution to the equations of motion. Other contributions
 are nonlagrangian, and we analyze them by decomposing the coefficients
$w4a$ and $w4b$  in Laurent series in $f_5$. 
\vskip 0.5cm

{\bf I. 4-derivative terms with $1/f_5$. }

These terms give the 
following contribution to the equations of motion
\bea
\nonumber
(\n_a^2-m_1^2)s_1&=&(w4ad)_{1234}\n^as_2\n^b s_3\n_a\n_bs_4
+(w4bd)_{1234}s_2\n^a\n^bs_3\n_a\n_bs_4.
\eea
We decompose $w4ad$ into parts symmetric and antisymmetric in 3, 4,
and shift its symmetric part to the 4b-structure by using 
the field redefinition
\bea
s_1\to s_1+J_{1234}s_2\n^as_3\n_as_4.\la{shift2}
\eea
The resulting 4-derivative vertices can be written in the form:
\bea
\nonumber
(\n_a^2-m_1^2)s_1&=&\frac{2(f_1-f_2)(f_3-f_4)}
{\rho f_5}
a_{125}a_{345}\n^as_2\n^b s_3\n_a\n_b s_4 \\
\nonumber
&-&\frac{(f_1-f_2)(f_3-f_4)}
{\rho f_5}
a_{125}a_{345}\n^a\n^b s_2 s_3\n_a\n_b s_4.
\eea
Finally performing the shift  
\bea
s_1\to s_1+J_{1234}\n^bs_2 s_3 \n_b s_4
\eea
and using the symmetry of the vertex w.r.t. 3, 4, we represent
the final result for the 4-derivative vertex as follows 
\bea
\nonumber
(\n_a^2-m_1^2)s_1&=&\frac{3(f_1-f_2)(f_3-f_4)}{\rho f_5}
a_{125}a_{345}\n^as_2\n^b s_3\n_a\n_b s_4 \\
\nonumber
&+&\frac{(f_1-f_2)(f_3-f_4)}{2\rho f_5}
(m_2^2+m_3^2+m_4^2-m_1^2-8)\n^a s_2  s_3\n_a s_4.
\eea
The 4-derivative term represents a Lagrangian contribution to 
the equation of motion, which can be derived from the Lagrangian
of the form
\bea
{\cal L}=A^{(4)}_{1234}\int s_1\n_as_2 \n_b^2(s_3\n^a s_4),
\la{AA4}
\eea
where the quartic coupling $A^4_{1234}$ is $antisymmetric$ 
in 1, 2 and 3, 4, and $symmetric$ under the interchange (1,2) and (3,4),
and is given by
$$
A^{(4)}_{1234}=-\kappa \frac{3}{8\rho f_5}(f_1-f_2)(f_3-f_4).
$$
The equations of motion that follow from the Lagrangian are 
\bea
\nonumber
\kappa(\n_a^2-m_1^2)s_1&=&-8A^{(4)}_{1234}\n^as_2\n^b s_3\n_a\n_b s_4 \\
\nonumber
&-&4(m_3^2+m_4^2-4)A^{(4)}_{1234}
\n^as_2 s_3\n_a  s_4 \\
\nonumber
&-&2(m_4^2-m_3^2)A^{(4)}_{1234}
s_2 \n^as_3\n_a  s_4 \\
\nonumber
&-&(m_3^2+m_4^2-4)(m_4^2-m_3^2)A^{(4)}_{1234} s_2s_3 s_4.
\eea
 It is clear that the 4-derivative term cannot be removed by 
any field redefinition.
\vskip 0.5cm

{\bf II. 4-derivative terms with $f_5^3$.}

The contribution of the terms with $f_5^3$ is given by
\bea
\nonumber
(\n_a^2-m_1^2)s_1&=&-\frac{f_5^3}{64\rho}(3l+8m)\n^as_2\n^b s_3\n_a\n_b s_4 \\
\nonumber
&-&\frac{f_5^3}{128\rho}(3l+4m+4n) s_2\n^a \n^b s_3\n_a\n_b s_4. 
\eea
Performing the shift 
\bea
s_1\to s_1+J_{1234}s_2 \n^b s_3 \n_b s_4,
\eea
where $2J_{1234}=-\frac{f_5^3}{128\rho}(3l+4m+4n)$ is symmetric in $3,4$,
we obtain the Lagrangian  4-derivative term
\bea
\nonumber
(\n_a^2-m_1^2)s_1&=&\frac{f_5^3}{16\rho}(n-m)\n^as_2\n^b s_3\n_a\n_b s_4\\
\nonumber
&+& \frac{f_5^3}{256\rho}(3l+4n+4m)(m_2^2+m_3^2+m_4^2-m_1^2-8) 
s_2\n^a s_3\n_a s_4,
\eea
that can be again derived from a vertex of the form (\ref{AA4}).
\vskip 0.5cm

{\bf III. 4-derivative terms with $f_5^2$.}

Here we first consider the term of the 4a-type. The term of the 
4b-type is also nonzero and we consider it later.
\bea
\nonumber
(\n_a^2-m_1^2)s_1&=&(h^n_{1234}n+h^m_{1234}m)\n^as_2\n^b s_3\n_a\n_b s_4.
\eea
Here $h^n,h^m $ denote the coefficients of the corresponding 
structures $n,m$.
We can rewrite this equation as follows
\bea
\nonumber
(\n_a^2-m_1^2)s_1&=&( h^n_{1234}n+h^n_{1243}m )\n^as_2\n^b s_3\n_a\n_b s_4\\
\nonumber
&+&\omega_{1432}a_{125}a_{345}\n^a\n^b s_2\n_as_3 \n_b s_4,
\eea
where 
$$
\omega_{1234}=(h^m_{1234}-h^n_{1243}).
$$
To convert the equation to the one containing only the 4b-type 
structure $s_2\n^a\n^bs_3\n_a\n_bs_4$ we make the field redefinition
$$
s_1\to s_1+\frac12\omega_{1432}a_{125}a_{345}\n^b s_2 s_3 \n_b s_4
+j_{1234}s_2 \n^b s_3 \n_b s_4,
$$
where
$$
j_{1234}=\frac{1}{4}( h^n_{1234}n+h^n_{1243}m-\omega_{1432} l ).
$$
Then the equation transforms as follows
\bea
\nonumber
(\n_a^2-m_1^2)s_1&=&-\frac{1}{2}
(\omega_{1423} n +\omega_{1324} m+ h^n_{1234}n+h^n_{1243}m-\omega_{1432} l ) 
s_2\n^a\n^b s_3 \n_a \n_b s_4 \\
\nonumber
&-&j_{1234} (m_2^2+m_3^2+m_4^2-m_1^2-8) s_2\n^a s_3 \n_a s_4 \\
\nonumber
&-& \frac{1}{2}\omega_{1432}l  
(m_2^2+m_3^2+m_4^2-m_1^2-8)\n^a s_2 s_3 \n_a s_4.
\eea
Now we sum the r.h.s. of the equation with the contribution of 4b-type 
and get a Lagrangian 4-derivative term
\bea
\nonumber
(\n_a^2-m_1^2)s_1&=&\frac{(2l-n-m)f_5^2}{32\rho}(-28+3f_1+3f_2+3f_3+3f_4)
s_2\n^a \n^b s_3\n_a\n_b s_4 \\
\nonumber
&-&j_{1234} (m_2^2+m_3^2+m_4^2-m_1^2-8) s_2\n^a s_3 \n_a s_4 \\
\nonumber
&-& \frac{1}{2}\omega_{1432}l  
(m_2^2+m_3^2+m_4^2-m_1^2-8)\n^a s_2 s_3 \n_a s_4.
\eea
It is convenient, however, to reduce the 4-derivative term to the term
of 4a-type by means of a field redefinition of the form, 
$$s_1 \to s_1+J_{1234}s_2\n^as_3\n_a s_4$$
and by using the 
symmetry of the 4a-type term under the permutation of the indices 2, 3.
The resulting equation looks as
\bea
\nonumber
&&(\n_a^2-m_1^2)s_1=-\frac{f_5^2(n-m)}{16\rho }(3(f_1+f_2+f_3+f_4)-28)
\n^a s_2 \n^b s_3\n_a\n_b s_4 \\
\nonumber  
\qquad&&-\frac{f_5^2(2l-n-m)}{64\rho }
(3(f_1+f_2+f_3+f_4)-28)(m_2^2+m_3^2+m_4^2-m_1^2-8)
s_2\n^a s_3 \n_a s_4 \\
\nonumber
\qquad&&-j_{1234} (m_2^2+m_3^2+m_4^2-m_1^2-8) s_2\n^a s_3 \n_a s_4 \\
\nonumber
\qquad&&- \frac{1}{2}\omega_{1432}l  
(m_2^2+m_3^2+m_4^2-m_1^2-8)\n^a s_2 s_3 \n_a s_4,
\eea
and the 4-derivative term can be obtained from the vertex of the form
(\ref{AA4}).
\vskip 0.5cm

{\bf IV. 4-derivative terms with $f_5$.}

We can reduce the 4b-type term to the 4a-type one by means of the shift 
$$
s_1\to s_1+ j_{1234}s_2\n^bs_3\n_b s_4,
~~~~j_{1234}=\frac{1}{2}(w4b)_{1234}f_5.
$$
This results in 
\bea
\nonumber
(\n_a^2-m_1^2)s_1&=&(w4a-2\cdot w4b)_{1234}\n^as_2\n^b s_3\n_a\n_b s_4 \\
\nonumber
&-&j_{1234}(m_2^2+m_3^2+m_4^2-m_1^2-8)s_2\n_a s_3\n^a s_4.
\eea
Representing 
$$
(w4a-2\cdot w4b)_{1234}= P^l_{1234}l+ P^n_{1234}n+ P^m_{1234}m,
$$
using the identity (\ref{imre}), and changing the summation 
indices 2 and 3 we
rewrite the equation in the form
\bea
\nonumber
(\n_a^2-m_1^2)s_1&=&( P^l_{1234}+ P^n_{1324}- P^m_{1324} -  
P^m_{1234})f_5 a_{125}a_{345}
\n^as_2\n^b s_3\n_a\n_b s_4 \\
\nonumber
&+&P^m_{1324}(f_1+f_2+f_3+f_4)\n^as_2\n^b s_3\n_a\n_b s_4 \\
\nonumber
&-&j_{1234}f_5(m_2^2+m_3^2+m_4^2-m_1^2-8)s_2\n_a s_3\n^a s_4.
\eea 
The 4-derivative term represents a Lagrangian contribution as can be seen 
by decomposing the coefficients in front of $f_5a_{125}a_{345}$ in parts 
antisymmetric and symmetric in 3, 4:
\bea\nonumber
(\n_a^2-m_1^2)s_1&=&\l Y^a + Y^s \r_{1234} 
f_5a_{125}a_{345}\n^as_2\n^b s_3\n_a\n_b s_4 \\
\nonumber
&+&P^m_{1324}(f_1+f_2+f_3+f_4)\n^as_2\n^b s_3\n_a\n_b s_4 \\
\nonumber
&-&j_{1234}f_5(m_2^2+m_3^2+m_4^2-m_1^2-8)s_2\n_a s_3\n^a s_4,
\eea
where 
\bea
Y^a_{1234}&=&\frac{3(f_1-f_2)(f_3-f_4)}
{16\rho}\nonumber\\
Y^s_{1234}&=&\frac{3(f_1+f_2+f_3+f_4-2)(f_1+f_2+f_3+f_4-12)}
{8\rho}.\nonumber
\eea
It is convenient to get rid of the symmetric in 3, 4 4-derivative 
contribution by using the following identity 
\bea
\nonumber
w_{1234}f_5a_{125}a_{345}=
\frac{1}{3}w_{1234}( (f_1+f_2+f_3+f_4)a_{125}a_{345}+f_5(a_{135}a_{245}-a_{145}a_{235}) ),
\eea
valid for any function $w_{1234}$ symmetric in $2,3$.
The final contribution is given by
\bea
\nonumber
(\n_a^2-m_1^2)s_1&=&\left(Y^a_{1234}f_5l
 + \frac{1}{3}Y^s_{1234}f_5(n-m) \right)
\n^as_2\n^b s_3\n_a\n_b s_4 \\
\nonumber
&+&\l  \frac{1}{3}Y^s_{1234} + P^m_{1324}(f_1+f_2+f_3+f_4)\r \n^as_2\n^b s_3\n_a\n_b s_4 \\
\nonumber
&-&j_{1234}f_5(m_2^2+m_3^2+m_4^2-m_1^2-8)s_2\n_a s_3\n^a s_4.
\eea
Thus, we are left only with the antisymmetric 4-derivative 
Lagrangian contribution 
and the additional 4a-type terms without $f_5$. 
\vskip 0.5cm

{\bf V. 4-derivative terms without $f_5$. }

Just as above we use the shift
$$
s_1\to s_1+ j_{1234}s_2\n^bs_3\n_b s_4,
~~~~j_{1234}=\frac{1}{2}(w4b0)_{1234},
$$
to get rid of the $4b$ structure, and take into account
the additional contribution coming from the terms with $f_5$. 
Then we symmetrize the resulting 
coefficient $w4a0$ in 2 and 3, and decompose it into parts symmetric and antisymmetric in 3 and 4. Then we shift the symmetric part back to the 4b 
structure and get the equation 
\bea
(\n_a^2-m_1^2)s_1&=&(L4a0)_{1234}a_{125}a_{345}\n^as_2\n^b s_3\n_a\n_b s_4 \\
\nonumber
&+&(L4b0)_{1234}a_{125}a_{345}s_2\n^a \n^b s_3\n_a\n_b s_4 \\
\nonumber
&+&2(L4b0)_{1234}(m_2^2+m_3^2+m_4^2-m_1^2-8)a_{125}a_{345}
s_2\n_a s_3\n^a s_4\\
\nonumber
&-&j_{1234}(m_2^2+m_3^2+m_4^2-m_1^2-8)s_2\n_a s_3\n^a s_4,
\eea
where
\bea
(L4a0)_{1234}&=&-\frac{21(f_1-f_2)(f_3-f_4)}{16\rho}\nonumber\\
(L4b0)_{1234}&=&-\frac{7\left( 2f_1f_2+2f_3f_4-(f_1+f_2)(f_3+f_4)\right)}
{8\rho}.\nonumber
\eea
Both the 4-derivative terms are Lagrangian. The antisymmetric term
can be derived from a Lagrangian of the form (\ref{AA4}), and the symmetric
term can be obtain from the following Lagrangian
\bea
{\cal L}=S^{(4)}_{1234}\int s_1\n_as_2 \n_b^2(s_3\n^as_4),
\la{SS4}
\eea
where the quartic coupling $S^4_{1234}$ is $symmetric$ 
in 1, 2 and 3, 4, and $symmetric$ under the interchange (1,2) and (3,4),
and is given by
$$
S^{(4)}_{1234}=\frac{\kappa }{4}(L4b0)_{1234}.
$$
Equations of motion that follow from the Lagrangian are
\bea
\nonumber
\kappa(\n_a^2-m_1^2)s_1&=&4S^{(4)}_{1234}s_2\n^a\n^b s_3\n_a\n_b s_4 \\
\nonumber
&+&4(m_3^2+m_4^2-6)S^{(4)}_{1234}
s_2 \n^a s_3\n_a  s_4 \\
\nonumber
&+&(m_3^2+m_4^2-4)(m_3^2+m_4^2)S^{(4)}_{1234} s_2s_3 s_4.
\eea
This completes considering 4-derivative terms.
\subsection{2-derivative terms}
We proceed with 2-derivative terms for which, 
we should take into account all the additional contributions appeared 
 because of
 the shifts used in the previous subsections, and contributions which 
appear when one represents the 4-derivative terms as variations of 
the vertices of the types (\ref{AA4}) and (\ref{SS4}).
 
The coefficient $w2a$ contains four Lagrangian terms proportional to
$$f_5^2p_{125}p_{345}, \quad 
(f_5-1)^3t_{125}t_{345},\quad (f_5-1)^2t_{125}t_{345}
\quad \mbox{and}\quad
\frac{1}{f_5-5}a_{125}a_{345}$$ 
that can be found in Section 2.

We find convenient to represent the contribution of the other 
2-derivative terms in the form
\bea
\nonumber (\n_a^2-m_1^2)s_1&=&
(w2a)_{1234}a_{125}a_{345}\n^a s_2 s_3 \n_a s_4 \\
\nonumber
&+&(w2b)_{1234}a_{125}a_{345} s_2 \n^b s_3 \n_a s_4 ,
\eea
where the coefficients $w2a$ and $w2b$ may depend on $f_5$.

This equation is non-Lagrangian, 
and we again analyze it by decomposing the coefficients
$w2a$ and $w2b$ in Laurent series in $f_5$. 
\vskip 0.5cm

{\bf I. 2-derivative terms with $1/f_5$. }

Taking into account all additional contributions
we represent the 2-derivative contribution 
in the form
\bea\nonumber
(\n_a^2-m_1^2)s_1&=& (A2ad)_{1234}
\frac{1}{f_5}a_{125}a_{345} \n^as_2 s_3\n_a  s_4 \\
\nonumber
&+&(S2ad)_{1234}\frac{1}{f_5}a_{125}a_{345}\n^as_2 s_3\n_a  s_4 
+(L2bd)_{1234}\frac{1}{f_5}a_{125}a_{345}s_2 \n^a s_3\n_a  s_4 
\eea
Here we decompose the coefficient of the 2a type on the antisymmetric 
$A2ad$ and symmetric $S2ad$ parts with respect to permutation of the 
indices 3, 4.

The antisymmetric part is Lagrangian and the coefficient is given by 
\bea\nonumber
(A2ad)_{1234}&=&\frac{1}{2\rho }(f_1-f_2)(f_3-f_4)\\
&\times& (36+f_1+f_2+f_3+f_4-20(k_1+k_2+k_3+k_4) +10(k_1+k_2)(k_3+k_4) ).
\nonumber
\eea 
The corresponding 2-derivative term can be derived from 
the following Lagrangian
\bea
{\cal L}=A^{(2)}_{1234}\int s_1\n_as_2 s_3\n^a s_4 ,
\la{AA2}
\eea
where the quartic coupling $A^2_{1234}$ is $antisymmetric$ 
in 1, 2 and 3, 4, and $symmetric$ under the interchange (1,2) and (3,4). 
The equations of motion that follow from the Lagrangian are 
\bea\nonumber
\kappa(\n_a^2-m_1^2)s_1=-4A^{(2)}_{1234}\n^as_2 s_3\n_a s_4 
-(m_4^2-m_3^2)A^{(2)}_{1234} s_2s_3 s_4,
\eea
and, therefore, we have
$$
A^{(2)}_{1234}=-\frac{\kappa}{4}(A2ad)_{1234}\frac{1}{f_5}a_{125}a_{345}. 
$$
Now we shift the remaining type 2a structure to the type 2b one 
and get 
\bea
(\n_a^2-m_1^2)s_1&=& 
(A2ad)_{1234}\frac{1}{f_5}a_{125}a_{345} 
\n^as_2s_3\n_as_4 \\
\nonumber
&+&(L2bd-\frac12 S2ad)_{1234}\frac{1}{f_5}a_{125}a_{345} 
s_2\n^as_3\n_as_4 \\
\la{shden}
&-&\frac{1}{4}(S2ad)_{1234} (m_2^2+m_3^2+m_4^2-m_1^2)
\frac{1}{f_5}a_{125}a_{345}s_2s_3s_4. 
\eea
Now we see that the 2b structure turns out to be Lagrangian with
\bea
(S2bd)_{1234}=-\frac{1}{2 \rho}(k_1-k_2)(k_3-k_4)(f_1-f_2)(f_3-f_4).
\eea
The corresponding 2-derivative term can be derived from the Lagrangian
\bea
{\cal L}=S^{(2)}_{1234}\int s_1\n_as_2 s_3\n^a s_4 ,
\la{SS2}
\eea
where the quartic coupling $S^2_{1234}$ is $symmetric$ 
in 1, 2 and 3, 4, and $symmetric$ under the interchange (1,2) and (3,4). 
The equations of motion that follow from the Lagrangian are 
\bea\nonumber
\kappa(\n_a^2-m_1^2)s_1=2S^{(2)}_{1234} s_2 \n^a s_3\n_a s_4
+(m_4^2+m_3^2)S^{(2)}_{1234} s_2s_3 s_4,
\eea
and, therefore, we have
$$
S^{(2)}_{1234}=\frac{\kappa}{2}(S2bd)_{1234}\frac{1}{f_5}a_{125}a_{345}.
$$
\vskip 0.5cm
We omit considering terms with $f_5^4$, $f_5^3$ and $f_5^2$,
because their analisys goes the same line as before. We just remark 
that we used the shift
$$s_1\to s_1+J_{1234}s_2s_3s_4$$  
to remove the terms completely symmetric with respect
to permutation of indices 2, 3 and 4 from the equations of motion. 
\vskip 0.5cm

{\bf II. 2-derivative terms with $f_5$. }

By using a field redefinition, we shift the type 2b term 
to the type 2a one, and represent the equation in the form
\bea
(\n_a^2-m_1^2)s_1=((w2)^l_{1234}l+(w2)^n_{1234}n+(w2)^m_{1432}m) 
f_5\n^as_2 s_3 \n_a s_4. 
\nonumber 
\eea
This equation can be further rewritten as follows
\bea\nonumber
(\n_a^2-m_1^2)s_1&=&((w2)^l_{1234}-(w2)^n_{1234}
-(w2)^n_{1432}+(w2)^m_{1432}) 
f_5a_{125}a_{345}\n^as_2 s_3 \n_a s_4 \\
\nonumber 
&+&(w2)^n_{1234}(f_1+f_2+f_3+f_4)\n^as_2 s_3 \n_a s_4 ,
\eea
where we use identity (\ref{imre}) and the symmetry of the tensor
$\n^as_2 s_3 \n_a s_4 $ in 2 and 4. Introducing the notation
\bea
F_{1234}=(w2)^l_{1234}-(w2)^n_{1234}
-(w2)^n_{1432}+(w2)^m_{1432},\nonumber
\eea
and by using again (\ref{imre}) we rewrite the equation as follows
\bea
\nonumber
(\n_a^2-m_1^2)s_1&=& (F^a+\frac{1}{3}F^s)_{1234}f_5 a_{125}a_{345}
\n^as_2s_3\n_as_4 \\\nonumber
&-&\frac{1}{3}F^s_{1324}f_5 a_{125}a_{345} 
s_2\n^as_3\n_a s_4 \\
\nonumber
&+&
\l (w2)^n_{1234}+\frac{1}{3} F^s_{1234}\r 
(f_1+f_2+f_3+f_4)a_{125}a_{345} \n^as_2 s_3\n_a  s_4 ,
\eea
where  $F^a$ and $F^s$
denote the parts of $F$ antisymmetric and symmetric in 2 and 4 
respectively. Finally we use a field redefinition to shift the
2b structure to the 2a one, decompose the resulting 2a coefficient
into parts symmetric and antisymmetric in 3 and 4
$$(F^a+\frac{1}{3}F^s)_{1234}+\frac{2}{3}F^s_{1324}=S_{1234}+A_{1234},$$
 and shift the symmetric part to 
the 2b structure. The resulting equation with Lagrangian 2-derivative 
terms $A_{1234}$ and $S_{1234}$ looks as follows
\bea
(\n_a^2-m_1^2)s_1&=&A_{1234}f_5 a_{125}a_{345}
\n^as_2 s_3\n_a  s_4 \\
\nonumber
&-&\frac12 S_{1234}f_5 a_{125}a_{345}
 s_2 \n^as_3\n_a  s_4 \\
\nonumber
&+&\l (w2)^n+\frac{1}{3} F^s\r_{1234} 
(f_1+f_2+f_3+f_4)a_{125}a_{345} \n^as_2 s_3\n_a  s_4 \\
\nonumber 
&-&\frac{1}{4}S_{1234}(m_2^2+m_3^2+m_4^2-m_1^2)f_5 
a_{125}a_{345}s_2s_3s_4 \\
\nonumber 
&+&\frac{1}{6}F^s_{1234}(m_2^2+m_3^2+m_4^2-m_1^2)f_5
 a_{125}a_{345} s_2s_3s_4.
\eea
The consideration of the terms without $f_5$ is simple. We sum up all the
additional contributions, shift the 2b structure to the 2a one, and finally
we use a shift to remove the part symmetric in 2, 3 and 4. After these steps
we obtain a Lagrangian term. As before we shift the symmetric part 
to the 2b structure to have a simple Lagrangian.
\subsection{Non-derivative terms}
The consideration of non-derivative terms is the simplest one. Summing
up all contributions we immediately obtain Lagrangian terms for all 
cases except the case with $f_5$ and without $f_5$. The equation of motion
for the term with $f_5$ has the form
\bea\nonumber
(\n_a^2-m_1^2)s_1&=&Q_{1234}f_5a_{125}a_{345}s_2s_3s_4.
\eea
 Now we write the equation as 
\bea
(\n_a^2-m_1^2)s_1&=&\frac{f_5}{3}\l 2 Q_{1234}a_{125}a_{345}
+ Q_{1324}a_{135}a_{245} \r s_2s_3s_4
\eea
and then apply the identity (\ref{imre}). We get
\bea
(\n_a^2-m_1^2)s_1&=&\frac{f_5}{3}\l 2Q_{1234} 
- Q_{1324}- Q_{1342} \r a_{125}a_{345}s_2s_3s_4 \\
\nonumber
&+&\frac{1}{3}Q_{1324}(f_1+f_2+f_3+f_4)s_2s_3s_4.
\eea
Here the term:
$$\frac{f_5}{3}\l 2Q_{1234} - Q_{1324}- Q_{1342} \r a_{125}a_{345}
$$
appears to be Lagrangian, and the additional term without $f_5$
makes the total contribution to the term without $f_5$ Lagrangian
as well.

Thus we showed  that the equations of motion for the scalars $s^I$
can be reduced to the Lagrangian form by means of a number of 
field redefinitions.
\section{Conclusion}
In this paper we derived all quartic couplings of the scalars dual to 
extended chiral primary operators in $\N =4$ SYM$_4$ by using the 
covariant equations of motion for type IIB supergravity. 
The quartic terms 
appeared to contain vertices with two and four derivatives. 
The appearance of 2-derivative vertices was of course expected. 
Some of the 4-derivative
vertices may be removed by such a field redefinition that changes the
structure of cubic terms, namely, one gets scalar cubic terms with two 
derivatives, and cubic terms describing non-minimal interaction 
of two scalars with vector
fields of the form $V_{IJK}\n^a s^I \n^b s^J F_{ab}^K$. However, we do 
not know if all of the 4-derivative terms can be removed in such a way.
It would be interesting to clarify this point because the derivation of
the Callan-Symanzik equations in the AdS/CFT framework performed in 
\cite{BVV} was based on a gravity action which does not contain terms
with four or more derivatives. 

Since we know the gravity action for the scalars $s^I$ up to the 
fourth order, we can start computing 4-point functions of CPOs.
In general this will require calculating two new 
types of Feynman diagrams:
$(i)$ contact diagrams with 4-derivative vertices, 
and $(ii)$ exchange diagrams
involving massive gravitons. 
It is not difficult to show that all contact diagrams with 4-derivative
vertices can be reduced to a sum of terms corresponding to 
simple non-derivative quartic couplings, just as this was done 
in  \cite{HFMMR} for the case of
contact diagrams with 2-derivative vertices. Thus the only real problem 
is to compute the exchange diagrams involving massive gravitons.
However, the 4-point
functions of CPOs $O_2$ can be easily found because 
all necessary diagrams have been already
calculated. This problem is now under consideration. 

We proved that, as was conjectured in \cite{AF5}, 
the quartic couplings obtained vanish in the extremal case, for which
$k_1=k_2+k_3+k_4$. This also implies the non-renormalization of
extremal 4-point functions of single-trace CPOs. 
The vanishing of the quartic couplings is not manifest, and requires
an additional field redefinition. Although the quartic couplings can be 
easily used for computing any 4-point function of CPOs, 
it would be useful to find such a 
representation for the quartic couplings that makes the vanishing 
in the extremal case explicit.

We showed that the quartic couplings admit the consistent KK truncation, 
and argued that the consistency of the KK reduction implies a
non-renormalization theorem of $n$-point functions of $n-1$ 
single-trace operators dual to the fields from the massless multiplet
and one single-trace operator dual to a field from a massive multiplet.
It would be interesting to check the non-renormalization of the 5-point
function of four CPOs $O_2$ and one CPO $O_4$ in perturbation theory.

The simplest example of the 4-point 
function of three CPOs $O_2$ and a CPO $O_4$ belongs, actually, to
the class of so-called "next-to-extremal" 4-point
functions, for which $k_1=k_2+k_3+k_4-2$. 
The non-renormalization of such correlation functions 
was proven in \cite{EHSSW2},
and very recently checked to first order in perturbation theory in
\cite{EP}.
The non-renormalization theorem also implies the vanishing of 
the corresponding functions of extended CPOs and, since
it is not difficult to show that there is no exchange diagram 
in this case, the corresponding "next-to-extremal" quartic 
couplings of scalars $s^I$ have to vanish too. It would be 
interesting to check this.

\vskip 1cm
{\bf Note added.}

We have recently shown that the relevant part of the gauged
${\cal N}=8$ 5-dimensional 
supergravity action coincides with the 
action for the scalar $s_2$ we found 
in the paper.

\section{Appendix A}
\setcounter{equation}{0}
Here we collect the quartic couplings of the scalars $s^I$ representing
our main result. The couplings are given by sums of terms depending on
various independent $SO(6)$ tensors. To simplify 
the presentation we sometimes use the following notations
$$x\equiv k_1,\quad y\equiv k_2,\quad t\equiv k_3,\quad 
w\equiv k_4,\quad z\equiv k_5,$$
$$\d =(x+1)(y+1)(t+1)(w+1).$$
All the $SO(6)$ tensors are given by tensors 
of the form $F(f_5) a_{I_1I_2I_5}a_{I_3I_4I_5}$,
$(f_5-1)^nt_{I_1I_2I_5}t_{I_3I_4I_5}$ and
$f_5^n p_{I_1I_2I_5}p_{I_3I_4I_5}$, where $F(f_5)$ is a function of $f_5$,
and summation over the index $I_5$ is assumed. To distinguish the couplings 
with different functions $F$ we use an additional subscript in  notation 
of a coupling.

\noindent {\bf Quartic couplings of 4-derivative vertices}
\bea
(A_3)^{(4)}_{I_1I_2I_3I_4}&=&\frac{1}{4\d}
f_5^3\l a_{145}a_{235}-a_{135}a_{245}\r .\nonumber\\
(A_2)^{(4)}_{I_1I_2I_3I_4}&=&-\frac{1}{4\d}(3(f_1+f_2+f_3+f_4)-28)
f_5^2\l a_{145}a_{235}-a_{135}a_{245}\r .\nonumber\\
(A_1)^{(4)}_{I_1I_2I_3I_4}&=&-\frac{3}{4\d}(f_1-f_2)(f_3-f_4)
f_5 a_{125}a_{345} \nonumber\\
&-&\frac{1}{\d}(f_1+f_2+f_3+f_4-2)(f_1+f_2+f_3+f_4-12)
f_5\l a_{145}a_{235}-a_{135}a_{245}\r .\nonumber\\
(A_0)^{(4)}_{I_1I_2I_3I_4}&=&\frac{21}{4\d}(f_1-f_2)(f_3-f_4)
 a_{125}a_{345}.\nonumber\\
(S_0)^{(4)}_{I_1I_2I_3I_4}&=&
\frac{7}{4\d}\left( 2f_1f_2+2f_3f_4-(f_1+f_2)(f_3+f_4)\right)
 a_{125}a_{345}.\nonumber\\
(A_{-1})^{(4)}_{I_1I_2I_3I_4}&=&-\frac{12}{\d}(f_1-f_2)(f_3-f_4)
 f_5^{-1}a_{125}a_{345}.\nonumber\\
(A_{t2})^{(4)}_{I_1I_2I_3I_4}&=&-\frac{3}{\d}(f_5-1)^2t_{125}t_{345}.
\nonumber
\eea
{\bf Quartic couplings of 2-derivative vertices}
\bea
(A_4)^{(2)}_{I_1I_2I_3I_4}&=&
\frac{5}{48\d}
 f_5^4 \l a_{145}a_{235}-a_{135}a_{245}\r .\nonumber\\
(A_3)^{(2)}_{I_1I_2I_3I_4}&=&-\frac{1}{2\d}(k_1-k_2)(k_3-k_4)
f_5^3 a_{125}a_{345} .\nonumber\\
(S_3)^{(2)}_{I_1I_2I_3I_4}&=&\frac{1}{16\d}\biggl( 137 -
80(k_1+k_2+k_3+k_4)+2(f_1+f_2+f_3+f_4)\cr
&+&
32(k_1k_2+k_3k_4)+24(k_1+k_2)(k_3+k_4)\biggl)
f_5^3 a_{125}a_{345} .\nonumber\\
(A_2)^{(2)}_{I_1I_2I_3I_4}&=&\frac{(k_1-k_2)(k_3-k_4)}{4\d}
\biggl( 40 -
12(k_1+k_2+k_3+k_4)+2(f_1+f_2+f_3+f_4)\cr
&+&
16(k_1k_2+k_3k_4)+(k_1+k_2)(k_3+k_4)\biggr)
f_5^2 a_{125}a_{345} .\nonumber\\
(S_2)^{(2)}_{I_1I_2I_3I_4}&=&-\frac{1}{16\d}
 \biggl( -3741 + 2984t - 342{t^2} - 
56{t^3} + 31{t^4} + 2984w - 2272tw + 376{t^2}w  \cr
&+& 
128{t^3}w - 342{w^2} 
+ 376t{w^2} +  42{t^2}{w^2} - 56{w^3} + 128t{w^3} + 31{w^4} + 
2984x \cr
&-& 
1760tx + 144{t^2}x + 88{t^3}x - 1760wx + 832twx + 
88{t^2}wx +  144{w^2}x 
+88t{w^2}x \cr
&+&
 88{w^3}x - 342{x^2} + 144t{x^2} 
+40{t^2}{x^2} + 144w{x^2} + 192tw{x^2} 
+40{w^2}{x^2} - 56{x^3} \cr
&+& 
88t{x^3} +88w{x^3} +31{x^4} + 2984y - 1760ty + 144{t^2}y +
 88{t^3}y - 1760wy \cr
&+& 
832twy + 88{t^2}wy 
+ 144{w^2}y 
+ 88t{w^2}y + 88{w^3}y - 2272xy + 832txy 
+ 192{t^2}xy \cr
&+&  
832wxy - 128twxy + 192{w^2}xy 
+376{x^2}y +  88t{x^2}y + 88w{x^2}y + 128{x^3}y - 
342{y^2}\cr
&+&
 144t{y^2} +  40{t^2}{y^2} + 144w{y^2} 
+ 192tw{y^2} + 
40{w^2}{y^2} +  376x{y^2} + 88tx{y^2} + 88wx{y^2}\cr
&+& 
42{x^2}{y^2} 
- 56{y^3}+ 
88t{y^3} + 88w{y^3} 
+  128x{y^3} + 31{y^4} \biggr)
f_5^2 a_{125}a_{345}  .\nonumber\\
(A_1)^{(2)}_{I_1I_2I_3I_4}&=&
\frac{(t-w)(x-y)}{48\d}
\biggl(  -1840 - 1964t + 160{t^2} + 156{t^3} + 16{t^4} - 1964w \cr
&+&
 1312tw -  388{t^2}w - 128{t^3}w + 160{w^2} - 388t{w^2} - 
120{t^2}{w^2} +  156{w^3} - 128t{w^3}\cr
&+&
 16{w^4} - 1964x + 645tx - 
48{t^2}x -  25{t^3}x + 645wx - 952twx - 73{t^2}wx \cr
&-& 
48{w^2}x - 73t{w^2}x - 25{w^3}x + 160{x^2} - 48t{x^2} - 
56{t^2}{x^2} -  48w{x^2} - 328tw{x^2}\cr
&-& 
56{w^2}{x^2} + 
156{x^3} - 25t{x^3} - 25w{x^3} + 16{x^4} - 1964y + 645ty - 48{t^2}y \cr
&-& 
25{t^3}y + 645wy - 952twy - 73{t^2}wy - 48{w^2}y - 
73t{w^2}y -  25{w^3}y + 1312xy\cr
&-& 
952txy - 328{t^2}xy 
- 952wxy -  656twxy - 328{w^2}xy - 388{x^2}y - 
73t{x^2}y\cr
&-& 73w{x^2}y - 
128{x^3}y + 160{y^2} - 48t{y^2} - 
56{t^2}{y^2} - 48w{y^2} - 328tw{y^2} - 56{w^2}{y^2}\cr
&-& 
388x{y^2}-
 73tx{y^2} - 73wx{y^2} - 120{x^2}{y^2} + 156{y^3} - 
25t{y^3} - 25w{y^3}\cr
&-& 
128x{y^3} + 16{y^4}\biggr)
f_5 a_{125}a_{345} \nonumber\\
(S_1)^{(2)}_{I_1I_2I_3I_4}&=&\frac{1}{48\d}\biggl(
20979 - 53784t + 18666{t^2} + 4056{t^3} - 
1197{t^4} + 192{t^5} 
+72{t^6} \cr
&-& 53784w + 59648tw - 17792{t^2}w - 
2816{t^3}w + 
       1896{t^4}w + 256{t^5}w + 18666{w^2} \cr
&-&
 17792t{w^2} 
+ 2736{t^2}{w^2} + 
       1344{t^3}{w^2} + 98{t^4}{w^2} + 4056{w^3} - 
2816t{w^3} + 
       1344{t^2}{w^3} \cr
&+& 256{t^3}{w^3} - 1197{w^4} + 
1896t{w^4} + 
       98{t^2}{w^4} + 192{w^5} + 256t{w^5} + 72{w^6} \cr
&-&
53784x + 65168tx - 
       11900{t^2}x - 3296{t^3}x + 1428{t^4}x + 
208{t^5}x + 65168wx\cr
&-& 
 53760twx + 7296{t^2}wx + 4000{t^3}wx + 
144{t^4}wx - 
       11900{w^2}x + 7296t{w^2}x \cr
&+& 1760{t^2}{w^2}x 
+ 
104{t^3}{w^2}x - 
       3296{w^3}x + 4000t{w^3}x + 104{t^2}{w^3}x + 
1428{w^4}x \cr
&+& 
       144t{w^4}x + 208{w^5}x 
+ 18666{x^2} - 
11900t{x^2} + 
       801{t^2}{x^2} + 1488{t^3}{x^2} \cr
&+& 
173{t^4}{x^2} - 
11900w{x^2} + 
       3840tw{x^2}
+ 4472{t^2}w{x^2} + 
704{t^3}w{x^2} + 801{w^2}{x^2} \cr
&+& 
       4472t{w^2}{x^2} + 252{t^2}{w^2}{x^2} + 
1488{w^3}{x^2} + 
       704t{w^3}{x^2} 
+
 173{w^4}{x^2} + 4056{x^3} \cr
&-& 
3296t{x^3} + 
       1488{t^2}{x^3} + 424{t^3}{x^3} - 3296w{x^3} + 
5632tw{x^3} 
+ 
       464{t^2}w{x^3} \cr
&+& 1488{w^2}{x^3} + 
464t{w^2}{x^3} + 424{w^3}{x^3} - 
       1197{x^4} + 1428t{x^4} + 173{t^2}{x^4} \cr
&+& 
1428w{x^4} + 576tw{x^4} + 
       173{w^2}{x^4} + 192{x^5} + 208t{x^5} + 208w{x^5} 
+ 72{x^6} \cr
&-& 
       53784y + 65168ty - 11900{t^2}y - 3296{t^3}y + 
1428{t^4}y + 
       208{t^5}y + 65168wy \cr
&-& 
53760twy + 7296{t^2}wy + 4000{t^3}wy + 
       144{t^4}wy - 11900{w^2}y + 7296t{w^2}y \cr
&+& 
1760{t^2}{w^2}y + 
       104{t^3}{w^2}y - 3296{w^3}y + 4000t{w^3}y + 
104{t^2}{w^3}y + 
       1428{w^4}y \cr
&+&
144t{w^4}y + 208{w^5}y +
59648xy - 53760txy + 
       3840{t^2}xy + 5632{t^3}xy \cr
&+& 576{t^4}xy - 
53760wxy + 
       23040twxy 
+ 3264{t^2}wxy - 384{t^3}wxy + 3840{w^2}xy \cr
&+& 
       3264t{w^2}xy - 128{t^2}{w^2}xy + 
5632{w^3}xy - 
       384t{w^3}xy 
+ 576{w^4}xy - 17792{x^2}y \cr
&+& 
7296t{x^2}y + 
       4472{t^2}{x^2}y + 464{t^3}{x^2}y + 
7296w{x^2}y + 
       3264tw{x^2}y 
+40{t^2}w{x^2}y \cr
&+& 
4472{w^2}{x^2}y + 
       40t{w^2}{x^2}y + 464{w^3}{x^2}y - 
2816{x^3}y + 4000t{x^3}y + 
       704{t^2}{x^3}y \cr
&+& 4000w{x^3}y - 
384tw{x^3}y + 
       704{w^2}{x^3}y + 1896{x^4}y + 144t{x^4}y + 144w{x^4}y + 
       256{x^5}y \cr
&+&
 18666{y^2} - 11900t{y^2} + 
801{t^2}{y^2} + 
       1488{t^3}{y^2} + 173{t^4}{y^2} - 11900w{y^2} + 
3840tw{y^2} \cr
&+& 
       4472{t^2}w{y^2} + 704{t^3}w{y^2} + 
801{w^2}{y^2} + 
       4472t{w^2}{y^2} + 252{t^2}{w^2}{y^2} + 
1488{w^3}{y^2} \cr
&+& 
       704t{w^3}{y^2} 
+173{w^4}{y^2} - 17792x{y^2} + 
7296tx{y^2} + 
       4472{t^2}x{y^2} + 464{t^3}x{y^2} \cr
&+& 
7296wx{y^2} + 
       3264twx{y^2} 
+ 40{t^2}wx{y^2} + 
4472{w^2}x{y^2} + 
       40t{w^2}x{y^2} + 464{w^3}x{y^2} \cr
&+& 
2736{x^2}{y^2} + 
1760w{x^2}{y^2} - 
       128tw{x^2}{y^2} +
 252{w^2}{x^2}{y^2} + 
1344{x^3}{y^2} + 
       104t{x^3}{y^2} \cr
&+& 104w{x^3}{y^2} + 98{x^4}{y^2} 
+ 4056{y^3} - 
       3296t{y^3} 
+ 1488{t^2}{y^3} + 424{t^3}{y^3} \cr
&-& 
3296w{y^3} + 
       5632tw{y^3} + 464{t^2}w{y^3} + 1488{w^2}{y^3} 
+ 464t{w^2}{y^3} +
       424{w^3}{y^3} \cr
&-& 2816x{y^3} + 4000tx{y^3} + 
704{t^2}x{y^3} + 
       4000wx{y^3} - 384twx{y^3} + 
704{w^2}x{y^3} \cr
&+& 1344{x^2}{y^3} + 
       104t{x^2}{y^3} + 104w{x^2}{y^3} + 
256{x^3}{y^3} - 1197{y^4} + 
       1428t{y^4} + 173{t^2}{y^4} \cr
&+& 1428w{y^4} + 
576tw{y^4} + 
       173{w^2}{y^4} + 1896x{y^4} + 144tx{y^4} + 144wx{y^4} + 
       98{x^2}{y^4} \cr
&+& 192{y^5} + 208t{y^5} + 208w{y^5} + 
256x{y^5} + 
       72{y^6}
\biggr) f_5a_{125}a_{345}.\nonumber\\
(A_0)^{(2)}_{I_1I_2I_3I_4}&=&-\frac{(x-y)(t-w)}{192\d}\biggl(
 -144288 + 74776t + 10752{t^2} - 5264{t^3} \cr
&+& 
992{t^4} + 440{t^5} + 
         32{t^6} + 74776w + 37504tw - 11664{t^2}w \cr
&+& 
2016{t^3}w + 
         1400{t^4}w + 128{t^5}w + 10752{w^2} - 
11664t{w^2} + 
         4512{t^2}{w^2} \cr
&+& 2088{t^3}{w^2} + 176{t^4}{w^2} - 
5264{w^3} + 
         2016t{w^3} + 2088{t^2}{w^3} + 256{t^3}{w^3} \cr
&+& 
992{w^4} + 
         1400t{w^4} + 176{t^2}{w^4} + 440{w^5} + 
128t{w^5} + 32{w^6} \cr
&+& 
         74776x - 26042tx - 5888{t^2}x + 3948{t^3}x + 
888{t^4}x + 
         46{t^5}x \cr
&-& 26042wx - 7648twx + 6380{t^2}wx + 1784{t^3}wx + 
         142{t^4}wx - 5888{w^2}x \cr
&+& 6380t{w^2}x + 1880{t^2}{w^2}x + 
         170{t^3}{w^2}x + 3948{w^3}x + 1784t{w^3}x + 
170{t^2}{w^3}x \cr
&+& 
         888{w^4}x + 142t{w^4}x + 46{w^5}x + 
10752{x^2} - 5888t{x^2} + 
         832{t^2}{x^2} \cr
&+& 1272{t^3}{x^2} + 160{t^4}{x^2} - 
5888w{x^2} + 
         768tw{x^2} + 1784{t^2}w{x^2} + 
144{t^3}w{x^2} \cr
&+& 
         832{w^2}{x^2} + 1784t{w^2}{x^2} + 
16{t^2}{w^2}{x^2} + 
         1272{w^3}{x^2} + 144t{w^3}{x^2} + 160{w^4}{x^2}\cr 
&-& 5264{x^3} + 
         3948t{x^3} + 1272{t^2}{x^3} + 5{t^3}{x^3} + 
3948w{x^3} + 
         1480tw{x^3} \cr
&-& 91{t^2}w{x^3} + 1272{w^2}{x^3} 
- 91t{w^2}{x^3} + 
         5{w^3}{x^3} + 992{x^4} + 888t{x^4} \cr
&+& 
160{t^2}{x^4} + 888w{x^4} + 
         208tw{x^4} + 160{w^2}{x^4} + 440{x^5} + 
46t{x^5} \cr
&+& 46w{x^5} + 
         32{x^6} + 74776y - 26042ty - 5888{t^2}y + 
3948{t^3}y \cr
&+& 
         888{t^4}y + 46{t^5}y - 26042wy - 7648twy 
+ 6380{t^2}wy + 
         1784{t^3}wy \cr
&+& 142{t^4}wy - 5888{w^2}y + 6380t{w^2}y + 
         1880{t^2}{w^2}y + 170{t^3}{w^2}y + 
3948{w^3}y \cr
&+& 1784t{w^3}y + 
         170{t^2}{w^3}y + 888{w^4}y + 142t{w^4}y + 46{w^5}y + 
         37504xy \cr
&-& 7648txy + 768{t^2}xy + 
1480{t^3}xy + 
         208{t^4}xy - 7648wxy + 448twxy \cr
&+& 
1608{t^2}wxy + 
         32{t^3}wxy + 768{w^2}xy + 
1608t{w^2}xy - 
         352{t^2}{w^2}xy + 1480{w^3}xy \cr
&+& 
32t{w^3}xy + 
         208{w^4}xy - 11664{x^2}y + 6380t{x^2}y + 
1784{t^2}{x^2}y - 
         91{t^3}{x^2}y \cr
&+& 6380w{x^2}y + 1608tw{x^2}y - 
         571{t^2}w{x^2}y + 1784{w^2}{x^2}y - 
571t{w^2}{x^2}y - 
         91{w^3}{x^2}y \cr
&+& 2016{x^3}y + 1784t{x^3}y + 
144{t^2}{x^3}y + 
         1784w{x^3}y + 32tw{x^3}y + 
144{w^2}{x^3}y \cr
&+& 1400{x^4}y + 
         142t{x^4}y + 142w{x^4}y + 128{x^5}y + 
10752{y^2} - 
         5888t{y^2} \cr
&+& 832{t^2}{y^2} + 1272{t^3}{y^2} + 
160{t^4}{y^2} - 
         5888w{y^2} + 768tw{y^2} + 1784{t^2}w{y^2} \cr
&+& 
144{t^3}w{y^2} + 
         832{w^2}{y^2} + 1784t{w^2}{y^2} + 
16{t^2}{w^2}{y^2} + 
         1272{w^3}{y^2} + 144t{w^3}{y^2} \cr
&+& 160{w^4}{y^2} 
- 11664x{y^2} + 
         6380tx{y^2} + 1784{t^2}x{y^2} - 
91{t^3}x{y^2} + 
         6380wx{y^2} \cr
&+& 1608twx{y^2} - 
571{t^2}wx{y^2} + 
         1784{w^2}x{y^2} - 571t{w^2}x{y^2} - 
91{w^3}x{y^2} + 
         4512{x^2}{y^2} \cr
&+& 1880t{x^2}{y^2} + 16{t^2}{x^2}{y^2} + 
         1880w{x^2}{y^2} - 352tw{x^2}{y^2} + 
16{w^2}{x^2}{y^2} + 
         2088{x^3}{y^2} \cr
&+& 170t{x^3}{y^2} + 
170w{x^3}{y^2} + 
         176{x^4}{y^2} - 5264{y^3} + 3948t{y^3} + 1272{t^2}{y^3} \cr
&+& 
         5{t^3}{y^3} + 3948w{y^3} + 1480tw{y^3} - 
91{t^2}w{y^3} + 
         1272{w^2}{y^3} - 91t{w^2}{y^3} \cr
&+& 5{w^3}{y^3} + 
2016x{y^3} + 
         1784tx{y^3} + 144{t^2}x{y^3} + 
1784wx{y^3} + 32twx{y^3} \cr
&+& 
         144{w^2}x{y^3} + 2088{x^2}{y^3} + 
170t{x^2}{y^3} + 
         170w{x^2}{y^3} + 256{x^3}{y^3} + 992{y^4} \cr
&+& 
888t{y^4} + 
         160{t^2}{y^4} + 888w{y^4} + 208tw{y^4} + 
160{w^2}{y^4} + 
         1400x{y^4} \cr
&+& 142tx{y^4} + 142wx{y^4} + 
176{x^2}{y^4} + 
         440{y^5} + 46t{y^5} + 46w{y^5} \cr
&+& 128x{y^5} + 32{y^6} 
\biggr) a_{125}a_{345}.\nonumber\\
(S_0)^{(2)}_{I_1I_2I_3I_4}&=&\frac{1}{576\d}\biggl(
-288576tw + 149552{t^2}w + 21504{t^3}w - 
10528{t^4}w \cr
&+& 
     1984{t^5}w + 880{t^6}w + 64{t^7}w + 149552t{w^2} 
- 52084{t^2}{w^2} - 
     11776{t^3}{w^2} \cr
&+& 7896{t^4}{w^2} + 1776{t^5}{w^2} + 92{t^6}{w^2} + 
     21504t{w^3} - 11776{t^2}{w^3} + 1664{t^3}{w^3} \cr
&+& 
2544{t^4}{w^3} + 
     320{t^5}{w^3} - 10528t{w^4} + 7896{t^2}{w^4} + 
2544{t^3}{w^4} + 
     10{t^4}{w^4} \cr
&+& 1984t{w^5} + 1776{t^2}{w^5} + 
320{t^3}{w^5} + 
     880t{w^6} + 92{t^2}{w^6} + 64t{w^7} \cr
&+& 144288tx - 
74776{t^2}x - 
     10752{t^3}x + 5264{t^4}x - 992{t^5}x - 440{t^6}x \cr
&-& 32{t^7}x + 
     144288wx + 26752{t^2}wx - 6400{t^3}wx + 
1024{t^4}wx + 
     960{t^5}wx \cr
&+& 96{t^6}wx - 74776{w^2}x + 26752t{w^2}x - 
     3520{t^2}{w^2}x + 2368{t^3}{w^2}x + 
1104{t^4}{w^2}x \cr
&+& 
     144{t^5}{w^2}x - 10752{w^3}x - 6400t{w^3}x + 
2368{t^2}{w^3}x + 
     1024{t^3}{w^3}x - 112{t^4}{w^3}x \cr
&+& 5264{w^4}x + 
1024t{w^4}x + 
     1104{t^2}{w^4}x - 112{t^3}{w^4}x - 992{w^5}x + 
960t{w^5}x \cr
&+& 
     144{t^2}{w^5}x - 440{w^6}x + 96t{w^6}x - 
32{w^7}x - 
     74776t{x^2} + 26042{t^2}{x^2} \cr
&+& 5888{t^3}{x^2} - 
3948{t^4}{x^2} - 
     888{t^5}{x^2} - 46{t^6}{x^2} - 74776w{x^2} - 53504tw{x^2} \cr
&+& 
     1760{t^2}w{x^2} + 2560{t^3}w{x^2} + 
480{t^4}w{x^2} + 
     26042{w^2}{x^2} + 1760t{w^2}{x^2} + 
96{t^3}{w^2}{x^2} \cr
&+& 
     28{t^4}{w^2}{x^2} + 5888{w^3}{x^2} + 
2560t{w^3}{x^2} + 
     96{t^2}{w^3}{x^2} - 256{t^3}{w^3}{x^2} - 
3948{w^4}{x^2} \cr
&+& 
     480t{w^4}{x^2} + 28{t^2}{w^4}{x^2} - 
888{w^5}{x^2} - 46{w^6}{x^2} - 
     10752t{x^3} + 5888{t^2}{x^3} \cr
&-& 832{t^3}{x^3} - 
1272{t^4}{x^3} - 
     160{t^5}{x^3} - 10752w{x^3} + 12800tw{x^3} - 
4928{t^2}w{x^3} \cr
&-& 
     512{t^3}w{x^3} + 176{t^4}w{x^3} + 5888{w^2}{x^3} 
- 
     4928t{w^2}{x^3} - 192{t^2}{w^2}{x^3} + 
128{t^3}{w^2}{x^3} \cr
&-& 
     832{w^3}{x^3} - 512t{w^3}{x^3} + 
128{t^2}{w^3}{x^3} - 
     1272{w^4}{x^3} + 176t{w^4}{x^3} - 160{w^5}{x^3} \cr
&+& 
5264t{x^4} - 
     3948{t^2}{x^4} - 1272{t^3}{x^4} - 5{t^4}{x^4} + 
5264w{x^4} - 
     2048tw{x^4} \cr
&-& 1584{t^2}w{x^4} - 64{t^3}w{x^4} 
- 3948{w^2}{x^4} - 
     1584t{w^2}{x^4} - 56{t^2}{w^2}{x^4} - 
1272{w^3}{x^4} \cr
&-& 
     64t{w^3}{x^4} - 5{w^4}{x^4} - 992t{x^5} - 
888{t^2}{x^5} - 
     160{t^3}{x^5} - 992w{x^5} \cr
&-& 1920tw{x^5} - 
144{t^2}w{x^5} - 
     888{w^2}{x^5} - 144t{w^2}{x^5} - 160{w^3}{x^5} - 
440t{x^6} \cr
&-& 
     46{t^2}{x^6} - 440w{x^6} - 192tw{x^6} - 
46{w^2}{x^6} - 32t{x^7} - 
     32w{x^7} \cr
&+& 144288ty - 74776{t^2}y - 10752{t^3}y 
+ 5264{t^4}y - 
     992{t^5}y - 440{t^6}y \cr
&-& 32{t^7}y + 144288wy + 
26752{t^2}wy - 
     6400{t^3}wy + 1024{t^4}wy + 960{t^5}wy \cr
&+& 
96{t^6}wy - 
     74776{w^2}y + 26752t{w^2}y - 3520{t^2}{w^2}y + 
2368{t^3}{w^2}y + 
     1104{t^4}{w^2}y \cr
&+& 144{t^5}{w^2}y - 10752{w^3}y - 
6400t{w^3}y + 
     2368{t^2}{w^3}y + 1024{t^3}{w^3}y - 
112{t^4}{w^3}y \cr
&+& 5264{w^4}y + 
     1024t{w^4}y + 1104{t^2}{w^4}y - 
112{t^3}{w^4}y - 992{w^5}y + 
     960t{w^5}y \cr
&+& 144{t^2}{w^5}y - 440{w^6}y + 96t{w^6}y - 
     32{w^7}y - 288576xy - 53504{t^2}xy \cr
&+& 
12800{t^3}xy - 
     2048{t^4}xy - 1920{t^5}xy - 192{t^6}xy - 
53504{w^2}xy - 
     512{t^2}{w^2}xy \cr
&-& 768{t^3}{w^2}xy - 
320{t^4}{w^2}xy + 
     12800{w^3}xy - 768{t^2}{w^3}xy - 
768{t^3}{w^3}xy - 
     2048{w^4}xy \cr
&-& 320{t^2}{w^4}xy - 
1920{w^5}xy - 192{w^6}xy + 
     149552{x^2}y + 26752t{x^2}y + 1760{t^2}{x^2}y \cr
&-& 
4928{t^3}{x^2}y - 
     1584{t^4}{x^2}y - 144{t^5}{x^2}y + 
26752w{x^2}y + 
     256{t^2}w{x^2}y + 384{t^3}w{x^2}y \cr
&+& 
160{t^4}w{x^2}y + 
     1760{w^2}{x^2}y + 256t{w^2}{x^2}y - 
256{t^3}{w^2}{x^2}y - 
     4928{w^3}{x^2}y + 384t{w^3}{x^2}y \cr
&-& 
256{t^2}{w^3}{x^2}y - 
     1584{w^4}{x^2}y + 160t{w^4}{x^2}y - 
144{w^5}{x^2}y + 
     21504{x^3}y - 6400t{x^3}y \cr
&+& 2560{t^2}{x^3}y - 
512{t^3}{x^3}y - 
     64{t^4}{x^3}y - 6400w{x^3}y + 
384{t^2}w{x^3}y + 
     384{t^3}w{x^3}y \cr
&+& 2560{w^2}{x^3}y + 384t{w^2}{x^3}y + 
     512{t^2}{w^2}{x^3}y - 512{w^3}{x^3}y + 
384t{w^3}{x^3}y - 
     64{w^4}{x^3}y \cr
&-& 10528{x^4}y + 1024t{x^4}y + 
480{t^2}{x^4}y + 
     176{t^3}{x^4}y + 1024w{x^4}y + 160{t^2}w{x^4}y \cr
&+& 
     480{w^2}{x^4}y + 160t{w^2}{x^4}y + 
176{w^3}{x^4}y + 1984{x^5}y + 
     960t{x^5}y + 960w{x^5}y \cr
&+& 880{x^6}y + 
96t{x^6}y + 
     96w{x^6}y + 64{x^7}y - 74776t{y^2} + 
26042{t^2}{y^2} \cr
&+& 
     5888{t^3}{y^2} - 3948{t^4}{y^2} - 888{t^5}{y^2} - 
46{t^6}{y^2} - 
     74776w{y^2} - 53504tw{y^2} \cr
&+& 1760{t^2}w{y^2} + 
2560{t^3}w{y^2} + 
     480{t^4}w{y^2} + 26042{w^2}{y^2} + 
1760t{w^2}{y^2} + 
     96{t^3}{w^2}{y^2} \cr
&+& 28{t^4}{w^2}{y^2} + 
5888{w^3}{y^2} + 
     2560t{w^3}{y^2} + 96{t^2}{w^3}{y^2} - 
256{t^3}{w^3}{y^2} - 
     3948{w^4}{y^2} \cr
&+& 480t{w^4}{y^2} + 
28{t^2}{w^4}{y^2} - 
     888{w^5}{y^2} - 46{w^6}{y^2} + 149552x{y^2} + 
26752tx{y^2} \cr
&+& 
     1760{t^2}x{y^2} - 4928{t^3}x{y^2} - 
1584{t^4}x{y^2} - 
     144{t^5}x{y^2} + 26752wx{y^2} + 
256{t^2}wx{y^2} \cr
&+& 
     384{t^3}wx{y^2} + 160{t^4}wx{y^2} + 1760{w^2}x{y^2} + 
     256t{w^2}x{y^2} - 256{t^3}{w^2}x{y^2} - 
4928{w^3}x{y^2} \cr
&+& 
     384t{w^3}x{y^2} - 256{t^2}{w^3}x{y^2} - 
1584{w^4}x{y^2} + 
     160t{w^4}x{y^2} - 144{w^5}x{y^2} - 
52084{x^2}{y^2} \cr
&-& 
     3520t{x^2}{y^2} - 192{t^3}{x^2}{y^2} - 
56{t^4}{x^2}{y^2} - 
     3520w{x^2}{y^2} - 512tw{x^2}{y^2} + 512{t^3}w{x^2}{y^2} \cr
&-& 
     192{w^3}{x^2}{y^2} + 512t{w^3}{x^2}{y^2} - 
56{w^4}{x^2}{y^2} - 
     11776{x^3}{y^2} + 2368t{x^3}{y^2} + 
96{t^2}{x^3}{y^2} \cr
&+& 
     128{t^3}{x^3}{y^2} + 2368w{x^3}{y^2} - 
768tw{x^3}{y^2} - 
     256{t^2}w{x^3}{y^2} + 96{w^2}{x^3}{y^2} - 
256t{w^2}{x^3}{y^2} \cr
&+& 
     128{w^3}{x^3}{y^2} + 7896{x^4}{y^2} + 
1104t{x^4}{y^2} + 
     28{t^2}{x^4}{y^2} + 1104w{x^4}{y^2} - 
320tw{x^4}{y^2} \cr
&+& 
     28{w^2}{x^4}{y^2} + 1776{x^5}{y^2} + 
144t{x^5}{y^2} + 
     144w{x^5}{y^2} + 92{x^6}{y^2} \cr
&-& 10752t{y^3} + 
5888{t^2}{y^3} - 
     832{t^3}{y^3} - 1272{t^4}{y^3} - 160{t^5}{y^3} - 10752w{y^3} \cr
&+& 
     12800tw{y^3} - 4928{t^2}w{y^3} - 
512{t^3}w{y^3} + 
     176{t^4}w{y^3} + 5888{w^2}{y^3} - 
4928t{w^2}{y^3} \cr
&-& 
     192{t^2}{w^2}{y^3} + 128{t^3}{w^2}{y^3} - 
832{w^3}{y^3} - 
     512t{w^3}{y^3} + 128{t^2}{w^3}{y^3} - 
1272{w^4}{y^3} \cr
&+& 
     176t{w^4}{y^3} - 160{w^5}{y^3} + 21504x{y^3} - 6400tx{y^3} + 
     2560{t^2}x{y^3} - 512{t^3}x{y^3} \cr
&-& 
64{t^4}x{y^3} - 6400wx{y^3} + 
     384{t^2}wx{y^3} + 384{t^3}wx{y^3} + 2560{w^2}x{y^3} + 
     384t{w^2}x{y^3} \cr
&+& 512{t^2}{w^2}x{y^3} - 
512{w^3}x{y^3} + 
     384t{w^3}x{y^3} - 64{w^4}x{y^3} - 
11776{x^2}{y^3} + 
     2368t{x^2}{y^3} \cr
&+& 96{t^2}{x^2}{y^3} + 
128{t^3}{x^2}{y^3} + 
     2368w{x^2}{y^3} - 768tw{x^2}{y^3} - 256{t^2}w{x^2}{y^3} + 
     96{w^2}{x^2}{y^3} \cr
&-& 256t{w^2}{x^2}{y^3} + 
128{w^3}{x^2}{y^3} + 
     1664{x^3}{y^3} + 1024t{x^3}{y^3} - 
256{t^2}{x^3}{y^3} + 
     1024w{x^3}{y^3} \cr
&-& 768tw{x^3}{y^3} - 
256{w^2}{x^3}{y^3} + 
     2544{x^4}{y^3} - 112t{x^4}{y^3} - 112w{x^4}{y^3} 
+ 320{x^5}{y^3} \cr
&+& 
     5264t{y^4} - 3948{t^2}{y^4} - 1272{t^3}{y^4} - 
5{t^4}{y^4} + 
     5264w{y^4} - 2048tw{y^4} \cr
&-& 1584{t^2}w{y^4} - 
64{t^3}w{y^4} - 
     3948{w^2}{y^4} - 1584t{w^2}{y^4} - 
56{t^2}{w^2}{y^4} - 
     1272{w^3}{y^4} \cr
&-& 64t{w^3}{y^4} - 5{w^4}{y^4} - 
10528x{y^4} + 
     1024tx{y^4} + 480{t^2}x{y^4} + 176{t^3}x{y^4}\cr 
&+& 1024wx{y^4} + 
     160{t^2}wx{y^4} + 480{w^2}x{y^4} + 
160t{w^2}x{y^4} + 
     176{w^3}x{y^4} + 7896{x^2}{y^4} \cr
&+& 
1104t{x^2}{y^4} + 
     28{t^2}{x^2}{y^4} + 1104w{x^2}{y^4} - 
320tw{x^2}{y^4} + 
     28{w^2}{x^2}{y^4} + 2544{x^3}{y^4} \cr
&-& 
112t{x^3}{y^4} - 
     112w{x^3}{y^4} + 10{x^4}{y^4} - 992t{y^5} - 
888{t^2}{y^5} - 
     160{t^3}{y^5} \cr
&-& 992w{y^5} - 1920tw{y^5} - 
144{t^2}w{y^5} - 
     888{w^2}{y^5} - 144t{w^2}{y^5} - 160{w^3}{y^5} \cr
&+& 
1984x{y^5} + 
     960tx{y^5} + 960wx{y^5} + 1776{x^2}{y^5} + 
144t{x^2}{y^5} + 
     144w{x^2}{y^5} \cr
&+& 320{x^3}{y^5} - 440t{y^6} - 
46{t^2}{y^6} - 
     440w{y^6} - 192tw{y^6} - 46{w^2}{y^6} \cr
&+& 
880x{y^6} + 96tx{y^6} + 
     96wx{y^6} + 92{x^2}{y^6} - 32t{y^7} - 
32w{y^7} \cr
&+& 64x{y^7}
\biggr) a_{125}a_{345}.\nonumber\\
(A_{-1})^{(2)}_{I_1I_2I_3I_4}&=&-\frac{4}{\d}(f_1-f_2)(f_3-f_4)
\biggl( 36 -20(k_1+k_2+k_3+k_4)+f_1+f_2+f_3+f_4\cr
&+&10(k_1+k_2)(k_3+k_4)\biggr)
 f_5^{-1}a_{125}a_{345}.\nonumber\\
(S_{-1})^{(2)}_{I_1I_2I_3I_4}&=&-\frac{8}{\d}(f_1-f_2)(f_3-f_4)
(k_1-k_2)(k_3-k_4)
 f_5^{-1}a_{125}a_{345}.\nonumber\\
(A_{t3})^{(2)}_{I_1I_2I_3I_4}&=&-\frac{2}{\d}(f_5-1)^3t_{125}t_{345}.
\nonumber\\
\nonumber
(A_{t2})^{(2)}_{I_1I_2I_3I_4}&=&-\frac{1}{\d }(f_5-1)^2t_{125}t_{345}(k_1^2+k_2^2+k_3^2+k_4^2 \\
\nonumber
&-&16(k_1+k_2+k_3+k_4)+10(k_1+k_2)(k_3+k_4)+44)
.\nonumber\\
(S_{t2})^{(2)}_{I_1I_2I_3I_4}&=&
-\frac{2}{\d }(k_1-k_2)(k_3-k_4)(f_5-1)^2t_{125}t_{345} 
.\nonumber\\
(S_{p2})^{(2)}_{I_1I_2I_3I_4}&=&\frac{4}{\d}f_5^2p_{125}p_{345}.\nonumber\\
(S_d)^{(2)}_{I_1I_2I_3I_4}&=&\frac{9a_{125}a_{345}}{16\d (f_5-5)}
( -1 + k_1 - k_2 )(1+k_1-k_2)(3+k_1+k_2)(5+k_1+k_2) \nonumber \\
&\times & ( -1 + k_3 - k_4 )(1+k_3-k_4)(3+k_3+k_4)(5+k_3+k_4).\nonumber
\eea
\noindent {\bf Quartic couplings of non-derivative vertices}
\bea
(S_5)^{(0)}_{I_1I_2I_3I_4}&=&\frac{3}{64\d}
f_5^5 a_{125}a_{345}.\nonumber\\
(S_4)^{(0)}_{I_1I_2I_3I_4}&=&-\frac{1}{192\d}
f_5^4 a_{125}a_{345}\biggl( 747-368(k_1+k_2+k_3+k_4)+
65(k_1^2+k_2^2+k_3^2+k_4^2)\cr
&+& 132(k_1k_2+k_3k_4)-96(k_1+k_2)(k_3+k_4)
\biggr) .\nonumber\\
(S_3)^{(0)}_{I_1I_2I_3I_4}&=&\frac{1}{64\d}
f_5^3 a_{125}a_{345}\biggl(
 -3293 + 4036t - 1012{t^2} - 96{t^3} \cr
&+& 
35{t^4} + 4036w - 2428tw + 
       48{t^2}w + 88{t^3}w - 1012{w^2} + 48t{w^2} \cr
&+& 
122{t^2}{w^2} - 
       96{w^3} + 88t{w^3} + 35{w^4} + 4036x - 2976tx + 
360{t^2}x \cr
&+& 
       40{t^3}x - 2976wx + 1056twx - 8{t^2}wx + 
360{w^2}x - 
       8t{w^2}x + 40{w^3}x \cr
&-& 1012{x^2} + 360t{x^2} + 
44{t^2}{x^2} + 
       360w{x^2} - 8tw{x^2} + 44{w^2}{x^2} - 96{x^3} \cr
&+& 40t{x^3} + 
       40w{x^3} + 35{x^4} + 4036y - 2976ty + 
360{t^2}y + 40{t^3}y \cr
&-& 
       2976wy + 1056twy - 8{t^2}wy + 360{w^2}y 
- 8t{w^2}y + 
       40{w^3}y - 2428xy \cr
&+& 1056txy - 8{t^2}xy + 
1056wxy - 
       8{w^2}xy + 48{x^2}y - 8t{x^2}y - 8w{x^2}y \cr
&+& 88{x^3}y - 
       1012{y^2} + 360t{y^2} + 44{t^2}{y^2} + 360w{y^2} 
- 8tw{y^2} + 
       44{w^2}{y^2} \cr
&+& 48x{y^2} - 8tx{y^2} - 
8wx{y^2} + 
       122{x^2}{y^2} - 96{y^3} + 40t{y^3} \cr
&+& 40w{y^3} + 
88x{y^3} + 35{y^4}
\biggr) .\nonumber\\
(S_2)^{(0)}_{I_1I_2I_3I_4}&=&-\frac{1}{64\d}
f_5^2 a_{125}a_{345}\biggl(
 8273 - 20116t + 9396{t^2} + 
1008{t^3} \cr
&-& 1227{t^4} + 36{t^5} + 
         26{t^6} - 20116w + 25644tw - 2688{t^2}w \cr
&-& 
3544{t^3}w + 
         356{t^4}w + 76{t^5}w + 9396{w^2} - 2688t{w^2} - 
2778{t^2}{w^2} \cr
&+& 
         664{t^3}{w^2} + 46{t^4}{w^2} + 1008{w^3} - 
3544t{w^3} + 
         664{t^2}{w^3} + 104{t^3}{w^3} \cr
&-& 1227{w^4} + 
356t{w^4} + 
         46{t^2}{w^4} + 36{w^5} + 76t{w^5} + 26{w^6} - 
20116x \cr
&+& 32384tx - 
         9032{t^2}x - 1696{t^3}x + 492{t^4}x + 
8{t^5}x + 32384wx \cr
&-& 
         23776twx + 224{t^2}wx + 1152{t^3}wx - 104{t^4}wx - 
         9032{w^2}x + 224t{w^2}x \cr
&+& 1096{t^2}{w^2}x - 
224{t^3}{w^2}x - 
         1696{w^3}x + 1152t{w^3}x - 224{t^2}{w^3}x + 
492{w^4}x \cr
&-& 
         104t{w^4}x + 8{w^5}x + 9396{x^2} - 
9032t{x^2} + 332{t^2}{x^2} + 
         288{t^3}{x^2} \cr
&+& 60{t^4}{x^2} - 9032w{x^2} + 
2152tw{x^2} - 
         96{t^2}w{x^2} + 144{t^3}w{x^2} + 
332{w^2}{x^2} \cr
&-& 
         96t{w^2}{x^2} + 96{t^2}{w^2}{x^2} + 
288{w^3}{x^2} + 
         144t{w^3}{x^2} + 60{w^4}{x^2} + 1008{x^3} \cr
&-& 
1696t{x^3} + 
         288{t^2}{x^3} + 80{t^3}{x^3} - 1696w{x^3} + 
608tw{x^3} + 
         32{t^2}w{x^3} \cr
&+& 288{w^2}{x^3} + 
32t{w^2}{x^3} + 80{w^3}{x^3} - 
         1227{x^4} + 492t{x^4} + 60{t^2}{x^4} \cr
&+& 
492w{x^4} - 20tw{x^4} + 
         60{w^2}{x^4} + 36{x^5} + 8t{x^5} + 8w{x^5} + 26{x^6} \cr
&-& 20116y + 
         32384ty - 9032{t^2}y - 1696{t^3}y + 
492{t^4}y + 8{t^5}y \cr
&+& 
         32384wy - 23776twy + 224{t^2}wy + 
1152{t^3}wy - 
         104{t^4}wy - 9032{w^2}y \cr
&+& 224t{w^2}y + 
1096{t^2}{w^2}y - 
         224{t^3}{w^2}y - 1696{w^3}y + 1152t{w^3}y - 
224{t^2}{w^3}y \cr
&+& 
         492{w^4}y - 104t{w^4}y + 8{w^5}y + 
25644xy - 23776txy + 
         2152{t^2}xy \cr
&+& 608{t^3}xy - 20{t^4}xy - 
23776wxy + 
         10496twxy + 32{t^2}wxy - 128{t^3}wxy \cr
&+& 2152{w^2}xy + 
         32t{w^2}xy + 168{t^2}{w^2}xy + 
608{w^3}xy - 
         128t{w^3}xy - 20{w^4}xy \cr
&-& 2688{x^2}y + 
224t{x^2}y - 
         96{t^2}{x^2}y + 32{t^3}{x^2}y + 
224w{x^2}y + 32tw{x^2}y - 
         16{t^2}w{x^2}y \cr
&-& 96{w^2}{x^2}y - 16t{w^2}{x^2}y + 
         32{w^3}{x^2}y - 3544{x^3}y + 1152t{x^3}y + 
144{t^2}{x^3}y \cr
&+& 
         1152w{x^3}y - 128tw{x^3}y + 144{w^2}{x^3}y + 356{x^4}y - 
         104t{x^4}y - 104w{x^4}y \cr
&+& 76{x^5}y + 
9396{y^2} - 9032t{y^2} + 
         332{t^2}{y^2} + 288{t^3}{y^2} + 60{t^4}{y^2} \cr
&-& 
9032w{y^2} + 
         2152tw{y^2} - 96{t^2}w{y^2} + 
144{t^3}w{y^2} + 332{w^2}{y^2} - 
         96t{w^2}{y^2} \cr
&+& 96{t^2}{w^2}{y^2} + 
288{w^3}{y^2} + 
         144t{w^3}{y^2} + 60{w^4}{y^2} - 2688x{y^2} + 
224tx{y^2} \cr
&-& 
         96{t^2}x{y^2} + 32{t^3}x{y^2} + 
224wx{y^2} + 32twx{y^2} - 
         16{t^2}wx{y^2} - 96{w^2}x{y^2} - 16t{w^2}x{y^2} \cr
&+& 
         32{w^3}x{y^2} - 2778{x^2}{y^2} + 
1096t{x^2}{y^2} + 
         96{t^2}{x^2}{y^2} + 1096w{x^2}{y^2} + 
168tw{x^2}{y^2} \cr
&+& 
         96{w^2}{x^2}{y^2} + 664{x^3}{y^2} - 
224t{x^3}{y^2} - 
         224w{x^3}{y^2} + 46{x^4}{y^2} + 1008{y^3} \cr
&-& 
1696t{y^3} + 
         288{t^2}{y^3} + 80{t^3}{y^3} - 1696w{y^3} + 
608tw{y^3} + 
         32{t^2}w{y^3} \cr
&+& 288{w^2}{y^3} + 
32t{w^2}{y^3} + 80{w^3}{y^3} - 
         3544x{y^3} + 1152tx{y^3} + 144{t^2}x{y^3} \cr
&+& 
1152wx{y^3} - 
         128twx{y^3} + 144{w^2}x{y^3} + 
664{x^2}{y^3} - 
         224t{x^2}{y^3} - 224w{x^2}{y^3} \cr
&+& 
104{x^3}{y^3} - 1227{y^4} + 
         492t{y^4} + 60{t^2}{y^4} + 492w{y^4} - 
20tw{y^4} \cr
&+& 
         60{w^2}{y^4} + 356x{y^4} - 104tx{y^4} - 104wx{y^4} + 
         46{x^2}{y^4} + 36{y^5} \cr
&+& 8t{y^5} + 8w{y^5} + 76x{y^5} + 26{y^6}
\biggr) .\nonumber\\
(S_1)^{(0)}_{I_1I_2I_3I_4}&=&\frac{(w-x)(t-y)}{288\d}
f_5 a_{125}a_{345}\biggl(
163692 - 128440t + 28616{t^2} + 2052{t^3} \cr
&-& 
3460{t^4} + 484{t^5} + 
       80{t^6} - 128440w + 72314tw - 3096{t^2}w \cr
&+& 
393{t^3}w + 468{t^4}w - 
       119{t^5}w + 28616{w^2} - 3096t{w^2} - 
1208{t^2}{w^2} \cr
&-& 
       864{t^3}{w^2} + 184{t^4}{w^2} + 2052{w^3} + 
393t{w^3} - 
       864{t^2}{w^3} + 41{t^3}{w^3} \cr
&-& 3460{w^4} + 
468t{w^4} + 
       184{t^2}{w^4} + 484{w^5} - 119t{w^5} + 80{w^6} - 
128440x \cr
&+& 72314tx - 
       3096{t^2}x + 393{t^3}x + 468{t^4}x - 119{t^5}x 
+ 88352wx \cr
&-& 
       31064twx + 13912{t^2}wx - 2360{t^3}wx - 392{t^4}wx - 
       18716{w^2}x + 15049t{w^2}x \cr
&-& 5472{t^2}{w^2}x + 
641{t^3}{w^2}x - 
       416{w^3}x - 1416t{w^3}x + 312{t^2}{w^3}x + 
1380{w^4}x \cr
&-& 
       335t{w^4}x - 256{w^5}x + 28616{x^2} - 
3096t{x^2} - 
       1208{t^2}{x^2} - 864{t^3}{x^2} \cr
&+& 184{t^4}{x^2} - 
18716w{x^2} + 
       15049tw{x^2} - 5472{t^2}w{x^2} + 
641{t^3}w{x^2} - 
       880{w^2}{x^2} \cr
&-& 4496t{w^2}{x^2} + 
2392{t^2}{w^2}{x^2} + 
       992{w^3}{x^2} + 524t{w^3}{x^2} + 104{w^4}{x^2} + 
2052{x^3} + 
       393t{x^3} \cr
&-& 864{t^2}{x^3} + 41{t^3}{x^3} - 
416w{x^3} - 
       1416tw{x^3} + 312{t^2}w{x^3} + 992{w^2}{x^3} \cr
&+& 
524t{w^2}{x^3} - 
       368{w^3}{x^3} - 3460{x^4} + 468t{x^4} + 
184{t^2}{x^4} + 1380w{x^4} \cr
&-& 
       335tw{x^4} + 104{w^2}{x^4} + 484{x^5} - 
119t{x^5} - 256w{x^5} + 
       80{x^6} \cr
&-& 128440y + 88352ty - 18716{t^2}y - 
416{t^3}y + 
       1380{t^4}y - 256{t^5}y \cr
&+& 72314wy - 31064twy 
+ 15049{t^2}wy - 
       1416{t^3}wy - 335{t^4}wy - 3096{w^2}y \cr
&+& 
13912t{w^2}y - 
       5472{t^2}{w^2}y + 312{t^3}{w^2}y + 393{w^3}y - 
2360t{w^3}y + 
       641{t^2}{w^3}y \cr
&+& 468{w^4}y - 392t{w^4}y - 
119{w^5}y + 
       72314xy - 31064txy + 15049{t^2}xy \cr
&-& 
1416{t^3}xy - 
       335{t^4}xy - 31064wxy + 34928twxy - 
15928{t^2}wxy + 
       1664{t^3}wxy \cr
&+& 15049{w^2}xy - 
15928t{w^2}xy + 
       4313{t^2}{w^2}xy - 1416{w^3}xy + 1664t{w^3}xy - 
       335{w^4}xy \cr
&-& 3096{x^2}y + 13912t{x^2}y - 
5472{t^2}{x^2}y + 
       312{t^3}{x^2}y + 15049w{x^2}y - 
15928tw{x^2}y \cr
&+& 
       4313{t^2}w{x^2}y - 4496{w^2}{x^2}y + 
3488t{w^2}{x^2}y + 
       524{w^3}{x^2}y + 393{x^3}y - 2360t{x^3}y \cr
&+& 641{t^2}{x^3}y - 
       1416w{x^3}y + 1664tw{x^3}y + 
524{w^2}{x^3}y + 468{x^4}y - 
       392t{x^4}y \cr
&-& 335w{x^4}y - 119{x^5}y + 
28616{y^2} - 18716t{y^2} - 
       880{t^2}{y^2} + 992{t^3}{y^2} \cr
&+& 104{t^4}{y^2} - 3096w{y^2} + 
       15049tw{y^2} - 4496{t^2}w{y^2} + 
524{t^3}w{y^2} - 
       1208{w^2}{y^2} \cr
&-& 5472t{w^2}{y^2} + 2392{t^2}{w^2}{y^2} - 
       864{w^3}{y^2} + 641t{w^3}{y^2} + 184{w^4}{y^2} - 
3096x{y^2} \cr
&+& 
       15049tx{y^2} - 4496{t^2}x{y^2} + 
524{t^3}x{y^2} + 
       13912wx{y^2} - 15928twx{y^2} + 
3488{t^2}wx{y^2} \cr
&-& 
       5472{w^2}x{y^2} + 4313t{w^2}x{y^2} + 
312{w^3}x{y^2} - 
       1208{x^2}{y^2} - 5472t{x^2}{y^2} + 2392{t^2}{x^2}{y^2} \cr
&-& 
       5472w{x^2}{y^2} + 4313tw{x^2}{y^2} + 2392{w^2}{x^2}{y^2} - 
       864{x^3}{y^2} + 641t{x^3}{y^2} + 
312w{x^3}{y^2} \cr
&+& 184{x^4}{y^2} + 
       2052{y^3} - 416t{y^3} + 992{t^2}{y^3} - 
368{t^3}{y^3} + 393w{y^3} \cr
&-& 
       1416tw{y^3} + 524{t^2}w{y^3} - 864{w^2}{y^3} + 
312t{w^2}{y^3} + 
       41{w^3}{y^3} + 393x{y^3} \cr
&-& 1416tx{y^3} + 
524{t^2}x{y^3} - 
       2360wx{y^3} + 1664twx{y^3} + 
641{w^2}x{y^3} - 864{x^2}{y^3} \cr
&+& 
       312t{x^2}{y^3} + 641w{x^2}{y^3} + 41{x^3}{y^3} 
- 3460{y^4} + 
       1380t{y^4} + 104{t^2}{y^4} \cr
&+& 468w{y^4} - 335tw{y^4} + 
       184{w^2}{y^4} + 468x{y^4} - 335tx{y^4} - 
392wx{y^4} \cr
&+& 
       184{x^2}{y^4} + 484{y^5} - 256t{y^5} - 119w{y^5} 
- 119x{y^5} + 
       80{y^6}
\biggr) .\nonumber\\
(S_0)^{(0)}_{I_1I_2I_3I_4}&=&\frac{1}{576\d}
a_{125}a_{345}\biggl(
-18225 - 24300t + 292830{t^2} - 71028{t^3} \cr
&-& 
111795{t^4} + 33444{t^5} + 
     8640{t^6} - 5124{t^7} - 618{t^8} + 192{t^9} \cr
&+& 24{t^{10}} 
- 24300w - 
     256068tw - 66756{t^2}w - 120628{t^3}w + 177012{t^4}w \cr
&-& 17836{t^5}w - 
     26900{t^6}w + 132{t^7}w + 800{t^8}w + 32{t^9}w + 
292830{w^2} \cr
&-& 
     66756t{w^2} + 177178{t^2}{w^2} + 160520{t^3}{w^2} - 
94964{t^4}{w^2} - 
     31788{t^5}{w^2} + 1416{t^6}{w^2} \cr
&+& 1136{t^7}{w^2} + 
112{t^8}{w^2} - 
     71028{w^3} - 120628t{w^3} + 160520{t^2}{w^3} - 
11080{t^3}{w^3} \cr
&-& 
     46764{t^4}{w^3} - 3076{t^5}{w^3} + 2640{t^6}{w^3} + 
336{t^7}{w^3} - 
     111795{w^4} + 177012t{w^4} \cr
&-& 94964{t^2}{w^4} - 
46764{t^3}{w^4} + 
     6638{t^4}{w^4} + 2920{t^5}{w^4} + 224{t^6}{w^4} + 
33444{w^5} \cr
&-& 
     17836t{w^5} - 31788{t^2}{w^5} - 3076{t^3}{w^5} + 
2920{t^4}{w^5} + 
     416{t^5}{w^5} + 8640{w^6} \cr
&-& 26900t{w^6} + 
1416{t^2}{w^6} + 
     2640{t^3}{w^6} + 224{t^4}{w^6} - 5124{w^7} + 
132t{w^7} \cr
&+& 
     1136{t^2}{w^7} + 336{t^3}{w^7} - 618{w^8} + 
800t{w^8} + 
     112{t^2}{w^8} + 192{w^9} \cr
&+& 32t{w^9} + 24{w^{10}} - 
24300x - 256068tx - 
     66756{t^2}x - 120628{t^3}x \cr
&+& 177012{t^4}x - 
17836{t^5}x - 
     26900{t^6}x + 132{t^7}x + 800{t^8}x + 32{t^9}x \cr
&-& 
256068wx + 
     12816twx + 319756{t^2}wx + 370976{t^3}wx \cr
&-& 
220300{t^4}wx - 
     57456{t^5}wx 
+ 5828{t^6}wx + 1024{t^7}wx - 96{t^8}wx \cr
&-& 
     66756{w^2}x + 319756t{w^2}x + 184632{t^2}{w^2}x 
- 
     247264{t^3}{w^2}x \cr
&-& 68020{t^4}{w^2}x - 
2724{t^5}{w^2}x + 
     1792{t^6}{w^2}x + 272{t^7}{w^2}x - 120628{w^3}x \cr
&+& 370976t{w^3}x - 
     247264{t^2}{w^3}x - 124960{t^3}{w^3}x + 
6940{t^4}{w^3}x + 
     4000{t^5}{w^3}x \cr
&+& 352{t^6}{w^3}x 
+ 177012{w^4}x 
- 220300t{w^4}x - 
     68020{t^2}{w^4}x + 6940{t^3}{w^4}x \cr
&+& 
1672{t^4}{w^4}x + 
     268{t^5}{w^4}x 
-17836{w^5}x - 57456t{w^5}x - 
2724{t^2}{w^5}x + 
     4000{t^3}{w^5}x \cr
&+& 268{t^4}{w^5}x - 26900{w^6}x + 
5828t{w^6}x + 
     1792{t^2}{w^6}x + 352{t^3}{w^6}x + 132{w^7}x \cr
&+& 
1024t{w^7}x + 
     272{t^2}{w^7}x + 800{w^8}x - 96t{w^8}x + 
32{w^9}x + 292830{x^2} \cr
&-& 
     66756t{x^2} + 177178{t^2}{x^2} + 160520{t^3}{x^2} - 
94964{t^4}{x^2} - 
     31788{t^5}{x^2} \cr
&+& 1416{t^6}{x^2} + 1136{t^7}{x^2} + 
112{t^8}{x^2} - 
     66756w{x^2} + 319756tw{x^2} + 184632{t^2}w{x^2}\cr 
&-&     247264{t^3}w{x^2} - 68020{t^4}w{x^2} - 
2724{t^5}w{x^2} + 
     1792{t^6}w{x^2} \cr
&+& 272{t^7}w{x^2} + 
177178{w^2}{x^2} + 
     184632t{w^2}{x^2} - 335076{t^2}{w^2}{x^2} \cr
&-& 
134456{t^3}{w^2}{x^2} - 
     1458{t^4}{w^2}{x^2} + 4296{t^5}{w^2}{x^2} \cr
&+& 
848{t^6}{w^2}{x^2} +
     160520{w^3}{x^2} - 247264t{w^3}{x^2} - 
134456{t^2}{w^3}{x^2} + 
     7688{t^3}{w^3}{x^2}\cr 
&+& 4056{t^4}{w^3}{x^2} +
592{t^5}{w^3}{x^2} - 
     94964{w^4}{x^2} - 68020t{w^4}{x^2} - 1458{t^2}{w^4}{x^2} \cr
&+& 
     4056{t^3}{w^4}{x^2} - 56{t^4}{w^4}{x^2} - 
31788{w^5}{x^2} -
     2724t{w^5}{x^2} + 4296{t^2}{w^5}{x^2} + 
592{t^3}{w^5}{x^2} \cr
&+& 
     1416{w^6}{x^2} + 1792t{w^6}{x^2} + 
848{t^2}{w^6}{x^2} + 
     1136{w^7}{x^2} + 272t{w^7}{x^2} + 112{w^8}{x^2} \cr
&-& 
71028{x^3} - 
     120628t{x^3} + 160520{t^2}{x^3} 
- 11080{t^3}{x^3} - 
46764{t^4}{x^3} \cr
&-& 
     3076{t^5}{x^3} + 2640{t^6}{x^3} + 336{t^7}{x^3} - 
120628w{x^3} 
+ 370976tw{x^3} \cr
&-& 247264{t^2}w{x^3} - 124960{t^3}w{x^3} + 
     6940{t^4}w{x^3} + 4000{t^5}w{x^3} + 
352{t^6}w{x^3} \cr
&+& 
     160520{w^2}{x^3} - 247264t{w^2}{x^3} - 
134456{t^2}{w^2}{x^3} + 
     7688{t^3}{w^2}{x^3} + 4056{t^4}{w^2}{x^3} \cr
&+& 
592{t^5}{w^2}{x^3} -
     11080{w^3}{x^3} - 124960t{w^3}{x^3} + 
7688{t^2}{w^3}{x^3} + 
     6720{t^3}{w^3}{x^3} \cr
&-& 1096{t^4}{w^3}{x^3} - 
46764{w^4}{x^3} +
     6940t{w^4}{x^3} + 4056{t^2}{w^4}{x^3} \cr
&-& 
1096{t^3}{w^4}{x^3} - 
     3076{w^5}{x^3} 
+ 4000t{w^5}{x^3} + 
592{t^2}{w^5}{x^3} \cr
&+& 
     2640{w^6}{x^3} + 352t{w^6}{x^3} + 336{w^7}{x^3} - 
111795{x^4} + 
     177012t{x^4} - 94964{t^2}{x^4} \cr
&-& 46764{t^3}{x^4} + 
6638{t^4}{x^4} + 
     2920{t^5}{x^4} + 224{t^6}{x^4} + 177012w{x^4} - 220300tw{x^4} \cr
&-& 
     68020{t^2}w{x^4} + 6940{t^3}w{x^4} + 
1672{t^4}w{x^4} + 
     268{t^5}w{x^4} - 94964{w^2}{x^4} - 
68020t{w^2}{x^4} \cr
&-& 
     1458{t^2}{w^2}{x^4} + 4056{t^3}{w^2}{x^4} - 56{t^4}{w^2}{x^4} - 
     46764{w^3}{x^4} + 6940t{w^3}{x^4} + 
4056{t^2}{w^3}{x^4} \cr
&-& 
     1096{t^3}{w^3}{x^4} + 6638{w^4}{x^4} + 
1672t{w^4}{x^4} - 
     56{t^2}{w^4}{x^4} + 2920{w^5}{x^4} + 
268t{w^5}{x^4} \cr
&+& 
     224{w^6}{x^4} + 33444{x^5} - 17836t{x^5} - 
31788{t^2}{x^5} - 
    3076{t^3}{x^5} + 2920{t^4}{x^5} \cr
&+& 416{t^5}{x^5} - 
17836w{x^5} - 
     57456tw{x^5} - 2724{t^2}w{x^5} + 
4000{t^3}w{x^5} + 
     268{t^4}w{x^5} \cr
&-& 31788{w^2}{x^5} - 
2724t{w^2}{x^5} + 
     4296{t^2}{w^2}{x^5} + 592{t^3}{w^2}{x^5} - 
3076{w^3}{x^5} + 
     4000t{w^3}{x^5} \cr
&+& 592{t^2}{w^3}{x^5} + 
2920{w^4}{x^5} + 
     268t{w^4}{x^5} + 416{w^5}{x^5} + 8640{x^6} - 26900t{x^6} \cr
&+& 
     1416{t^2}{x^6} + 2640{t^3}{x^6} + 224{t^4}{x^6} - 
26900w{x^6} + 
     5828tw{x^6} + 1792{t^2}w{x^6} \cr
&+& 
352{t^3}w{x^6} + 1416{w^2}{x^6} + 
     1792t{w^2}{x^6} + 848{t^2}{w^2}{x^6} + 
2640{w^3}{x^6} + 
     352t{w^3}{x^6} \cr
&+& 224{w^4}{x^6} - 5124{x^7} + 
132t{x^7} + 
     1136{t^2}{x^7} + 336{t^3}{x^7} + 132w{x^7} \cr
&+& 
1024tw{x^7} + 
     272{t^2}w{x^7} + 1136{w^2}{x^7} + 272t{w^2}{x^7} 
+ 336{w^3}{x^7} - 
     618{x^8} \cr
&+& 800t{x^8} + 112{t^2}{x^8} + 800w{x^8} - 
96tw{x^8} + 
     112{w^2}{x^8} + 192{x^9} + 32t{x^9} \cr
&+&
 32w{x^9} + 24{x^{10}} - 24300y - 
     256068ty - 66756{t^2}y - 120628{t^3}y \cr
&+& 177012{t^4}y - 17836{t^5}y - 
     26900{t^6}y + 132{t^7}y + 800{t^8}y + 32{t^9}y \cr
&-& 
256068wy + 
     12816twy + 319756{t^2}wy + 370976{t^3}wy - 
220300{t^4}wy \cr
&-& 
     57456{t^5}wy + 5828{t^6}wy + 1024{t^7}wy - 96{t^8}wy - 
     66756{w^2}y \cr
&+& 319756t{w^2}y + 184632{t^2}{w^2}y 
-      247264{t^3}{w^2}y - 68020{t^4}{w^2}y - 
2724{t^5}{w^2}y \cr
&+& 
     1792{t^6}{w^2}y + 272{t^7}{w^2}y - 120628{w^3}y 
+ 370976t{w^3}y - 
     247264{t^2}{w^3}y \cr
&-& 124960{t^3}{w^3}y + 
6940{t^4}{w^3}y + 
    4000{t^5}{w^3}y + 352{t^6}{w^3}y + 177012{w^4}y 
- 220300t{w^4}y \cr
&-&     68020{t^2}{w^4}y + 6940{t^3}{w^4}y + 
1672{t^4}{w^4}y + 
     268{t^5}{w^4}y - 17836{w^5}y \cr
&-& 57456t{w^5}y - 
2724{t^2}{w^5}y + 
     4000{t^3}{w^5}y + 268{t^4}{w^5}y - 26900{w^6}y \cr
&+& 
5828t{w^6}y + 
     1792{t^2}{w^6}y 
+ 352{t^3}{w^6}y + 132{w^7}y \cr
&+& 
1024t{w^7}y + 
     272{t^2}{w^7}y + 800{w^8}y - 96t{w^8}y \cr
&+& 
32{w^9}y - 256068xy + 
     12816txy + 319756{t^2}xy + 370976{t^3}xy - 
220300{t^4}xy \cr
&-&     57456{t^5}xy + 5828{t^6}xy + 1024{t^7}xy - 96{t^8}xy + 
     12816wxy + 2404608twxy \cr
&-& 48096{t^2}wxy 
- 749568{t^3}wxy - 
     41136{t^4}wxy \cr
&+& 4992{t^5}wxy - 
2112{t^6}wxy - 
     384{t^7}wxy \cr
&+&
 319756{w^2}xy - 48096t{w^2}xy - 
     403336{t^2}{w^2}xy - 128736{t^3}{w^2}xy - 
11588{t^4}{w^2}xy \cr
&+& 
     2432{t^5}{w^2}xy + 688{t^6}{w^2}xy + 
370976{w^3}xy - 
     749568t{w^3}xy - 128736{t^2}{w^3}xy \cr
&+& 
39680{t^3}{w^3}xy + 
     896{t^4}{w^3}xy + 384{t^5}{w^3}xy - 
220300{w^4}xy - 
     41136t{w^4}xy \cr
&-& 11588{t^2}{w^4}xy + 
896{t^3}{w^4}xy + 
     608{t^4}{w^4}xy \cr
&-& 57456{w^5}xy + 
4992t{w^5}xy + 
     2432{t^2}{w^5}xy \cr
&+& 384{t^3}{w^5}xy + 
5828{w^6}xy - 
     2112t{w^6}xy + 688{t^2}{w^6}xy + 
1024{w^7}xy - 
     384t{w^7}xy \cr
&-& 96{w^8}xy - 66756{x^2}y + 
319756t{x^2}y + 
     184632{t^2}{x^2}y - 247264{t^3}{x^2}y - 
68020{t^4}{x^2}y \cr
&-& 
    2724{t^5}{x^2}y + 1792{t^6}{x^2}y + 
272{t^7}{x^2}y + 
     319756w{x^2}y - 48096tw{x^2}y \cr
&-& 
403336{t^2}w{x^2}y - 
     128736{t^3}w{x^2}y - 11588{t^4}w{x^2}y + 
2432{t^5}w{x^2}y \cr
&+&     688{t^6}w{x^2}y + 184632{w^2}{x^2}y - 
403336t{w^2}{x^2}y - 
     169560{t^2}{w^2}{x^2}y \cr
&+& 17288{t^3}{w^2}{x^2}y + 
   4136{t^4}{w^2}{x^2}y + 1112{t^5}{w^2}{x^2}y - 
247264{w^3}{x^2}y \cr
&-& 
     128736t{w^3}{x^2}y + 17288{t^2}{w^3}{x^2}y + 
     2496{t^3}{w^3}{x^2}y \cr
&-& 1856{t^4}{w^3}{x^2}y -
68020{w^4}{x^2}y - 
     11588t{w^4}{x^2}y + 4136{t^2}{w^4}{x^2}y \cr
&-&      1856{t^3}{w^4}{x^2}y 
- 2724{w^5}{x^2}y + 
2432t{w^5}{x^2}y \cr
&+& 1112{t^2}{w^5}{x^2}y + 1792{w^6}{x^2}y + 
688t{w^6}{x^2}y + 
     272{w^7}{x^2}y - 120628{x^3}y \cr
&+& 370976t{x^3}y - 
     247264{t^2}{x^3}y - 124960{t^3}{x^3}y + 
6940{t^4}{x^3}y + 
     4000{t^5}{x^3}y \cr
&+& 352{t^6}{x^3}y + 
370976w{x^3}y - 
     749568tw{x^3}y - 128736{t^2}w{x^3}y + 
39680{t^3}w{x^3}y \cr
&+& 
     896{t^4}w{x^3}y + 384{t^5}w{x^3}y - 
247264{w^2}{x^3}y - 
     128736t{w^2}{x^3}y + 17288{t^2}{w^2}{x^3}y \cr
&+& 
     2496{t^3}{w^2}{x^3}y - 1856{t^4}{w^2}{x^3}y - 
124960{w^3}{x^3}y + 
     39680t{w^3}{x^3}y + 2496{t^2}{w^3}{x^3}y \cr
&-& 
     6912{t^3}{w^3}{x^3}y + 6940{w^4}{x^3}y + 
896t{w^4}{x^3}y \cr
&-&     1856{t^2}{w^4}{x^3}y + 4000{w^5}{x^3}y + 
384t{w^5}{x^3}y \cr
&+& 
     352{w^6}{x^3}y + 177012{x^4}y - 220300t{x^4}y - 
68020{t^2}{x^4}y + 
     6940{t^3}{x^4}y + 1672{t^4}{x^4}y \cr
&+& 
268{t^5}{x^4}y - 
     220300w{x^4}y - 41136tw{x^4}y \cr
&-& 
11588{t^2}w{x^4}y + 
    896{t^3}w{x^4}y + 608{t^4}w{x^4}y \cr
&-& 
68020{w^2}{x^4}y - 
     11588t{w^2}{x^4}y + 4136{t^2}{w^2}{x^4}y - 
     1856{t^3}{w^2}{x^4}y + 6940{w^3}{x^4}y \cr
&+& 
896t{w^3}{x^4}y - 
     1856{t^2}{w^3}{x^4}y + 1672{w^4}{x^4}y + 
608t{w^4}{x^4}y + 
     268{w^5}{x^4}y - 17836{x^5}y \cr
&-& 57456t{x^5}y - 
2724{t^2}{x^5}y + 
     4000{t^3}{x^5}y + 268{t^4}{x^5}y - 
57456w{x^5}y + 
     4992tw{x^5}y \cr
&+& 2432{t^2}w{x^5}y + 
384{t^3}w{x^5}y - 
     2724{w^2}{x^5}y \cr
&+& 2432t{w^2}{x^5}y + 
1112{t^2}{w^2}{x^5}y + 
     4000{w^3}{x^5}y \cr
&+& 384t{w^3}{x^5}y + 
268{w^4}{x^5}y - 
     26900{x^6}y + 5828t{x^6}y + 1792{t^2}{x^6}y + 
352{t^3}{x^6}y \cr
&+& 
     5828w{x^6}y - 2112tw{x^6}y + 
688{t^2}w{x^6}y + 
     1792{w^2}{x^6}y + 688t{w^2}{x^6}y + 
352{w^3}{x^6}y \cr
&+& 132{x^7}y + 
     1024t{x^7}y + 272{t^2}{x^7}y + 1024w{x^7}y - 
384tw{x^7}y + 
     272{w^2}{x^7}y \cr
&+& 800{x^8}y - 96t{x^8}y - 
96w{x^8}y + 
     32{x^9}y + 292830{y^2} - 66756t{y^2} \cr
&+& 
177178{t^2}{y^2} + 
     160520{t^3}{y^2} - 94964{t^4}{y^2} - 31788{t^5}{y^2} + 
1416{t^6}{y^2} \cr
&+& 
     1136{t^7}{y^2} + 112{t^8}{y^2} - 66756w{y^2} + 
319756tw{y^2} + 
     184632{t^2}w{y^2} \cr
&-& 247264{t^3}w{y^2} - 
68020{t^4}w{y^2} - 
     2724{t^5}w{y^2} + 1792{t^6}w{y^2} + 
272{t^7}w{y^2} \cr
&+& 
     177178{w^2}{y^2} + 184632t{w^2}{y^2} - 
335076{t^2}{w^2}{y^2} - 
     134456{t^3}{w^2}{y^2} - 1458{t^4}{w^2}{y^2} \cr
&+& 
4296{t^5}{w^2}{y^2} + 
     848{t^6}{w^2}{y^2} + 160520{w^3}{y^2} - 
247264t{w^3}{y^2} - 
     134456{t^2}{w^3}{y^2} \cr
&+& 7688{t^3}{w^3}{y^2} + 
4056{t^4}{w^3}{y^2} + 
     592{t^5}{w^3}{y^2} - 94964{w^4}{y^2} \cr
&-& 
68020t{w^4}{y^2} - 
     1458{t^2}{w^4}{y^2} + 
 4056{t^3}{w^4}{y^2} - 56{t^4}{w^4}{y^2} \cr
&-& 
     31788{w^5}{y^2} - 2724t{w^5}{y^2} + 
4296{t^2}{w^5}{y^2} + 
     592{t^3}{w^5}{y^2} \cr
&+& 1416{w^6}{y^2} + 
1792t{w^6}{y^2} + 
     848{t^2}{w^6}{y^2} + 1136{w^7}{y^2} + 
272t{w^7}{y^2} + 
     112{w^8}{y^2} \cr
&-& 66756x{y^2} + 319756tx{y^2} + 
184632{t^2}x{y^2} - 
     247264{t^3}x{y^2} - 68020{t^4}x{y^2} \cr
&-& 
2724{t^5}x{y^2} + 
     1792{t^6}x{y^2} + 272{t^7}x{y^2} + 
319756wx{y^2} - 
     48096twx{y^2} \cr
&-& 403336{t^2}wx{y^2} - 
128736{t^3}wx{y^2} - 
     11588{t^4}wx{y^2} + 2432{t^5}wx{y^2} \cr
&+& 
688{t^6}wx{y^2} +
     184632{w^2}x{y^2} - 403336t{w^2}x{y^2} - 
169560{t^2}{w^2}x{y^2} \cr
&+& 
     17288{t^3}{w^2}x{y^2} + 4136{t^4}{w^2}x{y^2} + 
     1112{t^5}{w^2}x{y^2} - 247264{w^3}x{y^2} \cr
&-& 
128736t{w^3}x{y^2} + 
     17288{t^2}{w^3}x{y^2} + 2496{t^3}{w^3}x{y^2} \cr
&-& 
     1856{t^4}{w^3}x{y^2} - 68020{w^4}x{y^2} - 11588t{w^4}x{y^2} + 
     4136{t^2}{w^4}x{y^2} - 1856{t^3}{w^4}x{y^2} \cr
&-& 
2724{w^5}x{y^2} + 
     2432t{w^5}x{y^2} + 1112{t^2}{w^5}x{y^2} + 
1792{w^6}x{y^2} + 
     688t{w^6}x{y^2} \cr
&+& 272{w^7}x{y^2} + 
177178{x^2}{y^2} + 
     184632t{x^2}{y^2} - 335076{t^2}{x^2}{y^2} - 
134456{t^3}{x^2}{y^2} \cr
&-& 
     1458{t^4}{x^2}{y^2} + 4296{t^5}{x^2}{y^2} + 
848{t^6}{x^2}{y^2} + 
     184632w{x^2}{y^2} - 403336tw{x^2}{y^2} \cr
&-& 
169560{t^2}w{x^2}{y^2} + 
    17288{t^3}w{x^2}{y^2} + 4136{t^4}w{x^2}{y^2} + 
     1112{t^5}w{x^2}{y^2} \cr
&-& 335076{w^2}{x^2}{y^2} - 
     169560t{w^2}{x^2}{y^2} + 
62988{t^2}{w^2}{x^2}{y^2} + 
     10224{t^3}{w^2}{x^2}{y^2} \cr
&-& 
3408{t^4}{w^2}{x^2}{y^2} - 
     134456{w^3}{x^2}{y^2} 
+ 17288t{w^3}{x^2}{y^2} + 
     10224{t^2}{w^3}{x^2}{y^2} \cr
&-& 
9680{t^3}{w^3}{x^2}{y^2} - 
     1458{w^4}{x^2}{y^2} + 4136t{w^4}{x^2}{y^2} \cr
&-& 
     3408{t^2}{w^4}{x^2}{y^2} + 4296{w^5}{x^2}{y^2} + 
     1112t{w^5}{x^2}{y^2} + 848{w^6}{x^2}{y^2} + 
160520{x^3}{y^2} \cr
&-& 
     247264t{x^3}{y^2} - 134456{t^2}{x^3}{y^2} + 
7688{t^3}{x^3}{y^2} + 
     4056{t^4}{x^3}{y^2} + 592{t^5}{x^3}{y^2} \cr
&-& 
247264w{x^3}{y^2} - 
     128736tw{x^3}{y^2} + 17288{t^2}w{x^3}{y^2} + 
     2496{t^3}w{x^3}{y^2} \cr
&-& 1856{t^4}w{x^3}{y^2} - 
     134456{w^2}{x^3}{y^2} + 17288t{w^2}{x^3}{y^2} + 
     10224{t^2}{w^2}{x^3}{y^2} \cr
&-& 
9680{t^3}{w^2}{x^3}{y^2} + 
     7688{w^3}{x^3}{y^2} + 2496t{w^3}{x^3}{y^2} - 
     9680{t^2}{w^3}{x^3}{y^2} \cr
&+& 4056{w^4}{x^3}{y^2} - 
     1856t{w^4}{x^3}{y^2} + 592{w^5}{x^3}{y^2} \cr
&-& 
94964{x^4}{y^2} - 
     68020t{x^4}{y^2} - 1458{t^2}{x^4}{y^2} + 
4056{t^3}{x^4}{y^2} - 
     56{t^4}{x^4}{y^2} \cr
&-& 68020w{x^4}{y^2} - 
11588tw{x^4}{y^2} + 
     4136{t^2}w{x^4}{y^2} - 1856{t^3}w{x^4}{y^2} - 
     1458{w^2}{x^4}{y^2} \cr
&+& 4136t{w^2}{x^4}{y^2} - 
     3408{t^2}{w^2}{x^4}{y^2} + 4056{w^3}{x^4}{y^2} - 
     1856t{w^3}{x^4}{y^2} - 56{w^4}{x^4}{y^2} \cr
&-& 
31788{x^5}{y^2} - 
     2724t{x^5}{y^2} + 4296{t^2}{x^5}{y^2} + 
592{t^3}{x^5}{y^2} - 
     2724w{x^5}{y^2} \cr
&+& 2432tw{x^5}{y^2} + 
1112{t^2}w{x^5}{y^2} + 
     4296{w^2}{x^5}{y^2} + 1112t{w^2}{x^5}{y^2} + 
592{w^3}{x^5}{y^2} \cr
&+& 
     1416{x^6}{y^2} + 1792t{x^6}{y^2} + 
848{t^2}{x^6}{y^2} + 
     1792w{x^6}{y^2} + 688tw{x^6}{y^2} \cr
&+& 
848{w^2}{x^6}{y^2} + 
     1136{x^7}{y^2} + 272t{x^7}{y^2} + 272w{x^7}{y^2} 
+ 112{x^8}{y^2} \cr
&-& 
     71028{y^3} - 120628t{y^3} + 160520{t^2}{y^3} - 
11080{t^3}{y^3} - 
     46764{t^4}{y^3} \cr
&-& 3076{t^5}{y^3} + 2640{t^6}{y^3} + 
336{t^7}{y^3} - 
     120628w{y^3} + 370976tw{y^3} \cr
&-& 247264{t^2}w{y^3} - 
     124960{t^3}w{y^3} + 6940{t^4}w{y^3} + 4000{t^5}w{y^3} + 
     352{t^6}w{y^3} \cr
&+& 160520{w^2}{y^3} - 
247264t{w^2}{y^3} - 
     134456{t^2}{w^2}{y^3} + 7688{t^3}{w^2}{y^3} + 
4056{t^4}{w^2}{y^3} \cr
&+& 
     592{t^5}{w^2}{y^3} - 11080{w^3}{y^3} - 124960t{w^3}{y^3} + 
     7688{t^2}{w^3}{y^3} + 6720{t^3}{w^3}{y^3} \cr
&-& 
1096{t^4}{w^3}{y^3} - 
     46764{w^4}{y^3} + 6940t{w^4}{y^3} + 
4056{t^2}{w^4}{y^3} - 
     1096{t^3}{w^4}{y^3} \cr
&-& 3076{w^5}{y^3} + 
4000t{w^5}{y^3} + 
     592{t^2}{w^5}{y^3} + 2640{w^6}{y^3} + 
352t{w^6}{y^3} \cr
&+& 
     336{w^7}{y^3} - 120628x{y^3} + 370976tx{y^3} - 
247264{t^2}x{y^3} - 
     124960{t^3}x{y^3} \cr
&+& 6940{t^4}x{y^3} + 4000{t^5}x{y^3} + 
     352{t^6}x{y^3} + 370976wx{y^3} - 
749568twx{y^3} \cr
&-& 
     128736{t^2}wx{y^3} + 39680{t^3}wx{y^3} + 
896{t^4}wx{y^3} + 
     384{t^5}wx{y^3} - 247264{w^2}x{y^3} \cr
&-& 
128736t{w^2}x{y^3} + 
     17288{t^2}{w^2}x{y^3} + 2496{t^3}{w^2}x{y^3} - 
     1856{t^4}{w^2}x{y^3} - 124960{w^3}x{y^3} \cr
&+& 
39680t{w^3}x{y^3} + 
     2496{t^2}{w^3}x{y^3} - 6912{t^3}{w^3}x{y^3} + 
6940{w^4}x{y^3} + 
     896t{w^4}x{y^3} \cr
&-& 1856{t^2}{w^4}x{y^3} + 
4000{w^5}x{y^3} + 
     384t{w^5}x{y^3} + 352{w^6}x{y^3} + 
160520{x^2}{y^3} \cr
&-& 
     247264t{x^2}{y^3} - 134456{t^2}{x^2}{y^3} + 
7688{t^3}{x^2}{y^3} + 
     4056{t^4}{x^2}{y^3} + 592{t^5}{x^2}{y^3} \cr
&-& 
247264w{x^2}{y^3} - 
     128736tw{x^2}{y^3} + 17288{t^2}w{x^2}{y^3} + 
     2496{t^3}w{x^2}{y^3} \cr
&-& 1856{t^4}w{x^2}{y^3} - 
     134456{w^2}{x^2}{y^3} + 17288t{w^2}{x^2}{y^3} + 
     10224{t^2}{w^2}{x^2}{y^3} \cr
&-& 
9680{t^3}{w^2}{x^2}{y^3} + 
     7688{w^3}{x^2}{y^3} +2496t{w^3}{x^2}{y^3} - 
     9680{t^2}{w^3}{x^2}{y^3} \cr
&+& 4056{w^4}{x^2}{y^3} - 
     1856t{w^4}{x^2}{y^3} + 592{w^5}{x^2}{y^3} \cr
&-& 
11080{x^3}{y^3} - 
     124960t{x^3}{y^3} + 7688{t^2}{x^3}{y^3} + 6720{t^3}{x^3}{y^3} - 
     1096{t^4}{x^3}{y^3} \cr
&-& 124960w{x^3}{y^3} + 
39680tw{x^3}{y^3} + 
     2496{t^2}w{x^3}{y^3} - 6912{t^3}w{x^3}{y^3} + 
     7688{w^2}{x^3}{y^3} \cr
&+& 2496t{w^2}{x^3}{y^3} - 
     9680{t^2}{w^2}{x^3}{y^3} + 6720{w^3}{x^3}{y^3} - 
     6912t{w^3}{x^3}{y^3} - 1096{w^4}{x^3}{y^3} \cr
&-& 
46764{x^4}{y^3} + 
     6940t{x^4}{y^3} + 4056{t^2}{x^4}{y^3} - 
1096{t^3}{x^4}{y^3} + 
     6940w{x^4}{y^3} \cr
&+& 896tw{x^4}{y^3} - \
1856{t^2}w{x^4}{y^3} + 
     4056{w^2}{x^4}{y^3} - 1856t{w^2}{x^4}{y^3} - 
1096{w^3}{x^4}{y^3} \cr
&-& 
     3076{x^5}{y^3} + 4000t{x^5}{y^3} + 
592{t^2}{x^5}{y^3} + 
     4000w{x^5}{y^3} + 384tw{x^5}{y^3} \cr
&+& 
592{w^2}{x^5}{y^3} + 
     2640{x^6}{y^3} + 352t{x^6}{y^3} + 352w{x^6}{y^3} 
+ 336{x^7}{y^3} \cr
&-& 
     111795{y^4} + 177012t{y^4} - 94964{t^2}{y^4} - 
46764{t^3}{y^4} + 
     6638{t^4}{y^4} \cr
&+& 2920{t^5}{y^4} + 224{t^6}{y^4} + 
177012w{y^4} - 
     220300tw{y^4} - 68020{t^2}w{y^4} \cr
&+& 
6940{t^3}w{y^4} + 
    1672{t^4}w{y^4} + 268{t^5}w{y^4} - 
94964{w^2}{y^4} - 
     68020t{w^2}{y^4} \cr
&-& 1458{t^2}{w^2}{y^4} + 
4056{t^3}{w^2}{y^4} - 
     56{t^4}{w^2}{y^4} - 46764{w^3}{y^4} + 
6940t{w^3}{y^4} \cr
&+& 
     4056{t^2}{w^3}{y^4} - 1096{t^3}{w^3}{y^4} + 
6638{w^4}{y^4} + 
     1672t{w^4}{y^4} - 56{t^2}{w^4}{y^4} + 
2920{w^5}{y^4} \cr
&+& 
     268t{w^5}{y^4} + 224{w^6}{y^4} + 177012x{y^4} - 
220300tx{y^4} - 
     68020{t^2}x{y^4} \cr
&+& 6940{t^3}x{y^4} + 
1672{t^4}x{y^4} + 
     268{t^5}x{y^4} - 220300wx{y^4} - 
41136twx{y^4} \cr
&-& 
     11588{t^2}wx{y^4} + 896{t^3}wx{y^4} + 
608{t^4}wx{y^4} - 
     68020{w^2}x{y^4} - 11588t{w^2}x{y^4} \cr
&+& 4136{t^2}{w^2}x{y^4} - 
     1856{t^3}{w^2}x{y^4} + 6940{w^3}x{y^4} + 
896t{w^3}x{y^4} - 
     1856{t^2}{w^3}x{y^4} \cr
&+& 1672{w^4}x{y^4} + 
608t{w^4}x{y^4} + 
     268{w^5}x{y^4} - 94964{x^2}{y^4} - 
68020t{x^2}{y^4} \cr
&-& 
     1458{t^2}{x^2}{y^4} + 4056{t^3}{x^2}{y^4} - 56{t^4}{x^2}{y^4} - 
     68020w{x^2}{y^4} - 11588tw{x^2}{y^4} \cr
&+& 4136{t^2}w{x^2}{y^4} - 
     1856{t^3}w{x^2}{y^4} - 1458{w^2}{x^2}{y^4} + 
     4136t{w^2}{x^2}{y^4} - 3408{t^2}{w^2}{x^2}{y^4}\cr 
&+& 
     4056{w^3}{x^2}{y^4} - 1856t{w^3}{x^2}{y^4} - 
56{w^4}{x^2}{y^4} - 
     46764{x^3}{y^4} + 6940t{x^3}{y^4} \cr
&+& 
4056{t^2}{x^3}{y^4} - 
     1096{t^3}{x^3}{y^4} + 6940w{x^3}{y^4} + 
896tw{x^3}{y^4} - 
     1856{t^2}w{x^3}{y^4} \cr
&+& 4056{w^2}{x^3}{y^4} - 
     1856t{w^2}{x^3}{y^4} - 1096{w^3}{x^3}{y^4} + 
6638{x^4}{y^4} + 
   1672t{x^4}{y^4} \cr
&-& 56{t^2}{x^4}{y^4} + 1672w{x^4}{y^4} + 
     608tw{x^4}{y^4} - 56{w^2}{x^4}{y^4} + 
2920{x^5}{y^4} \cr
&+& 
     268t{x^5}{y^4} + 268w{x^5}{y^4} + 224{x^6}{y^4} 
+ 33444{y^5} - 
     17836t{y^5} \cr
&-& 31788{t^2}{y^5} - 3076{t^3}{y^5} + 
2920{t^4}{y^5} + 
     416{t^5}{y^5} - 17836w{y^5} \cr
&-& 57456tw{y^5} - 
2724{t^2}w{y^5} + 
     4000{t^3}w{y^5} + 268{t^4}w{y^5} - 
31788{w^2}{y^5} \cr
&-& 
     2724t{w^2}{y^5} + 4296{t^2}{w^2}{y^5} + 
592{t^3}{w^2}{y^5} - 
     3076{w^3}{y^5} + 4000t{w^3}{y^5} \cr
&+& 
592{t^2}{w^3}{y^5} + 
     2920{w^4}{y^5} + 268t{w^4}{y^5} + 416{w^5}{y^5} - 
17836x{y^5} \cr
&-& 
     57456tx{y^5} - 2724{t^2}x{y^5} + 
4000{t^3}x{y^5} + 
     268{t^4}x{y^5} - 57456wx{y^5} \cr
&+& 
4992twx{y^5} + 
     2432{t^2}wx{y^5} + 384{t^3}wx{y^5} - 
2724{w^2}x{y^5} + 
     2432t{w^2}x{y^5} \cr
&+& 1112{t^2}{w^2}x{y^5} + 
4000{w^3}x{y^5} + 
     384t{w^3}x{y^5} + 268{w^4}x{y^5} - 
31788{x^2}{y^5} \cr
&-& 
     2724t{x^2}{y^5} + 4296{t^2}{x^2}{y^5} + 
592{t^3}{x^2}{y^5} - 
     2724w{x^2}{y^5} + 2432tw{x^2}{y^5} \cr
&+& 
1112{t^2}w{x^2}{y^5} + 
     4296{w^2}{x^2}{y^5} + 1112t{w^2}{x^2}{y^5} + 
592{w^3}{x^2}{y^5} - 
     3076{x^3}{y^5} \cr
&+& 4000t{x^3}{y^5} + 
592{t^2}{x^3}{y^5} + 
     4000w{x^3}{y^5} + 384tw{x^3}{y^5} + 
592{w^2}{x^3}{y^5} \cr
&+& 
     2920{x^4}{y^5} + 268t{x^4}{y^5} + 268w{x^4}{y^5} 
+ 416{x^5}{y^5} + 
     8640{y^6} \cr
&-& 26900t{y^6} + 1416{t^2}{y^6} + 
2640{t^3}{y^6} + 
  224{t^4}{y^6} - 26900w{y^6} \cr
&+& 5828tw{y^6} + 
1792{t^2}w{y^6} + 
     352{t^3}w{y^6} + 1416{w^2}{y^6} + 
1792t{w^2}{y^6} \cr
&+& 
     848{t^2}{w^2}{y^6} + 2640{w^3}{y^6} + 
352t{w^3}{y^6} + 
     224{w^4}{y^6} - 26900x{y^6} \cr
&+& 5828tx{y^6} + 
1792{t^2}x{y^6} + 
     352{t^3}x{y^6} + 5828wx{y^6} - 
2112twx{y^6} \cr
&+& 
     688{t^2}wx{y^6} + 1792{w^2}x{y^6} + 688t{w^2}x{y^6} + 
     352{w^3}x{y^6} + 1416{x^2}{y^6} \cr
&+& 
1792t{x^2}{y^6} + 
     848{t^2}{x^2}{y^6} + 1792w{x^2}{y^6} + 
688tw{x^2}{y^6} + 
     848{w^2}{x^2}{y^6} \cr
&+& 2640{x^3}{y^6} + 
352t{x^3}{y^6} + 
     352w{x^3}{y^6} + 224{x^4}{y^6} - 5124{y^7} + 
132t{y^7} \cr
&+& 
     1136{t^2}{y^7} + 336{t^3}{y^7} + 132w{y^7} + 
1024tw{y^7} + 
     272{t^2}w{y^7} \cr
&+& 1136{w^2}{y^7} + 272t{w^2}{y^7} 
+ 336{w^3}{y^7} + 
     132x{y^7} + 1024tx{y^7} + 272{t^2}x{y^7} \cr
&+& 
1024wx{y^7} - 
     384twx{y^7} + 272{w^2}x{y^7} + 1136{x^2}{y^7} 
+ 272t{x^2}{y^7} \cr
&+& 
     272w{x^2}{y^7} + 336{x^3}{y^7} - 618{y^8} + 
800t{y^8} + 
     112{t^2}{y^8} + 800w{y^8} \cr
&-& 96tw{y^8} + 112{w^2}{y^8} + 
     800x{y^8} - 96tx{y^8} - 96wx{y^8} + 
112{x^2}{y^8} \cr
&+& 192{y^9} + 
     32t{y^9} + 32w{y^9} + 32x{y^9} + 24{y^{10}}
\biggr) .\nonumber\\
(S_{-1})^{(0)}_{I_1I_2I_3I_4}&=&
\frac{2}{\d}
f_5^{-1} a_{125}a_{345}(k_1-k_2)(k_3-k_4)(f_1-f_2)(f_3-f_4)\cr
&\times&
\biggl(-36+2(k_1+k_2+k_3+k_4)+f_1+f_2+f_3+f_4-
2k_1k_2-2k_3k_4\biggr) .\nonumber\\
(S_{t2})^{(0)}_{I_1I_2I_3I_4}&=&\frac{(f_5-1)^2t_{125}t_{345}}{2\d}(k_1-k_2)(k_3-k_4)(f_1+f_2+f_3+f_4 \nonumber\\
&+&2(k_1+k_2+k_3+k_4)-2(k_1k_2+k_3k_4)-36). \nonumber\\
(S_{p3})^{(0)}_{I_1I_2I_3I_4}&=&-\frac{1}{\d}f_5^3p_{125}p_{345}. \nonumber\\ 
(S_{p2})^{(0)}_{I_1I_2I_3I_4}&=&\frac{2}{\d}f_5^2p_{125}p_{345}
(k_1^2+k_2^2+k_3^2+k_4^2 -2(k_1+k_2+k_3+k_4)+2(k_1k_2+k_3k_4)-4). \nonumber\\ 
(S_d)^{(0)}_{I_1I_2I_3I_4}&=&
\frac{9a_{125}a_{345}}{64\d (f_5-5)}
( -1 + k_1 - k_2 )(1+k_1-k_2)(3+k_1+k_2)(5+k_1+k_2)\nonumber \\
&\times & ( -1 + k_3 - k_4 )(1+k_3-k_4)(3+k_3+k_4)(5+k_3+k_4)\nonumber\\
&\times & (2k_1^2+2k_2^2+2k_3^2+2k_4^2+4k_1k_2+4k_3k_4-4(k_1+k_2+k_3+k_4)-5).\nonumber
\eea
\section{Appendix B}
\setcounter{equation}{0}
We need a number of relations between the structure constants $a_{123}$, 
$t_{123}$ and $p_{123}$, defined as\footnote{For a detailed description 
of spherical harmonics on $S^5$ see \cite{LMRS, AF5}.} 
\bea
\la{a123}
&&a_{123}=\int~ Y^{I_1}Y^{I_2}Y^{I_3}
\\\la{t123}
&&t_{123}=\int~ \n^\a Y^{I_1}Y^{I_2}Y_\a^{I_3}\\
&&p_{123}=\int~ \n^\a Y^{I_1}\n^\b Y^{I_2}Y_{(\a\b )}^{I_3}\la{p123}
\eea
In deriving the equations of motions for scalar fields $s_k$ and $t_k$
and for tensor $\varphi_{ab}^k$ one comes across a number 
of integrals of scalar spherical harmonics, all of them can be reduced 
to $a_{123}$. Introducing the concise notation $f_i=f(k_i)=k_i(k_i+4)$
we present below the corresponding formulae:
\bea
\nonumber
b_{123}&=&\int \n^{\a}Y^{I_1}\n_{\a}Y^{I_2}Y^{I_3}
=\frac{1}{2}(f_1+f_2-f_3)a_{123}, \\
\nonumber
c_{123}&=&\int \n^{\a}\n^{\b}Y^{I_1}\n_{\a}\n_{\b}Y^{I_2}Y^{I_3}=
\frac{1}{2}\l -f_1f_3-f_2f_3+\frac{3}{5}f_1f_2
+\frac{1}{2}f_1^2 \right. \\
\nonumber
\qquad\qquad&+&\left.\frac{1}{2}f_2^2+\frac{1}{2}f_3^2 
-4(f_1+f_2-f_3)\r a_{123}.
\eea
Since any scalar function on a sphere can be decomposed in scalar 
spherical harmonics, we have the following relations
\bea
Y^1Y^2=a_{123}Y^3,\quad \n^\a Y^1\n_\a Y^2=b_{123}Y^3,\quad
\n^\a\n^\b Y^1\n_\a\n_\b Y^2=c_{123}Y^3.
\la{rel1}
\eea
These relations allow one to calculate some integrals involving 4 scalar
spherical harmonics. In particular we have
\bea
a_{1234}&=&\int~ Y^{I_1}Y^{I_2}Y^{I_3}Y^{I_4}=
a_{125}a_{345}=a_{135}a_{245}=a_{145}a_{235},\nonumber\\
b_{1234}&=&\int~ Y^{I_1}Y^{I_2}\n^\a Y^{I_3}\n_\a Y^{I_4}=
a_{125}b_{345},\nonumber\\
c_{1234}&=&\int~ Y^{I_1}Y^{I_2}\n^\a\n^\b Y^{I_3}\n_\a\n_\b Y^{I_4}=
a_{125}c_{345},\nonumber
\eea
where the summation over the index 5 is assumed.

There is the following important relation 
\bea
f_5(a_{125}a_{345}+a_{135}a_{245}+a_{235}a_{145})=(f_1+f_2+f_3+f_4)
a_{125}a_{345},
\la{imre}
\eea
that shows that among the three tensors
$f_5a_{125}a_{345}$, $f_5a_{135}a_{245}$ and $f_5a_{235}a_{145}$
only the following two tensors are independent
$$f_5a_{125}a_{345} \quad \mbox{and} \quad 
f_5(a_{135}a_{245}-a_{235}a_{145}).$$

We also encounter tensors of the form $f_5^nt_{125}t_{345}$ and
$f_5^np_{125}p_{345}$. Some of them can be reduced to sums of tensors of
the form $f_5^na_{125}a_{345}$.
To see this we note that any vector and any traceless symmetric tensor 
on a sphere can be decomposed as follows
\bea
&&\n_{\a} Y^1 Y^2=t_{125}Y_{\a}^5+\frac{b_{152}}{f_5}\n_{\a }Y^5 ,
\la{vecdec}\\
&&\n_{(\a} Y^1\n_{\b )} Y^2=p_{125}Y_{(\a\b )}^5+
\mu_{125}\n_{(\a }Y_{\b )}^5 +\nu_{125}\n_{(\a }\n_{\b )}Y^5 ,
\la{tendec}
\eea
where
\bea
&&\mu_{125}=\frac{f_2-f_1}{f_5-5}t_{125},
\quad\nu_{125}=\frac{1}{q_5} d_{125} ,\nonumber\\
&&q_5=\frac45 f_5(f_5-5),\quad
d_{125}= c_{125}+(\frac15 f_5-f_2+4) b_{125}
+f_1b_{251}.
\nonumber
\eea
Note that in the paper we use the following normalizations
$$
\int~Y_{(\a\b )}^IY^{(\a\b )}_J=\d^I_J,\quad
\int~Y_{\a }^IY^{\a}_J=\d^I_J,
$$
where summation over $\a,\b$ is assumed.

By using the decompositions one can find the following relations
\bea
&&t_{125}t_{345}=-\frac{(f_1-f_2)(f_3-f_4)}{4f_5}a_{125}a_{345}
+\frac{1}{4}f_5(a_{145}a_{235}-a_{245}a_{135}),
\la{TT}\\
&&(1-f_5)t_{125}t_{345}=\frac{1}{4}(f_5^2-f_5(f_1+f_2+f_3+f_4-4))
(a_{145}a_{235}-a_{135}a_{245})\la{TTF}\\
&&\qquad-\frac{4-f_5}{4f_5}(f_1-f_2)(f_3-f_4)a_{125}a_{345}\nonumber\\
&&p_{125}p_{345}=
-\frac{(f_1-f_2)(f_3-f_4)}{2(f_5-5)}t_{125}t_{345}
-\frac{1}{q_5}d_{125}d_{345}\nonumber\\ 
&&\qquad-\frac{1}{20}(f_1+f_2-f_5)(f_3+f_4-f_5)a_{125}a_{345}
+\frac{1}{8}(f_1+f_3-f_5)(f_2+f_4-f_5)a_{135}a_{245}
\nonumber\\
\la{PP}
&&\qquad+\frac{1}{8}(f_1+f_4-f_5)(f_2+f_3-f_5)a_{145}a_{235},\\
&&p_{125}(2-f_5)p_{345}=\int \n_{\g}^2(\n^{\a}Y^1\n^{\b}Y^2)
\n^{(\a}Y^3\n^{\b )}Y^4
-\frac{10-f_5}{q_5}d_{125}d_{345} \nonumber\\
&&\qquad-\frac{(f_1-f_2)(f_3-f_4)}{(f_5-5)}t_{125}t_{345}
+\frac{1}{2}(f_1-f_2)(f_3-f_4)t_{125}t_{345}, \la{PPF}
\eea
where 
\bea
&& \int \n_{\g}^2(\n^{\a}Y^1\n^{\b}Y^2)
\n^{(\a}Y^3\n^{\b )}Y^4=
\frac{1}{8}(8-f_1-f_2)(f_1+f_3-f_5)(f_2+f_4-f_5)a_{135}a_{245}\nonumber \\
\nonumber
&&\qquad+\frac{1}{8}(8-f_1-f_2)(f_1+f_4-f_5)(f_2+f_3-f_5)a_{145}a_{235}
\\
\nonumber
&&\qquad+\frac{1}{20}(f_1+f_2-f_5)(f_3+f_4-f_5)f_5a_{125}a_{345}
+g_{1324}+g_{1423}
\eea
and
\bea
\nonumber
&&g_{1234}=\frac{1}{16f_5}(f_1+f_5-f_2)(f_1+f_2-f_5)
(f_3+f_5-f_4)(f_3+f_4-f_5)a_{125}a_{345}\\
\nonumber
&&\qquad+\frac{1}{4}(f_1+f_2-4)(f_3+f_4-4)t_{125}t_{345}
+\frac{1}{4}(f_1+f_2+f_3+f_4-8)t_{125}(1-f_5)t_{345}\\\nonumber
&&\qquad+\frac{1}{4}t_{125}(1-f_5)^2t_{345}.
\eea
The following integrals are also used in deriving
the equations of motion for scalars $s_k$:
\bea
\nonumber
\int \n^{(\a}\n^{\b)}Y^{I_1}Y^{I_2}\n_{\a}Y_{\b}^{I_3}&=&
\frac{1}{2}\l f_3+f_1-f_2-5 \r t_{123}, \\
\nonumber
\int \n^{(\a}\n^{\b)}Y^{I_1}\n_{\a}Y^{I_2}Y_{\b}^{I_3}
&=&\frac{1}{2}\l f_2+\frac{3}{5}f_1 - f_3-3  \r t_{123},\\
\nonumber
\int \n^{(\a}\n^{\g)}Y^{I_1}\n^{\b}\n_{\g} Y^{I_2}Y_{(\a\b)}^{I_3}&=&
\frac{1}{10}(3f_1+5f_2-5f_3-30)p_{123}, \\
\nonumber
\int \n^{(\a}\n^{\b)}Y^{I_1}\n_{\g} Y^{I_2}\n^{\g}Y_{(\a\b)}^{I_3}&=&
\frac{1}{2}(f_1-f_2-f_3-8) p_{123}.
\eea

Considering the action for the fields from the massive graviton multiplet 
we need the following explicit expression for the integral $a_{123}$ of scalar
spherical harmonics \cite{LMRS,AF5}:
\bea
\la{aexp}
a_{123}&=&(z(k_1)z(k_2)z(k_3))^{-1/2}
\frac{\pi^3}{(\frac12\S +2)!2^{\frac12 (\S -2)}}
\frac{k_1!k_2!k_3!}{\a_1!\a_2!\a_3!}\langle C^{I_1}C^{I_2}C^{I_3}\rangle 
\eea
Here 
$$
z(k)=\frac{\pi^3}{2^{k-1}(k+1)(k+2)}
$$
and $\a_i =\frac 12 (k_j+k_l-k_i)$, $j\neq l\neq i$.
Notation $\langle C^{I_1}C^{I_2}C^{I_3}\rangle $ is used to denote 
the unique $SO(6)$ invariant obtained 
by contracting $\a_1$ indices between $C^{I_2}$ and  $C^{I_3}$, 
$\a_2$ indices between $C^{I_3}$ and  $C^{I_1}$, and
$\a_3$ indices between $C^{I_2}$ and  $C^{I_1}$.

\vskip 1cm
{\bf ACKNOWLEDGMENT}
We are grateful to Prof. S. Theisen and to S. Kuzenko
for numerous valuable discussions. We would like to thank A. Tseytlin 
for valuable comments. The work of G.A. was
supported by the Alexander von Humboldt Foundation and in part by the
RFBI grant N99-01-00166, and the work of S.F. was supported by
the U.S. Department of Energy under grant No. DE-FG02-96ER40967.
 
\end{document}